\documentclass[a4paper,fleqn,usenatbib]{mnras}

\usepackage[T1]{fontenc}
\usepackage{ae,aecompl}

\usepackage{graphicx}	% Including figure files
\usepackage{amsmath}	% Advanced maths commands
\usepackage{amssymb}	% Extra maths symbols
\newcommand{\um}{$\mu$m}
\newcommand{\kms}{km s$^{-1}$}
\newcommand{\lsun}{\mbox{L$_\odot$}}% Lsun
\newcommand{\msun}{\mbox{M$_\odot$}}% Msun
\newcommand{\lm}{\mbox{$L/M$}} % luminosity over mass ratio
\newcommand{\ee}[1]{\mbox{${} \times 10^{#1}$}}% scientific number format
\newcommand{\eten}[1]{\mbox{$10^{#1}$}}% power of ten

\newcommand\arcdeg{\mbox{$^\circ$}}%

\newcommand{\cmv}{\mbox{cm$^{-3}$}}
\newcommand{\cmc}{\mbox{cm$^{-2}$}}
\newcommand{\tk}{\mbox{$T_{\textit{kin}}$}}
\newcommand{\td}{\mbox{$T_{\textit{dust}}$}}

\newcommand{\jyb}{\mbox{Jy beam$^{-1}$}}

\newcommand{\hii}{\mbox{\ion{H}{ii}}}

\newcommand{\ammonia}{\mbox{{\rm NH}$_3$}}
\newcommand{\hg}{\mbox{Hi-GAL}}
\newcommand{\water}{\mbox{H$_2$O}}
\newcommand{\chiammonia}{\mbox{$\chi$(NH$_3$)}}

\newcommand{\eff}{Effelsberg}
\newcommand{\goodsample}{final sample}

\newcommand{\arcsecword}{~arcsec}
\newcommand{\arcminword}{~arcmin}
\newcommand{\percentword}{~per~cent}

\newcommand{\revision}[1]{{#1}}

\title[Dust and gas properties of dense Hi-GAL clumps]{Thermal balance and comparison of gas and dust properties of dense clumps in the Hi-GAL survey}
\author[M. Merello et al.]{M. Merello$^{1,2}$\thanks{E-mail: manuel.merello@iaps.inaf.it}, 
S. Molinari$^{1}$, 
K. L. J. Rygl$^{3}$, 
N. J. Evans II$^{4,5,6}$,
D. Elia$^{1}$, 
\newauthor
E. Schisano$^{1}$,
A. Traficante$^{1}$, 
Y. Shirley$^{7,8}$, 
B. Svoboda$^{7}$, 
P. F. Goldsmith$^{9}$
\\
$^{1}$ INAF-Istituto di Astrofisica e Planetologia Spaziale, Via Fosso del Cavaliere 100, I-00133 Roma, Italy\\
$^{2}$ Universidade de S\~ ao Paulo, IAG Rua do Mat\~ ao, 1226, Cidade Universit\'aria, 05508-090, S\~ ao Paulo, Brazil\\
$^{3}$ Italian ALMA Regional Centre, INAF-IRA, Via P. Gobetti 101, 40129 Bologna, Italy\\
$^{4}$ Department of Astronomy, The University of Texas at Austin,
2515 Speedway, Stop C1400, Austin, Texas 78712-1205, U.S.A.\\
$^{5}$ Korea Astronomy and Space Science Institute, 776 Daedeokdaero, Daejeon 305-348, Korea\\
$^{6}$ Humanitas College, Global Campus, Kyung Hee University, Yongin-shi 17104, Korea\\
$^{7}$ Steward Observatory, University of Arizona, 933 North Cherry Avenue, Tucson, AZ 85721, USA\\
$^{8}$ Adjunct Astronomer, National Radio Astronomy Observatory, USA\\
$^{9}$ Jet Propulsion Laboratory, California Institute of Technology, 4800 Oak Grove Drive, Pasadena CA 91109, USA
}

\date{Accepted XXX. Received YYY; in original form ZZZ}

\pubyear{2018}

\begin{document}
\label{firstpage}
\pagerange{\pageref{firstpage}--\pageref{lastpage}}
\maketitle

\begin{abstract}

We present a comparative study of physical properties derived from gas and dust emission in a sample of 1068 dense Galactic clumps. The sources are selected from the crossmatch of the \hg\ survey with 16 catalogues of \ammonia\ line emission in its lowest inversion (1,1) and (2,2) transitions. The sample covers a large range in masses and bolometric luminosities, with surface densities above $\Sigma=0.1$~g~\cmc\ and with low virial parameters $\alpha<1$. The comparison between dust and gas properties shows an overall agreement between \tk\ and \td\ at volumetric densities $n\gtrsim1.2\times10^{4}$~\cmv, and a median fractional abundance \chiammonia$=1.46\times10^{-8}$. 
\revision{While the protostellar clumps in the sample have small differences between \tk\ and \td, 
prestellar clumps have a median ratio \tk/\td\ $=1.24$, suggesting that these sources are thermally decoupled.
}
A correlation is found between the evolutionary tracer \lm\ and the parameters \tk/\td\ and \chiammonia\ in prestellar sources and protostellar clumps with \lm$<1$~\lsun\,\msun$^{-1}$. In addition, a weak correlation is found between non-thermal velocity dispersion and the \lm\ parameter, possibly indicating an increase of turbulence with protostellar evolution in the interior of clumps. Finally, different processes are discussed to explain the differences between gas and dust temperatures in prestellar candidates, and the origin of non-thermal motions observed in the clumps.

\end{abstract}

\begin{keywords}
ISM: clouds -- ISM: evolution -- stars: formation -- stars: protostars -- catalogues
\end{keywords}

%%%%%%%%%%%%%%%%%%%%%%%%%%%%%%%%%%%%%%%%%%%%%%%%%%

%%%%%%%%%%%%%%%%% BODY OF PAPER %%%%%%%%%%%%%%%%%%

%%---------------------------
%% INTRODUCTION
%%---------------------------

\section{Introduction}

High-mass stars ($M\ge8$~\msun) play a key role in the evolution of the interstellar medium. They are the principal source of heavy elements and UV radiation, affecting the process of formation of stars and planets, and the physical, chemical, and morphological structure of galaxies~\citep{ken12}. In particular, high-mass stars can shape whole galaxies through supernova explosions~\citep{bol13}. Despite their importance, their formation process is still under debate, contrary to the successful model currently accepted for low-mass stars (i.e., inside-out collapse of gravitational bound gas in approximate hydrostatic equilibrium). In contrast, high-mass stars have very short Kelvin-Helmholtz timescales, meaning that they continue accreting significant mass even after reaching the main sequence.
This fact implies that feedback from the forming star itself could halt the infall process~\citep{wol87}. 
Furthermore, the outcomes in numerical simulations of high-mass star formation processes depend strongly in the appropriate set of initial conditions~\citep[e.g.,][]{gir11}. 
Consequently, it is essential to obtain an observational basis of the physical properties of high-mass star forming regions with a statistically robust sample of sources.

The primary sites for the formation of massive stars and stellar clusters are molecular clumps, dense ($\Sigma~\ge~0.1$~g~\cmc) and cold ($T\le30$~K) condensations of gas and dust with sizes between 0.1--few~pc~\citep{ber07}.
Significant progress has been made in our understanding of these high-mass star forming regions thanks to far-infrared and submillimeter Galactic plane surveys.

\revision{
Surveys of continuum emission have shown the distribution of dust across the Milky Way using ground-based facilities, identifying thousands of compact sources. 
Among them, the Bolocam Galactic Plane Survey~\citep[BGPS;][]{agu11,ros10,gin13} mapped the Northern Galactic plane in 1.1 mm continuum emission with a resolution of 33\arcsecword\ using the 10 m Caltech Submillimeter Observatory. In the Southern Galactic plane, the ATLASGAL survey~\citep{sch09} mapped the 0.87 mm emission at a resolution of 19\arcsecword\ using the APEX telescope. %Another of these surveys, the JCMT Plane %Survey~\citep{ede17}, observed six fieleds  .
In addition, sources in advanced stages of high-mass star formation have also been identified using infrared space telescopes. Using the MSX satellite (four bands between 8--21 \um), the RMS survey~\citep{lum13} has identified nearly 700 luminous sources in the young stellar objects
 and \hii\ region stage.
}

\revision{Line emission surveys have included large-scale mapping of both dense gas and tracers of high-mass star formation. Among these, the 
Galactic Ring Survey~\citep[GRS;][]{jac06} has observed the $^{13}$CO(1-0) emission in the range 18\arcdeg$<\ell<$55.7\arcdeg, $|b|<1$\arcdeg, while the SEDIGISM survey~\citep{sch17} has mapped the emission of $^{13}$CO(2-1) and C$^{18}$O(2-1) in the range -60\arcdeg$<\ell<$18\arcdeg, $|b|<0.5$\arcdeg.
Other surveys include the Methanol MultiBeam Survey~\citep{gre09}, observing the emission of the 6-GHz methanol maser emission, and the H$_{2}$O Southern Galactic Plane Survey~\citep[HOPS;][]{wal11, pur12} that has observed the 22 GHz H$_{2}$O maser emission and the ammonia (1,1) and (2,2) transitions, at a resolution of 2 arcmin, over 100 square degrees using the 22 m Mopra antenna. Line surveys and large-scale Galactic continuum surveys complement each other, revealing the properties of gas and dust emission toward star forming regions.
}

The \textit{Herschel} Infrared Galactic Plane Survey~\citep[\hg,][]{mol10} is an Open Time Key Project for the \textit{Herschel Space Observatory}~\citep{pil10}, that has as objective to deliver a complete and unbiased view of the continuum emission in the Galactic plane in five bands: 70 and 160~\um\ using PACS instrument~\citep{pog10}, and 250, 350 and 500~\um\ with SPIRE~\citep{gri10}. This range of wavelengths covers the peak of emission of the spectral energy distribution (SED) of the cold dust emission (T$<50$~K). The \hg\ survey provides a unique view of the diffuse emission across the Galactic plane and the most complete sample of dense filamentary features and compact structures, where star formation is taking place. The first public release of the high-quality products from the \hg\ survey has been presented by~\cite{mol16a}, in which $\sim\eten{5}$ compact sources are identified in each band.

This large number of compact sources allows us to study the formation of high-mass stars at different stages with statistical significance, and the derivation of physical properties from dust emission can be compared and contrasted with properties derived from molecular line observations from different surveys and ground-based facilities. 

Ammonia (\ammonia) emission is a common estimator of temperature in dense regions in molecular clouds. The lowest inversion transitions $(J,K)=(1,1)$ and $(2,2)$ at $\sim$23.7 GHz are sensitive to cold and dense gas, and the ratio of these emissions is commonly used to estimate rotational and kinetic temperatures of the gas~\citep{ho83}. The effective density of the lowest pure inversion transition is $n_{\textit{eff}}\sim\eten{3}$~\cmv~\citep{eva99,shi15}.

In regions of star formation, stellar photons are not well coupled with gas molecules, hence the dust particles are heated by these photons and then the gas is heated by collisions with the dust~\citep{gol74,eva99}. It is expected that in dense regions ($n$~$>$~$10^4$~\cmv) the gas and dust temperatures are coupled~\citep[e.g.,][]{gol78,gol01,cri10}. Nevertheless, observational work addressing the dust-gas temperature coupling relies on a small sample of objects~\citep[e.g.,][]{gia13,mer15}, or is done in particular regions or set of sources without considering different environments~\citep[e.g.,][]{mor10}. 

Recent studies have compared properties of \ammonia\ line observations and far-infrared/submm continuum emission in Galactic clumps at early stages of evolution~\citep[e.g.,][]{bat14, guz15,svo16}.
Both gas and dust observations provide features used as evidence of the early stages of dense, compact sources, and in this sense different parameters derived from SED fitting using \textit{Herschel} bands have been suggested as indicators of the evolutionary state on massive clumps: the ratio between the luminosity and mass (\lm, in units of [\lsun\,\msun$^{-1}$]), the ratio $L_{\textit{smm}}/L_{\textit{bol}}$ and the bolometric temperature~\citep[see e.g.,][and references therein]{eli16b}. 

The comparison between the diagnostics derived from dust and gas emission, in a common set of sources with a statistically significant number, will allow refinement of conventional evolutionary indicators and our overall picture of early stages of high-mass star formation. 
In the present work, we are interested in the following:
\begin{itemize}
\item The compatibility between temperature estimates from dust continuum emission derived from the Hi-GAL survey, and gas kinetic temperatures derived from large surveys of \ammonia. 
\item The relation between physical properties derived independently from continuum and molecular line emission. In particular, we aim to investigate the possible relation between the \lm\ parameter and proposed evolutionary indicators derived from \ammonia\ observations, such as linewidths and the kinetic temperature.
\item The virial stability of clumps, and the possible origin of non-thermal motions observed in these sources.
\end{itemize}

This work is organised as follows: 
in Section~\ref{sec:surveys}, we present the selection of \ammonia\ surveys gathered from literature, along with a description of the \hg\ catalogue of physical properties, and the crossmatching and selection criteria of sources.
In Section~\ref{sec:results}, we compare physical properties derived from gas and dust emission, address the comparison between \ammonia\ kinetic temperature and dust temperature.
In Section~\ref{sec:discussion}, we correlate gas and dust properties with indicators of evolutionary stages, discuss the thermal coupling between gas and dust, examine the stability of clumps in the sample and possible scenarios for the observed non-thermal motions, and address the distribution of \ammonia\ fractional abundance with Galactocentric distances.
Finally, Section~\ref{sec:conclusions} presents the conclusions and a summary of the most important results of our study.

In a companion article, we used the well-characterised sample presented here to establish a evolutionary classification of sources from the \hg\ survey using different parameters derived from gas and dust emission. In addition, we derived probability distributions with respect to the \lm\ parameter and timescales estimates for the different evolutionary stages of high-mass star forming clumps.

%%---------------------------
%% SURVEYS
%%---------------------------
\section{Surveys}
\label{sec:surveys}
\subsection{NH$_3$ catalogues}

%:Catalogs
\begin{table*}
\caption{\ammonia\ catalogues}             % title of Table
%\label{table:catalogslist}      % is used to refer this table in the text
\centering                          % used for centering table
\begin{tabular}{ccccccl}        % centered columns (4 columns)
\hline\hline                 % inserts double horizontal lines
Index & Type & Telescope & $\theta_{\textrm{FWHM}}$ &\ammonia\ sources&\hg\ clumps& References\\
& & & (arcsec)&\#&\#\\
\hline                        % inserts single horizontal line
1 & BGPS clumps & GBT & 30 & 1398 & 1469 & \cite{dun11},\\
   &&&&&&\cite{svo16} \\
2 & ATLASGAL clumps & Eff & 40 & 694 & 812 &  \cite{wie12} \\
3 & EGOs & NRO45 & 73 & 58 & 102 &  \cite{cyg13} \\ 
4 & HMPOs & Eff & 40 & 34 & 37 &  \cite{sri02} \\
5 & RMS sources& GBT & 30 & 274 & 286 &  \cite{urq11} \\
6 & UC\hii\ reg./precursors cand. & Eff & 40 & 28 & 31 &  \cite{mol96} \\
7 & Massive clumps & ATCA & ~20$^{\dagger}$ & 28 & 31 &  \cite{gia13} \\
8 & Bright rimmed clouds assoc. & GBT & 30 & 8 & 8 & \cite{mor10} \\
9 & High contrast IRDCs & Eff & 40 & 61 & 72 & \cite{chi13} \\
10 & Young MSFR & Par & 58 & 83 & 122 &  \cite{hil10} \\
11 & IRDCs & Eff & 40 & 15 & 19 & \cite{pil06} \\
12 & Dense structures in W3 & GBT & 30 & 31 & 33 & \cite{mor14} \\
13 & \water\ maser associations & Eff & 40 & 9 & 12 & \cite{wu06} \\
14 & High IR extinction clouds & Eff & 40 & 39 & 41 & \cite{ryg10}\\
15 & UC\hii\ regions cand. & Eff & 40 & 23 & 31 & \cite{chu90}\\
16 & Luminous IRAS sources & Eff & 40 & 9 & 12 & \cite{sch96}\\
\hline
  & One-beam sample & & & & 2277 & \\
%  &  Reduced sample & & & & 1514 & \\ 
%  &  Reduced sample & & & & 1586 & \\   %if no further constraints are taken
\hline                                   %inserts single line
\end{tabular}
%{\footnotesize 
Observatories: GBT: Robert F. Byrd Green Bank Telescope; Eff: Effelsberg-100m telescope; NRO45: Nobeyama Radio Observatory 45m telescope; ATCA: Australia Telescope Compact Array; Par: Australia Telescope National Facility Parkes radio telescope. 

$^{\dagger}$Primary beam of $\sim2.5$~arcmin.
% }
%\tablefoottext{a}{Primary beam of $\sim2$\farcm5. }
\label{tbl:catalogs}
\end{table*}

We collected 16 catalogues of \ammonia\ observations toward dense, clump-type star forming sources across the Galaxy. The two largest catalogues correspond to the ammonia surveys targeting sources from the Bolocam Galactic Plane Survey~\cite[BGPS;][]{dun11,svo16}, and sources from the ATLASGAL survey~\citep{wie12}. These ammonia surveys observed dense condensations identified in ground-based observations of dust continuum emission at (sub)millimeter wavelengths across the Galactic Plane in the northern sky, and they cover a large range of evolutionary stages across the clumps~\citep{svo16}.

The rest of the catalogues are usually smaller samples targeted toward regions with signposts of massive star formation, covering different stages. From high extinction clouds and obscured features at mid-IR wavelengths~\citep{pil06,ryg10,chi13}, regions hosting high-mass protostellar objects~\citep{sri02,hil10} or with signposts of high-mass stellar activity such as outflows~\citep{cyg13} and \water\ masers~\citep{wu06}, up to the very late stages of massive star formation corresponding to the detection of ultra compact \hii\ (UC\hii) regions~\citep{urq11,mol96,chu90}.

We note that a catalogue of \ammonia\ observations of the (1,1) to (3,3) inversion transitions of 354 dense ATLASGAL sources between 300\arcdeg~$<\ell<$359~\arcdeg\ using the Parkes 64m telescope, has been recently released by~\cite{wie18}. Although that catalogue is not considered in our analysis since it was published after the present work was completed, the IV Galactic Quadrant is not well covered by the \ammonia\ surveys previously described and therefore the association between \hg\ sources and those line observations could lead to complementary results to those presented in this work.

Table~\ref{tbl:catalogs} presents the list of catalogues used in the comparison with clumps from the \hg\ survey. We will refer to each catalogue as \textit{Cat-\#}, with \textit{\#} the index number shown in that table. A description of the catalogues is presented in Appendix~\ref{sec:catalogs}.

The physical parameters from the ammonia emission, such as the rotational temperature ($T_{\textit{rot}}$), the kinetic temperature (\tk), and the ammonia column density ($N$(\ammonia)), were derived in their respective catalogue following the standard formulation for ammonia \citep{ho83,ung86}.
We considered for each sample the sources detected in both (1,1) and (2,2) inversion transitions, and therefore the quoted values for kinetic temperatures should not represent upper limits. 

We took the values of physical quantities as given in each survey of Table~\ref{tbl:catalogs}, without corrections from our part. Of course, this can produce inconsistencies in the analysis due to the different instruments involved in the ammonia detections, along with possible differences in technique and precision of analysis for each survey. Nevertheless, for most of the sample the resolutions of the instruments are comparable, and the main \ammonia\ associations for our clumps, which are the line observations used for characterisation of gas properties, are obtained from three main catalogues (see section~\ref{sec:crossmatch}): \textit{Cat-1} (BGPS sources), \textit{Cat-2} (ATLASGAL sources), and \textit{Cat-5} (RMS sources).
For those surveys where only the rotational temperatures $T_{\textit{rot}}$ are quoted~\citep[e.g.,][]{sri02}, we estimated the kinetic temperature $T_{kin}$ using the approximation of~\cite{taf04}:
\begin{equation}
T_{kin} = \frac{T_{rot}}{1-\frac{T_{rot}}{T_0}\ \textrm{ln}\left[1+1.1\times \textrm{exp}\left(\frac{-15.7}{T_{rot}}\right)\right]}\ \ \ ,
\label{eq:tkin_trot}
\end{equation}

\noindent where $T_0=\frac{E_{(2,2)}-E_{(1,1)}}{k_{\textrm{B}}} \approx41.5$~K~\citep{ho83}.\\

Excepting the sample of~\cite{sri02}, all catalogues present column densities derived for \ammonia\ from their fit. We also use, when it is quoted, the respective measured error in the estimation of \tk\ and $N$(\ammonia).

\subsection{\hg\ catalogue of physical properties of Galactic clumps}
\label{sec:cat_hgprops}

The characteristics of dust emission were obtained from the \hg\ survey. Sources in the inner Galaxy (-71{\arcdeg} $\lesssim \ell \lesssim$ 67{\arcdeg}) are described by the \hg\ photometric catalogue~\citep{mol16a} and catalogue of physical properties \citep{eli16}. For sources in the outer Galaxy (67{\arcdeg} $\lesssim \ell \lesssim$ 289{\arcdeg}), the photometric catalogue and physical properties of clumps in that region are described by~Merello et al. (2018, in prep.). We direct the reader to those surveys for details in the bandmerging and derivation of physical parameters of dust emission from \hg\ bands at 70--500~\um\ bands.

Physical properties are derived from the SED fitting from a modified blackbody:
\begin{equation}
\label{eq:greybody}
F_\nu=\left( 1-e^{-\tau_\nu} \right)B_\nu(T_{\textit{dust}})\Omega\ ,
\end{equation}

\noindent where $B_\nu(T_{\textit{dust}})$ is the Planck function at the dust temperature \td, $\Omega$ is the source solid angle, $\tau_\nu$ is the optical depth, considered as $\tau_\nu=(\nu/\nu_0)^\beta$ with $\beta$ the spectral index and $\nu_0$ the frequency at which the optical depth is equal to 1. The solid angle is considered as the source size measured in the 250~\um\ band. The spectral index has been fixed to $\beta=2.0$, and then the SED fitting is performed by $\chi^2$ optimisation, varying the dust temperature in the range $5<$~\td~$<40$~K.

Finally, \hg\ sources are assigned with counterpart emissions at mid-IR (21--24~\um) and sub-mm/mm (870 and 1100~\um) wavelengths, if available. The mid-IR bands are not considered in the modified blackbody fitting, since the mid-IR fluxes are associated to the emission of a different and hotter dust, although they are used in protostellar sources for the estimation of bolometric luminosities.

The total \hg\ catalogue of physical properties consists of $\sim$150000 clumps distributed along the Galactic plane. The sources are classified as protostellar and starless candidates\footnote{
	We stress that even though in this work we use the terms ``starless" and ``prestellar" sources, these should be taken always as candidates.
	Some studies have proven the detection of protostellar activity in regions previously considered starless or progenitors of clusters~\citep[e.g.,][]{gin12,lon17}.
	The identification of clumps without stellar activity based only on the lack of 70~\um\ emission has shown its limitations~\citep{tra17a}, and it presents caveats in the completeness of the 70~\um\ detections at large Heliocentric distances~\citep{bal17}.
	},
according to their detection/non-detection of a counterpart at 70~\um, respectively~\citep{dun08,eli13}.
Starless objects are considered ``prestellar" sources if they satisfy the mass-radius requirement to be gravitationally bound~\citep{lar81}, and therefore they are candidates for future star formation. In addition, we singled out prestellar sources only detected in the 250-350-500~\um\ bands (``SPIRE-only" sources), and we caution about their temperature determination due to possible difficulties to constrain their SED peaks.  
The ratio between protostellar, reliable prestellar and SPIRE-only prestellar is 1.00:1.30:1.04 in the inner Galaxy.

\subsection{Crossmatching and sample selection}
\label{sec:crossmatch}

Initially, the associations between \hg\ clumps and \ammonia\ sources were obtained by requiring an angular distance of less than or equal to one ammonia beam between the observed position of line emission and the peak position of the dust feature, which is extracted from the shortest wavelength counterpart of each clump (70~\um\ for protostellar sources, 160~\um\ or 250~\um\ for prestellar sources). Table~\ref{tbl:catalogs} indicates the FWHM beam size of each line catalogue used for this crossmatch (column 4), and the resulting number of \ammonia\ sources and \hg\ clumps associated (columns 5 and 6, respectively). While 1571 clumps are associated with a single detection of \ammonia, there are 72 clumps with another \ammonia\ detection from the same catalogue, and 634 clumps have been observed in two or more of these catalogues. 

The determination of the main \ammonia\ association for clumps with line detections in two or more different surveys is described in Appendix~\ref{sec:multi}, understanding as main association the \ammonia\ detection considered for the derivation of gas properties. Then, the one-beam sample consists of a total of 2277 \hg\ clumps with \ammonia\ counterparts. Most of the objects in this one-beam sample (76\percentword) are classified as protostellar.

%:Restrictions on sample selection 
\begin{table}
\caption{Final sample selection}             % title of Table
%\label{table:catalogslist}      % is used to refer this table in the text
\centering                          % used for centering table
\begin{tabular}{lc}        % centered columns (4 columns)
\hline\hline                 % inserts double horizontal lines
Criteria&\hg\ clumps\\
%\\ 
%Index & Type & Telescope\tablefootmark{*} & $\theta_{FWHM}$ & \ammonia\ sources& \hg\ clumps & References\\
% & & CS96\tablefootmark{*} & CS15\tablefootmark{*}\\
\hline                        % inserts single horizontal line
One-beam angular separation& 2277 \\
Half-beam angular separation& 1585 \\
\td: 5 $-$ 40 K & 1563 \\
\tk: 5 $-$ 41.5 K & 1521 \\
uncertainty in \tk\ $<$\ \tk & 1514 \\
$\Delta V$: 0.3 $-$ 10 \kms & 1513 \\
$\tau (1,1)$: 0.1 $-$ 10 & 1450 \\
Single association clump $-$ \ammonia & 1219\\
Fraction of clump area covered $\geq50$\% & 1071 \\
Distance: $< 18$ kpc & 1068 \\
\hline
{\bf Final sample} & {\bf 1068} \\ 
\hline                                   %inserts single line
\end{tabular}
%\tablefoot{}
%\tablefoottext{a}{Primary beam of $\sim2$\farcm5. }
\label{tbl:selection}
\end{table}

The one-beam sample give us the association between dust condensations of the \hg\ survey and molecular line observations of dense gas from the different catalogues of Table~\ref{tbl:catalogs}, and therefore it is useful to relate a clump with a specific type of source or signpost of star formation. Nevertheless, for a better comparison between gas and dust properties we applied additional restrictions summarised in Table~\ref{tbl:selection}. First, we limited the crossmatching distance to half the FWHM beam size of \ammonia\ detections, ensuring that the line emission characterises the inner part of each dust structure. We excluded sources out of the range of reliable dust temperature estimation ($5<$~\td~$<40$~K) due to poor or not proper SED fitting, and sources with unreliable low gas temperature (\tk$<5$~K) or above the uncertainty limit considered for \ammonia\ estimation of kinetic temperature (\tk$>41.5$~K, the temperature associated with the energy difference between (1,1) and (2,2) inversion levels). We note that although \textit{Cat-1} and \textit{Cat-5} include the (3,3) inversion transition and therefore clumps with this detection are sensitive in \tk\ up to $\sim$100~K, the previous constraint discards only three sources with legitimately high temperature (\tk=45--64~K) from the sample.

We also excluded clumps with uncertainties in the estimation of kinetic temperatures larger than the measurement of \tk\ itself. This last constraint results on the exclusion of most clumps with \tk$<8$~K, though most of them were associated with larger estimates of dust temperature. From the information on their respective source catalogues, we consider only \ammonia\ sources with FWHM velocity linewidth $\Delta V$ between 0.3 and 10 \kms, and with an optical depth value $0.1<\tau(1,1)<10.0$ to eliminate sources with marginal fitting of \ammonia\ physical parameters. There are 1450 clumps that satisfy the above restrictions.

We found a large number of clumps with angular sizes, obtained from the 250~\um\ band, smaller than the \ammonia\ beam sizes (see section~\ref{sec:clump_sizes}). It is possible that in the association between clump-line observation, the ammonia beam is contaminated by another clump, which could introduce differences in temperature between the \tk\ and \td. 
Then, in our analysis we consider only unique associations between clumps and \ammonia\ detections (single peak emission of dust source per ammonia detection), and we explore the effects on the relation between ammonia beam solid angles and clump areas.
For single associations, 520 clumps have more than 90\percentword\ of their area covered by the \ammonia\ FWHM beam, with most of these sources being protostellar sources (485, 93\percentword). For our sample, we will consider clumps with at least half their total area lying inside the solid angle of the ammonia beam.

Finally, we took into consideration only clumps with kinematic distances below 18 kpc (see section~\ref{sec:distances}) in the estimation of physical parameters, discarding three sources with large heliocentric ($>21$ kpc) and Galactocentric distances ($>15$ kpc). The restricted or {\textit \goodsample} of Hi-GAL clumps with ammonia counterparts that will be considered for further analysis through this paper consists of 1068 sources. This \goodsample\ contains 909 protostellar sources, 157 prestellar sources (of them, 19 are SPIRE-only prestellar), and 2 unbound starless objects.

\revision{It is important to have in mind that the sample of clumps obtained after the above restrictions does not represent the overall \hg\ catalogue of compact sources across the Galactic plane. First,~\cite{eli16} showed that the protostellar population represents only the 24.4 per cent of the total number of sources, while in our sample protostellar clumps represent more than 85 per cent of the total. Second, as we will show later in Section~\ref{sec:comparisonhg} when physical properties of this sample are estimated, the clumps in our sample are denser and more massive than most of the \hg\ sources.~\citeauthor{eli16} have also shown that even considering different bins in distance, the fraction of prestellar sources decreases in the upper tail of the mass distribution. A similar result has been reported by~\cite{urq18}, showing that in clumps of the ATLASGAL survey the proportion of prestellar to protostellar sources decreases as the clump mass increases.
}

\revision{The higher median mass found in our sample compared with the general population of \hg\ sources does not necessarily imply more advanced evolutionary stages~\citep[e.g.,][]{urq18}. The sample covers different stages of high-mass star formation, from prestellar stage and initial protostellar evolution, up to late stages such as UC\hii\ regions, as shown by different evolutionary tracers and by the association with signposts of early-star formation activity such as detection of H$_{2}$O or CH$_{3}$OH maser emission (Merello et al, in prep).}

We note that most of the clumps have \ammonia\ counterparts, or main associations in the case of clumps crossmatched with more than one ammonia detection, obtained from three main catalogues: \textit{Cat-1} (592 clumps), \textit{Cat-2} (209), and \textit{Cat-5} (187).

We also note that some studies have found discrepancies between the \ammonia\ peak emission observations and peaks of submillimeter emission. For example, \cite{mor14} showed that the peak position of their sample of submillimeter sources detected in 850~\um\ continuum emission is not always correlated with the peak position of their \ammonia\ maps, suggesting also that ammonia abundance is anti-correlated with column density traced in the submillimeter within individual cores, although these differences could be due to sensitivity and angular resolution limitations.   
Our selection of isolated clumps within the beam of \ammonia\ observations suggests that the molecular line emission and derived physical properties are related to the dense gas of the source, diminishing possible environmental contamination.

\subsection{Beam filling factors}
\label{sec:beamff}

The estimates of column densities presented in the \ammonia\ catalogues assumed that the observed sources filled completely the beam of the line emission (filling factor $\eta_{ff}=1$), although these catalogues also presented estimates of $\eta_{ff}$ from the relation $T_{ex}=T_{rot}\times \eta_{ff}$. The excitation temperature $T_{ex}$ is used in the derivation of gas column densities considering a local thermodynamic equilibrium analysis, and it is obtained from a single transition. 

In this work we assumed that both transitions of ammonia (1,1) and (2,2) originate in the same area described by the \hg\ clump,
which is in general a region smaller than the beamsizes of the \ammonia\ catalogues, hence we evaluate 
the fraction of the beam solid angle covered by the clump angular size.

%:Restrictions on sample selection 
\begin{table}
\caption{Fraction of \ammonia\ beam solid angle covered by clump angular size, and filling factors. The columns shows the number of sources and median values for each sample. In parenthesis the median absolute deviation of the respective group.}             % title of Table
%\label{table:catalogslist}      % is used to refer this table in the text
\centering                          % used for centering table
\begin{tabular}{lcccccc}        % centered columns (4 columns)
\hline\hline                 % inserts double horizontal lines
Clumps&\multicolumn{3}{c}{\textit{Cat-1+Cat-5} (30\arcsecword)}&\multicolumn{3}{c}{\textit{Cat-2} (40\arcsecword)}\\
&$N$&fraction&$\eta_{ff}$&$N$&fraction&$\eta_{ff}$\\
\hline                        % inserts single horizontal line
Sample&779&0.51(0.09)&0.15&209&0.33(0.07)&0.13\\
Prestellar&108&0.65(0.13)&0.12&31&0.48(0.12)&0.14\\
Protostellar&669&0.50(0.07)&0.15&178&0.32(0.06)&0.12\\
\hline                                   %inserts single line
\end{tabular}
%\tablefoot{}
%\tablefoottext{a}{Primary beam of $\sim2$\farcm5. }
\label{tbl:fillingfraction}
\end{table}

Table~\ref{tbl:fillingfraction} presents the median of the ratios between the clump areas and the beam solid angle of their \ammonia\ counterparts for sources in the catalogues \textit{Cat-1+Cat-5} (779 sources, 30\arcsecword\ beam) and \textit{Cat-2} (209 sources, 40\arcsecword\ beam), along with the median value of the $\eta_{ff}$ given for them. 
The table shows that clumps typically cover one half and one third of the area of beamsizes of 30\arcsecword\ and 40\arcsecword, respectively. 
%\revision{Also, although the values of the beam filling factors of the \ammonia\ emission are typically $\eta_{ff}=0.13-0.15$, the region of gas emission should fill up a larger fraction of the clump if we consider that it should be located within the clump observed in the continuum. 
%}
The smaller filling factors are likely explained as line emission from smaller dense fragments within the clump, since the ammonia transitions are not expected to be subthermally excited.
\revision{Nevertheless, although the values of the filling factors are typically $\eta_{ff}=0.13-0.15$ with respect to the \ammonia\ beam, the region of gas emission should fill a larger fraction of the area of the clump if we consider that it should arise within its observed dust emission. 
}

While the estimation of $T_{\textit{rot}}$ and \tk\ is not affected by the beamsize of the observations since they are derived from the ratio of the \ammonia\ transitions, the difference between the clump size derived from continuum maps and the \ammonia\ beams needs to be taken into account in the estimation of fractional abundances.

%%---------------------------
%% RESULTS
%%---------------------------
\section{Results}
\label{sec:results}

\subsection{Distance-independent parameters derived from gas and dust emission}
 
\subsubsection{Comparison between \tk\ and \td}
\label{sec:tktd}
\begin{figure}
	\centering
	\includegraphics[width=0.95\hsize]{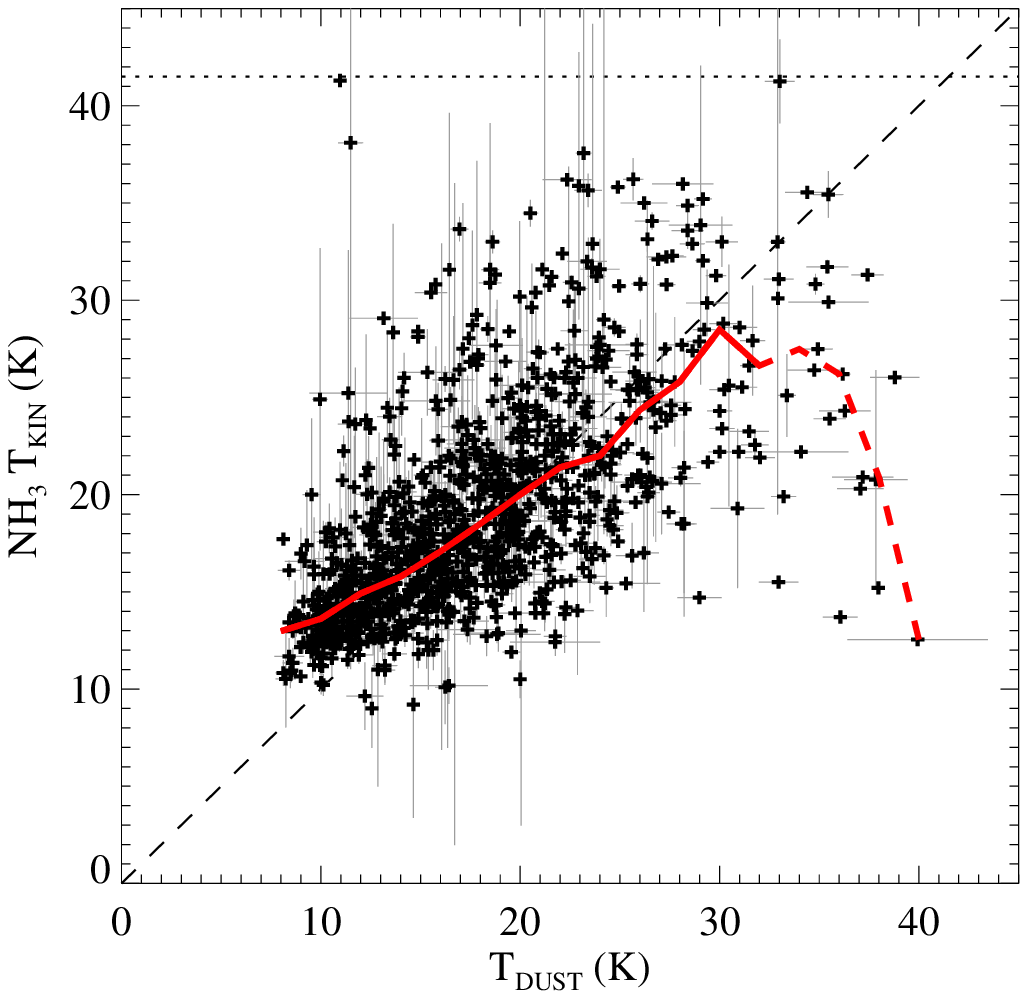}
	\includegraphics[width=0.95\hsize]{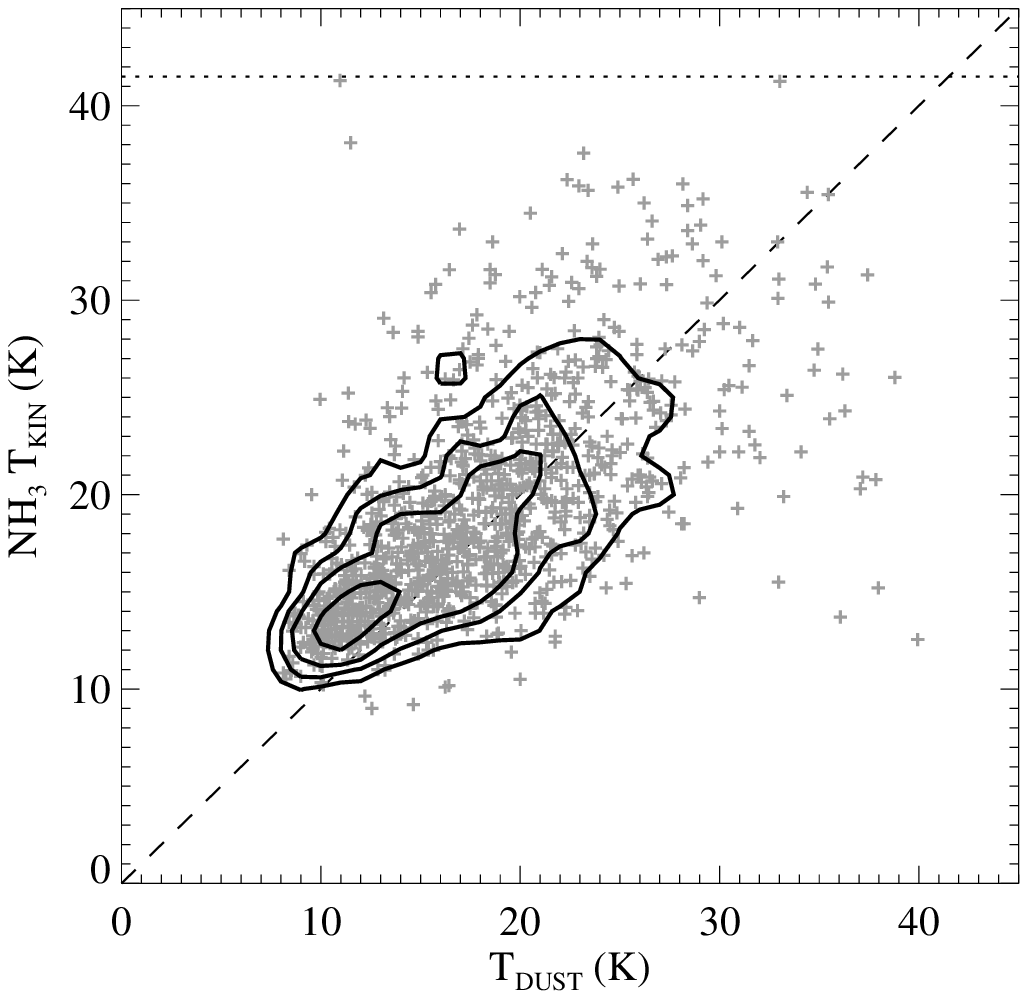}
	\caption{Comparison between kinetic temperature and dust temperature for the \goodsample\ of 1068 \hg\ clumps with \ammonia\ associations. The dashed grey line represents the line of equal temperatures, and the dotted line represents the uncertainty limit on the estimation of kinetic temperature (see section~\ref{sec:crossmatch}). The error bars are obtained from their source catalogue. The red line shows the median values of the sample in bins of 2~K in dust temperature. Bottom: Contours representing the 15\percentword, 30\percentword, 50\percentword\ and 90\percentword\ levels of number density of sources in the sample.
	}
	\label{fig:temp_singlematch}
\end{figure}

Figure~\ref{fig:temp_singlematch} shows the comparison between dust and kinetic temperatures found in the final sample of clumps. Errors on the kinetic temperatures are obtained from their respective source catalogue, while errors on dust temperatures are obtained from the \hg\ catalogue of physical properties.  
Contours of density of sources are shown in the bottom panel of Figure~\ref{fig:temp_singlematch}. The range of temperatures corresponding to the 30\percentword\ contour of density is $8<$~\td~$<24$~K, and $10<$~\tk~$<25$~K, and therefore most clumps lie in the optimal range of \tk\ determination. The figure shows that there is a large fraction of cold clumps (\td~$<20$~K) with \tk~$>$~\td. The red line in the figure represents the median value of the sample in bins of 2~K. The observed trend does not match the unity line, and increases until \td$\sim$30~K, then drops for last 42 clumps above that limit where \tk\ estimations only from the (1,1) and (2,2) transitions become less reliable.

%% Fig 1:
%% Tkin-tdust complete sample

\begin{figure}
	\centering
	\includegraphics[width=0.8\hsize]{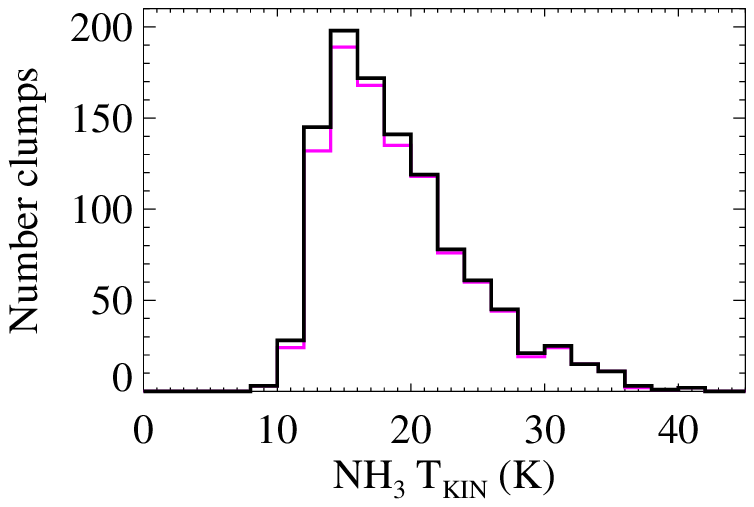}\\
	\includegraphics[width=0.8\hsize]{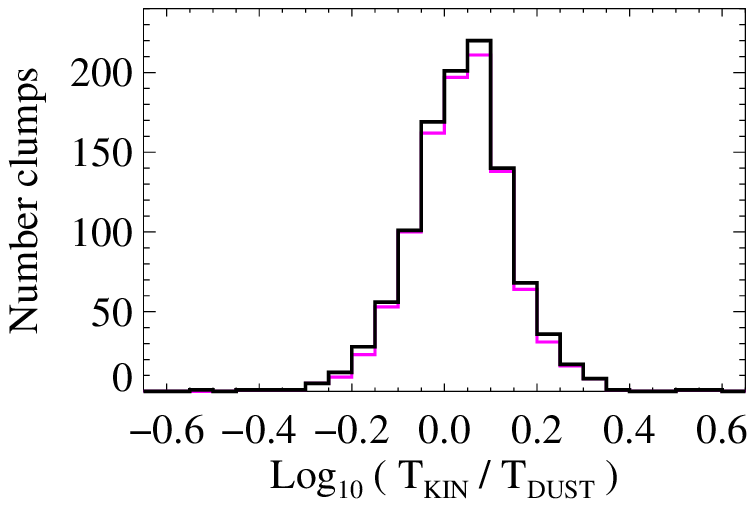}
	\caption{ Histogram of \tk\ and \tk/\td\ associated with \hg\ clumps. Black: \goodsample, consisting of 1068 sources. In magenta, clumps with fluxes at 160, 250, 350 and 500~\um\ bands (1023 sources). 
	}
	\label{fig:histo_tkin}
\end{figure}

The histogram of kinetic temperatures matched for the sample of clumps is shown in Figure~\ref{fig:histo_tkin}. The peak of this distribution lies in the 14$--$16~K range, with a steep rise at low temperatures.
The magenta line in the figure represents those clumps with emission at all 160, 250, 350, and 500~\um\ bands. The logarithmic ratio of \tk/\td\ is shown in the bottom panel of Figure~\ref{fig:histo_tkin}. The distribution has a peak at Log(\tk/\td)$=0.0-0.1$, and it appears symmetric around that value. The clumps with counterparts at all 160 to 500~\um\ bands have a distribution similar to the final sample.

%% Fig 2:
%% Histogram Tkin, tkin/tdust

\begin{figure}
	\centering
	\includegraphics[width=0.48\hsize]{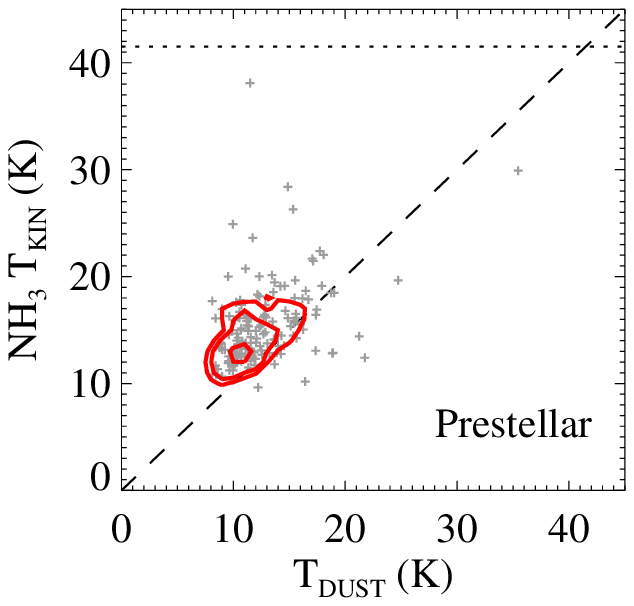}
	\includegraphics[width=0.48\hsize]{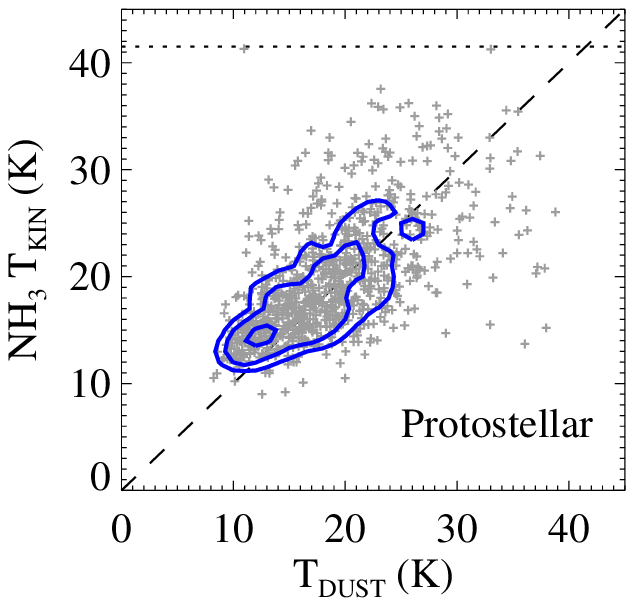}\\
	\includegraphics[width=0.48\hsize]{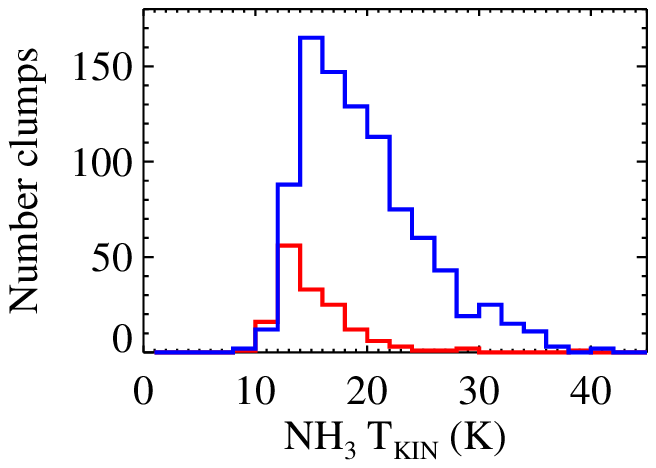}
	\includegraphics[width=0.48\hsize]{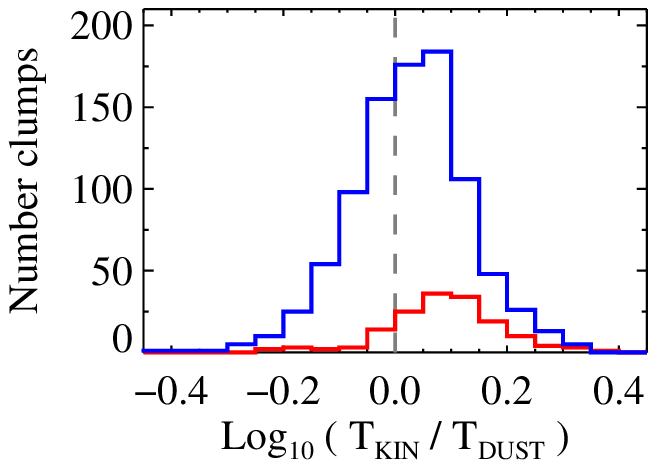}\\    
	\caption{Comparison between kinetic temperature and the dust temperature from prestellar (in red, 157 sources) and protostellar clumps (in blue, 909 sources). Contours representing the 30\percentword, 50\percentword\ and 90\percentword\ levels of number density of the sample. Bottom panels show the histograms of \tk\ and \tk/\td\ associated with prestellar and protostellar clumps.
	}
	\label{fig:temp_preprotostellar}
\end{figure}

The comparison between prestellar and protostellar clumps is shown in Figure~\ref{fig:temp_preprotostellar}. In general prestellar clumps are confined to a small range of low temperatures, with the 50\percentword\ contour of density of sources within $8<$~\td~$<13$~K, and $11<$~\tk~$<17$~K.
Protostellar candidates cover a wider range of temperatures, with the 50\percentword\ contour of density of sources within $9<$~\td~$<21$~K, and $12<$~\tk~$<23$~K. The median values of the ratio \tk/\td\ for prestellar and protostellar clumps are 1.24 and 1.06, respectively.
\revision{For the distributions of \tk/\td\ obtained for prestellar and protostellar clumps, we performed a KS test, and the null hypothesis that prestellar and protostellar samples are drawn from the same distribution is rejected by 5 sigma.}

%% Fig 3:
%% Temperatures pre/proto stellar

To test if the large difference in gas and dust temperatures of prestellar sources is due to inadequate or poor reliability of estimation of dust parameters, we performed a visual inspection of \hg\ images and the fitted SED curve on each prestellar clump. From the 110 sources with well-behaved grey body fitting across \hg\ bands, we found the same median value 1.24 for the \tk/\td\ ratio, and therefore bad SED fitting is not the cause of this temperature difference.
 	
Finally, we found one protostellar clump with \tk/\td$>3$ and after revision we consider this temperature difference to be caused by less dependable SED fitting.

\subsubsection{Column density and mass surface density}

The total column density of clumps is derived assuming optically thin and isothermal dust emission, following
\begin{equation}
N(\textrm{H}_2)= \tau(\lambda_\textrm{ref})/\left(\kappa(\lambda_{\textrm{ref}})\,\mu\, m_{\textrm{H}}\right) \ , 
\end{equation}

\noindent where $\kappa$ and $\tau$ are the opacity and the optical depth at a given wavelength $\lambda_{\textrm{ref}}$, and we assume a mean mass per particle $\mu=2.29$~\cite[e.g.,][]{eva99}. 
The value of $\kappa$ is chosen as 0.1 cm$^2$ g$^{-1}$ at $\lambda_{\textrm{ref}}=300$~\um, which includes a gas-to-dust ratio of 100~\citep{eli16}.
In comparison, the conventionally used OH5 model~\citep{oss94} gives a dust opacity at 300 $\mu$m of $\kappa=0.13$ cm$^2$ g$^{-1}$, and therefore our adopted model will produce higher column density and mass estimates by 30\percentword.
The median value of $N$(H$_2$) for the \goodsample\ is $1.37\times10^{23}$~\cmc.

Similarly, the mass surface density of a clump is estimated as: \begin{equation}
\Sigma=
\tau(\lambda_\textrm{ref})/\kappa(\lambda_{\textrm{ref}})\ .
\label{eq:sigma}
\end{equation}

\begin{figure}
	\centering
	\includegraphics[width=0.95\hsize]{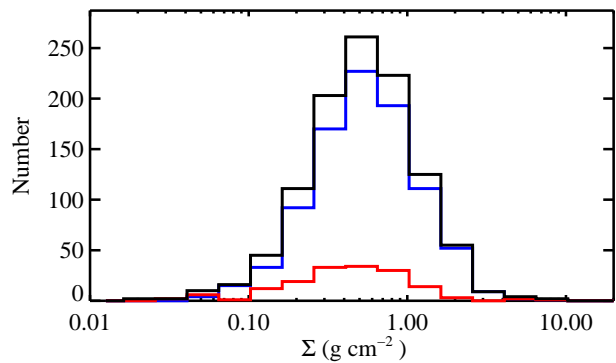}
	\caption{ Histogram of surface densities for \hg\ clumps (black line). The red and blue lines show the distributions for prestellar and protostellar sources, respectively.
	}
	\label{fig:histo_surface_restricted}
\end{figure}

Histograms of the distribution of surface densities for the \goodsample, and its prestellar and protostellar clumps are presented in Figure~\ref{fig:histo_surface_restricted}.
The sample has in general high surface densities, ranging between 0.02 and few $\times$ 1 g cm$^{-2}$, with a median $\Sigma_{\textrm{med}}=0.52$~g~\cmc\ and without large differences between the distribution of prestellar and protostellar clumps.

%% Fig 4:
%% Surface densities

\subsubsection{Fractional abundance of \ammonia}
 
The \ammonia\ column densities $N$(NH$_3$) are compared with the column densities of H$_2$ derived from the dust emission for \hg\ clumps to obtain the fractional abundance of ammonia in each source, defined as the ratio \chiammonia$=N$(NH$_3$)/$N$(H$_2$).

The values of $N$(\ammonia) were taken directly from the listed catalogues and therefore were evaluated over beamsizes typically larger than the clump area (see Section~\ref{sec:beamff}). To avoid effects of beam dilution in the estimation of \chiammonia\, we followed the assumption that line emission originates from the same clump area $\Omega$ measured at the 250~\um\ band, and therefore we scaled the values of the ammonia column density $N$(\ammonia) by a factor $(\Omega_{\textit{beam}}/\Omega)$, with $\Omega_{\textit{beam}}$ the beam area of the observation. 
This correction is applied only for sources with $\Omega<\Omega_{\textit{beam}}$.  

\begin{figure}
	\centering
	\includegraphics[width=0.95\hsize]{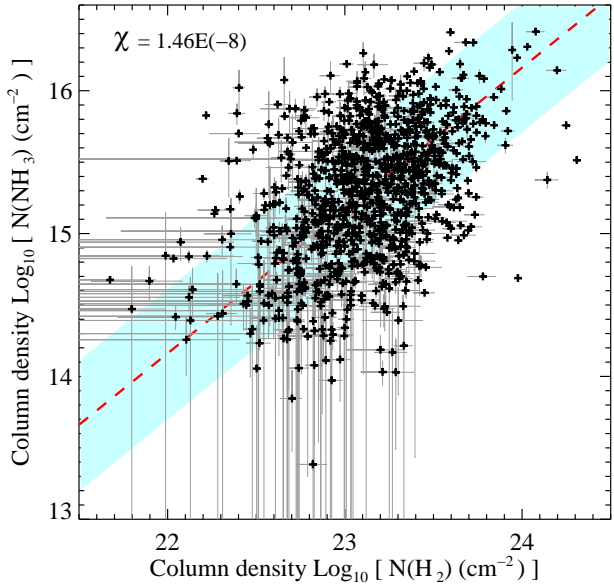}   
	\caption{Comparison between column densities of \ammonia, and column densities of H$_2$ derived from dust emission, for the final sample of \hg\ clumps. The dashed-red line represent the median fractional abundance \chiammonia$=N$(NH$_3$)/$N$(H$_2$) indicated in the upper left. The coloured area represents the median absolute deviation of the sample.
	}
	\label{fig:ngasnh2_restricted}
\end{figure}

Figure~\ref{fig:ngasnh2_restricted} shows $N$(NH$_3$) versus $N$(H$_2$) column density for the \goodsample\ of crossmatched \hg\ clumps. Errors in the column density of ammonia are plotted when available. 

%% Fig 5:
%% Ngas vs Nh2 from dust emission

\begin{figure}
	\centering
	\includegraphics[width=0.8\hsize]{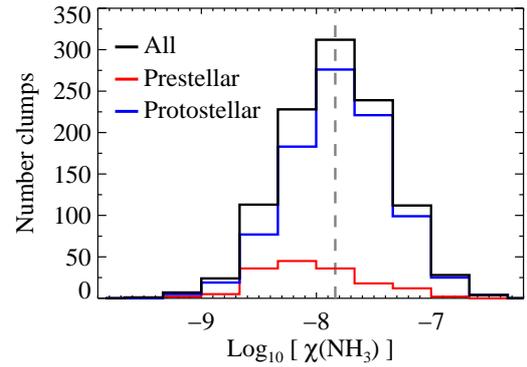}
	\caption{ Histogram of fractional abundance \chiammonia\ for \hg\ clumps. The black line represents the \goodsample, with a median value of 1.46\ee{-8} (dashed vertical line). The red and blue lines represent the prestellar and protostellar clumps, respectively.
	}
	\label{fig:histo_abund}
\end{figure}

We found a median value of \chiammonia$=1.46\times10^{-8}$ ($\log[\chi]=-7.84\pm0.31$, with the error representing the median absolute deviation of the sample). For prestellar and protostellar sources, the median values of \chiammonia\ are 8.33$\times10^{-9}$ and 1.60$\times10^{-8}$, respectively
\footnote{
If the 48 sources with less-reliable SED fitting after visual inspection are left out of this calculation, the median value of \chiammonia\ for prestellar clumps changes only by 12\percentword, and main results of fractional abundance for the \goodsample\ remain unaltered.
 }. 
The distribution of \chiammonia\ for the \goodsample\ and its prestellar/protostellar sources are shown in Figure~\ref{fig:histo_abund}.

%% Fig 6:
%:Histogram Abundance

It is commonly assumed that \td\ $\sim$ \tk\ in the estimation of the fractional abundance of ammonia for sources with high density. Among these estimates, median values of \chiammonia\ of $\sim2.5\ee{-8}$ are found toward high-mass star forming clumps~\citep{urq15}, $3.4\ee{-8}$ and $2.3\ee{-8}$ for protostellar and starless cores in Perseus~\citep{fos09}, and values of $\sim3\ee{-8}$ for two of starless cores in Taurus-Auriga~\citep{taf06}. Similarly, under this assumption some of the catalogues in Table~\ref{tbl:catalogs} gave values for \chiammonia\ of
 $4.6\times10^{-8}$~\citep[\textit{Cat-1}]{dun11}, $1.2\times10^{-7}$~\citep[\textit{Cat-2}]{wie12},
$\sim2\ee{-8}$~\citep[\textit{Cat-9}]{chi13}, and 
 $6.2\times10^{-7}$~\citep[\textit{Cat-12}]{mor14}. We note that considering only the clumps in our sample that have the above catalogues as their main associations, we found differences in our estimates with respect to the fractional abundances given by those authors: \textit{Cat-1} (592 sources), $1.09\ee{-8}$; \textit{Cat-2} (209), $2.96\ee{-8}$; \textit{Cat-9} (14), $4.40\ee{-9}$; \textit{Cat-12} (12), $8.34\ee{-9}$. 
 The discrepancy between our estimate of \chiammonia\ for sources from \textit{Cat-1} and the average value given by~\cite{dun11} is due mainly to different estimates of $N$(H$_2$). While our sub-sample of sources selected from \textit{Cat-1} has larger values of $N$(NH$_3$) before corrections by the factor $(\Omega_{\textit{beam}}/\Omega)$ (median of 8.53\ee{14}~\cmc, compared to the median value of 5.8\ee{14}~\cmc\ given by~\citeauthor{dun11} for the complete sample of BGPS objects), our values of $N$(H$_2$) are higher by a factor of $\sim6.1$ as a result of the larger angular sizes that BGPS have in general respect to \hg\ sources~\citep[see appendix D in][]{eli16}.

In a study of independent measurements of \td\ and \tk\ by~\cite{bat14} using interferometric ammonia observations with the Karl G. Jansky VLA telescope, and dust continuum emission from Hi-GAL maps, these authors determined a \chiammonia\ of $4.6\ee{-8}$ based on a pixel-by-pixel analysis toward two clumps within an IRDC, one showing signs of active star formation and one without these indicators.

Therefore, our results for the fractional abundance of \ammonia, based on a large sample of clumps and independent measurements of \td\ and \tk, are in general lower by a factor 2--3 with respect to values of \chiammonia\ found in the literature. \revision{The reported values of \chiammonia\ in different catalogues usually consider \tk$=$\td. Then, since we found that for a large part of our sample \tk/\td$>1$, the estimates of $N$(H$_2$) in these catalogues would result in an erroneously higher \td\ and thus $N$(H$_2$) would be underestimated. Hence, their values of ammonia abundances would necessarily be higher than \chiammonia\ estimates using independent measures of \tk\ and \td.}

\subsection{Distances}
\label{sec:distances}

Kinematic distances were estimated for the sample of clumps considering the $V_{lsr}$ associated for each \ammonia\ counterpart and following the rotation curve of~\cite{rei09}, with $\mathcal{R}_0=8.34$ kpc and $\Theta_0=240$ \kms .
 
In the longitude range 90\arcdeg$<\ell<$270\arcdeg, unambiguous distances were estimated for 54 clumps. To resolve the distance ambiguity in the inner Galaxy, we considered the following constraints: for clumps with ammonia counterparts from BGPS, ATLASGAL and RMS surveys we adopted the same near/far distance solution as given in their distance determination from~\cite{ell13,ell15} for BGPS,~\cite{wie15} for ATLASGAL, and~\cite{urq14} for RMS, favouring the main association in case of multiple \ammonia\ observations.

In addition, we adopted the near distances in sources associated with IRDCs samples (\textit{Cat-9}, \textit{Cat-11}, and \textit{Cat-14}), in those clumps found inside the equivalent radius of \textit{Spitzer} dark clouds in the catalogue presented by~\cite{per09}, and if a far kinematic distance would imply a scale-height larger than 150~pc~\citep[see, e.g.,][]{gia13}. For the rest of the sample, we use the near/far distance solution taken for sources in the \hg\ catalogue, if available. For those 88 clumps with no way to resolve the ambiguity, we adopted the near kinematic distance.

We performed further checks on the estimated distances. Clumps toward the Gemini OB1 molecular cloud ($\ell\sim189$\arcdeg$-190$\arcdeg) are revised as $d$=2.1 kpc~\citep{dun10}. Distances of a few others were estimated from literature: 
S106 molecular cloud, $d$=1.7 kpc~\citep{sch07}; 
IRAS 05439+3035, $d$=3.5 kpc~\citep{tap97}.

Although our estimates of distances may differ from the values provided by the \hg\ catalogue of physical properties which follow the method presented in~\cite{rus11} and uses the rotation curve of~\cite{bra93} ($\mathcal{R}_0=8.5$ kpc and $\Theta_0=220$ \kms), only 60 sources have differences above 20\percentword\ between the above method and our estimations of distances, and therefore overall results and conclusions are not expected to change significantly.
However, since differences between kinematic distances and distances obtained from parallax measurements can differ up to a factor of two~\citep{rei09}, a conservative approach should take the same factor of two for uncertainties in source distances.

\begin{figure}
	\centering
	\includegraphics[width=0.8\hsize]{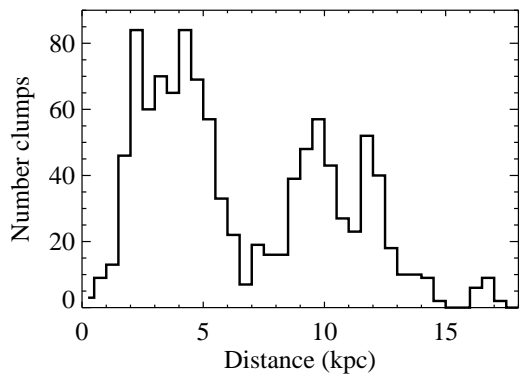}
	\caption{ Histogram of kinematic distances for \hg\ clumps estimated from their associated \ammonia\ line emission.
	}
	\label{fig:histo_dist_restricted}
\end{figure}

\begin{figure}
	\centering
	\includegraphics[width=0.9
	\hsize]{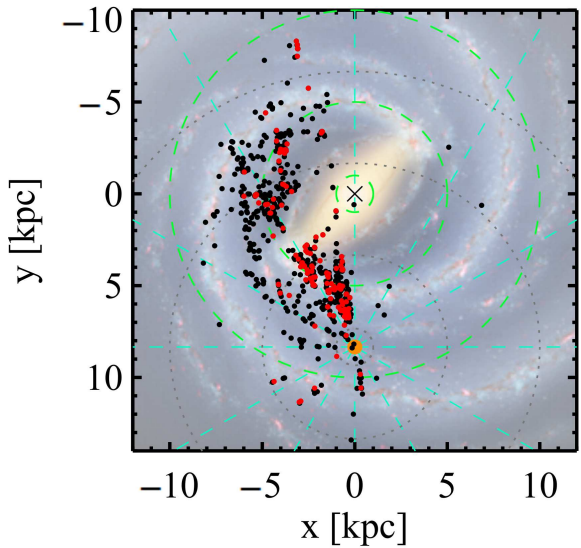}
	\caption{Plot of the position in the Galactic plane of \hg\ clumps associated with \ammonia\ emission. Red points represent those clumps considered prestellar. The position of the Sun is indicated as an orange dot at [x,y]=[0,8.34], while the Galactic centre is indicated with a X symbol. Cyan dashed lines indicate steps of 30\arcdeg\ in Galactic longitude. Green dashed circles show Galactocentric distances 1, 5 and 10 kpc, and grey dotted circles show heliocentric distances 1, 5, 10 and 15 kpc. In the background, artist's concept of the Milky Way [Credit: NASA/JPL-Caltech/R. Hurt (SSC/Caltech)].
	}
	\label{fig:distribution_distance}
\end{figure}

Figure~\ref{fig:histo_dist_restricted} shows the histogram of kinematic distances for clumps in the \goodsample. Nearly 56\percentword\ of the sample is at a distance less than 6 kpc, with a second large group (20\percentword) in the range 8.5-11 kpc. The distribution of sources in the Galactic plane is shown in Figure~\ref{fig:distribution_distance}.  
Large groups of sources are found in the I Galactic Quadrant toward the Scutum-Centaurus arm, and the Sagittarius and Perseus arms. 

\revision{A possible explanation for the scarcity of sources at 6--8 kpc is the complexity of allocating sources inside the Galactic bar and the 3-kpc arm, where strong radial streaming motions are present, and by the flatness of the projected rotation curve of the kinematic model in the range $ \ell=$30\arcdeg--60\arcdeg\ that produce large uncertainties in distances for small variations of measured $V_{lsr}$~\citep[see e.g.,][]{ell15}.
}

\revision{A second possibility is that the high values of $V_{lsr}$ required by the rotation curve of~\cite{rei09} to fill the region toward the tangent of the Galactic bar are in general not found. In the Galactic plane toward $\ell=30$\arcdeg, the rotation curve model requires $V_{lsr}>110$ \kms\ for kinematic distances in the range 6.2--8.3 kpc. As a test, we took the inner Galaxy \hg\ catalogue from~\cite{eli16} and considered the sources in the range 28\arcdeg$<\ell<$32\arcdeg\ with measured $V_{lsr}$ (2305 sources). Of them, 516 (22 per cent) have $V_{lsr}>100$ \kms, and only 64 have $V_{lsr}>110$ \kms. In our sample of clumps, 157 sources are in the same longitude range, with a similar fraction of sources with high radial velocities (42 clumps with $V_{lsr}>100$ \kms, and 4 with $V_{lsr}>110$ \kms). Then, the sample of sources with \ammonia\ detections in this region is not biased in favor of lower $V_{lsr}$ with respect to the rest of the \hg\ sources in this Galactic range. }

\revision{We do not discard the possibility that the adopted rotation curve is the reason of the dip at $\sim6$ kpc in the distribution of clump distances shown in Figure~\ref{fig:distribution_distance}. We note though that using the Bayesian approach for resolving the kinematic distance ambiguity, and in this way estimate a probability density function of the sources distances~\citep{ell13,rei16}, it is likely that clumps in the $\ell\sim30$\arcdeg\ region even with $V_{lsr}\sim100$ \kms\ will be located toward the tangent of the bar at a heliocentric distance of $\sim7.5$ kpc, considering the weight given to the spiral arm probability density function in the distance estimation.}

%% Fig 7:
%% Histogram revised distances
%	

%% Fig 8:
%% Distribution of sources

\subsection{Distance-dependent parameters}
\subsubsection{Mass and clump size}
\label{sec:clump_sizes}

Once the distance $d$ of a source is known, the clump mass obtained from dust emission is derived from the expression $M_{\textit{clump}}=\Omega\ d^2\times \tau(\lambda_\textrm{ref})/\kappa(\lambda_{\textrm{ref}})=\Omega\ d^2\times\Sigma$, with $\Omega$ the clump solid angle measured in the 250~\um\ band.

%: Median properties of the final sample
\begin{table}
\caption{{%\bf 
		Median values of distance-dependent physical properties of the sample considered in this work, and sources with known distance from the complete inner Galaxy \hg\ catalogue~\citep[$\sim$33600 prestellar and $\sim$15800 protostellar sources,][]{eli16}. In parenthesis, the median absolute deviation of the respective group. The asterisk indicates that only prestellar sources with reliable SED fitting after inspection are considered.}
	}   
%\label{table:catalogslist}      % is used to refer this table in the text
\centering                          % used for centering table
\begin{tabular}{lcccc}        % centered columns (4 columns)
\hline\hline                 % inserts double horizontal lines
Clumps& $M_{\textit{clump}}$ & $L_{\textit{bol}}$&$R_{250}$ & $n$\\
&(\msun)&(\lsun)&(pc)&($10^4$ \cmv)\\
%\\ 
%Index & Type & Telescope\tablefootmark{*} & $\theta_{FWHM}$ & \ammonia\ sources& \hg\ clumps & References\\
% & & CS96\tablefootmark{*} & CS15\tablefootmark{*}\\
\hline                        % inserts single horizontal line
Final sample & 850 (700) & 3080 & 0.33 & 9.6 \\
Prestellar* & 850 (690) & 130 & 0.34 & 9.3 \\
Protostellar & 870 (710) & 4600 & 0.32 & 9.8\\
\hline
\hg\ prestellar & 390 (300) & 120 & 0.48 & 1.8 \\
\hg\ protostellar & 450 (370) & 1020 & 0.37 & 4.3 \\
\hline 
%{\bf Final sample} & {\bf 1068} \\ 
%\hline                                   %inserts single line
\end{tabular}
%\tablefoot{}
%\tablefoottext{a}{Primary beam of $\sim2$\farcm5. }
\label{tbl:distanceprops}
\end{table}

\begin{figure}
	\centering
	\includegraphics[width=0.95\hsize]{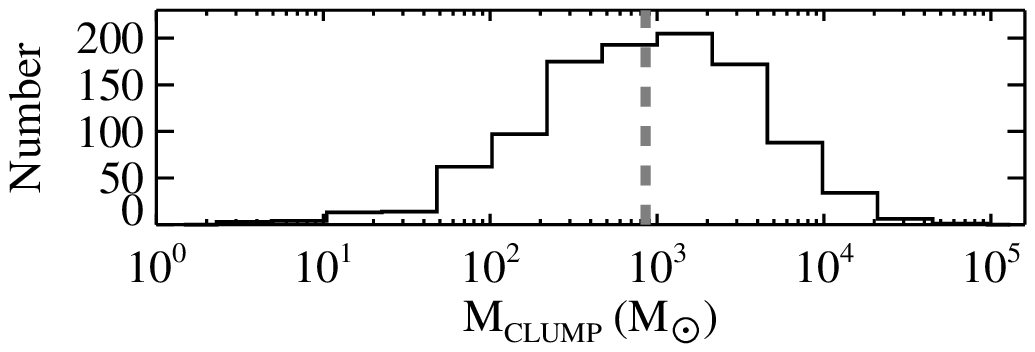}
	\includegraphics[width=0.95\hsize]{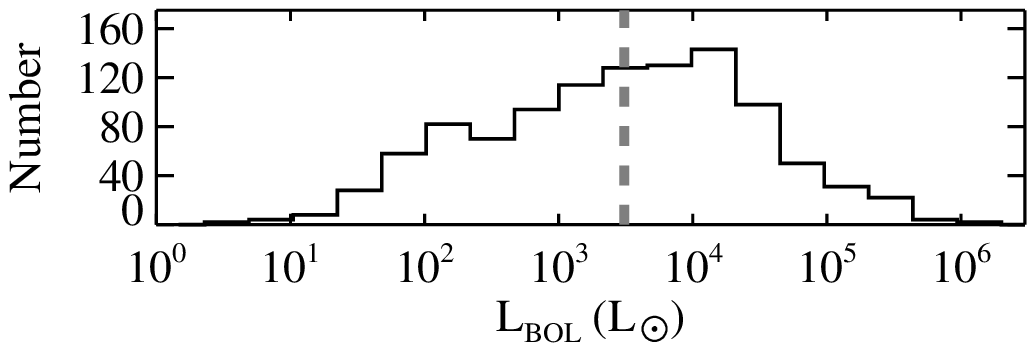}
	\includegraphics[width=0.95\hsize]{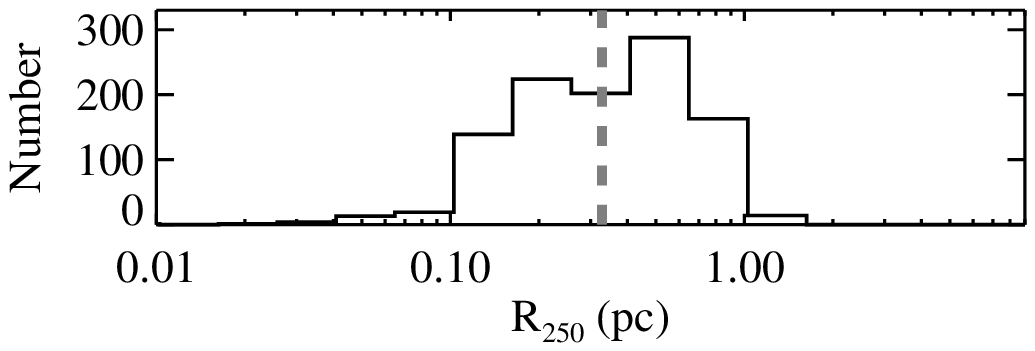}
	\includegraphics[width=0.95\hsize]{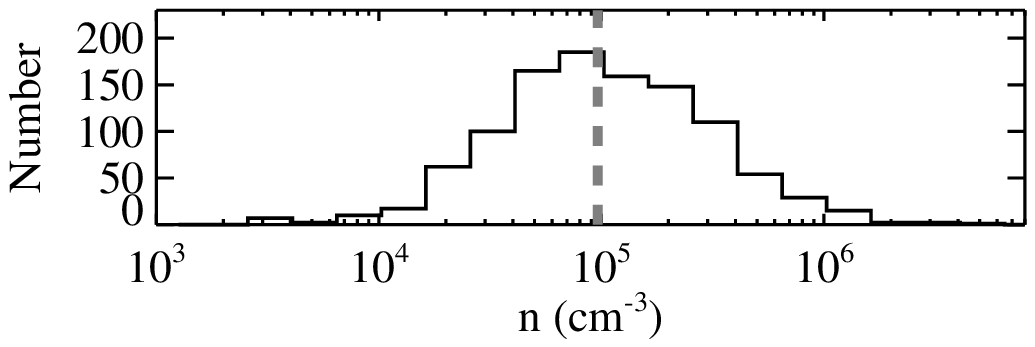}	
	\caption{ Histogram of distance-dependent parameters: clump mass $M_{\textit{clump}}$, bolometric luminosity $L_{\textit{bol}}$, radius $R_{250}$, and density $n$. The vertical line in each panel represents the median value of the distribution.
	}
	\label{fig:histo_distprop}
\end{figure}

The masses and bolometric luminosities of the \hg\ catalogue were rescaled to the new distances for our sample. Figure~\ref{fig:histo_distprop} shows the histograms with the distribution of distance-dependent physical parameters of sources, and the median values of these parameters are presented in Table~\ref{tbl:distanceprops}. 
The \goodsample\ of \hg\ clumps have a median mass of $M_{\textit{clump},\textrm{MED}}=850$~\msun\ (with average $M_{\textit{clump},\textrm{AVG}}=2000$~\msun\ and median absolute deviation $\textit{MAD}(M_{\textit{clump}})=700$~\msun), and median bolometric luminosities of $L_{\textit{bol},\textrm{MED}}=3080$~\lsun (with $L_{\textit{bol},\textrm{AVG}}=22400$~\lsun, $\textit{MAD}(L_{\textit{bol}})=2970$~\lsun). 
The sample covers a large range in masses for both prestellar and protostellar clumps, with similar dispersion range. 

%% Fig 9:
%% Histograms of distance-dependent parameters

Physical sizes are calculated from the estimated kinematic distances. The FWHM sizes of \hg\ sources presented in the photometric catalogue were measured using the extraction package \textit{CuTEx}~\citep{mol11} in the 160, 250, 350 and 500~\um\ bands. Because of the method followed in the creation of the band-merged catalogue and posterior estimation of physical properties, all sources in our sample have angular sizes measured at 250 and 350~\um, but a small fraction ($\sim$6\percentword) were not detected in the continuum maps at 160 or 500~\um. As shown in the photometric catalogue, most \hg\ sources are not PSF-like features, and a large majority of them show axis ratios below 1.5. We consider then for our analysis the angular sizes directly measured from the continuum images at 250~\um\ (geometric mean of minor and major axis of the ellipses estimated by \textit{CuTEx} in this band). The 250~\um\ band has a FWHM beam of 18\arcsecword, hence this is the highest angular resolution band that is common for all sources.

\begin{figure}
	\centering
	\includegraphics[width=1\hsize]{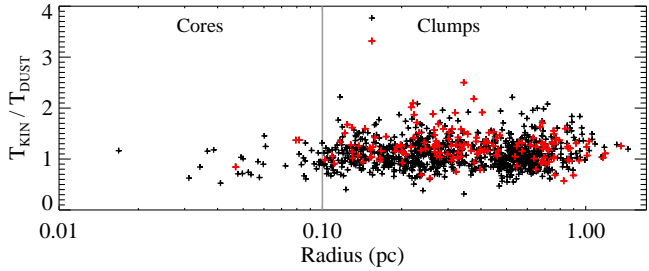}
	\caption{Ratio between kinetic temperature and dust temperature, as a function of the source radius estimated from the 250~\um\ band emission. Prestellar candidates are shown in red and protostellar sources are shown in black. The vertical line represents the boundary scale between cores and clumps.}
	\label{fig:tkintkin_angsize}
\end{figure}

Figure~\ref{fig:tkintkin_angsize} shows the physical radius $R_{250}$ for each clump, estimated from the angular size measured in the 250~\um\ band. The sample has a median $R_{250, \textrm{MED}}=0.33$~pc, with a median absolute deviation of 0.17~pc.
Most of the \hg\ sources have indeed sizes corresponding to clump-type objects, following the criteria from~\cite{ber07}, and only 38 sources have radii smaller than 0.1~pc, hence these sources are more consistent with the definition of cores rather than clumps. Prestellar clumps are distributed in the range $\sim$0.5--1.4~pc, without a preferential size for those clumps with high ratio \tk/\td.

%% Fig 10:
%% Tkin/Tdust vs angular radius

\subsubsection{Volumetric density}
\label{sec:vol_dens}

The volumetric densities $n$ are estimated from
\begin{equation}
n=\frac{M_{\textit{clump}}}{(4/3)\,\pi\, R_{250}^3\,\mu\, m_\textrm{H}}\ \ ,
\end{equation}

\noindent with $M_{\textit{clump}}$ the mass derived from SED fitting, and  $R_{250}$ the radius of the clump obtained from the 250~\um\ band, and $\mu=2.29$. 

The median value of the density distribution is $n_{\textrm{med}}=9.6\times10^4$ cm$^{-3}$, with
95\percentword\ of the sample between 1.2$\times10^4$ and 7.8$\times10^5$ cm$^{-3}$.
We do not find large differences in $n$ if only the upper 10\percentword\ of the sample mass distribution is considered ($M_{\textit{clump}}>4.8\times10^3$~\msun, $n_{\textrm{med}}=8.4\times10^4$ cm$^{-3}$).
The sample of prestellar clumps is in general less dense than protostellar sources (median densities of $6.9\times10^4$ and $9.8\times10^4$ cm$^{-3}$, respectively).

\begin{figure}
	\centering
	\includegraphics[width=0.95\hsize]{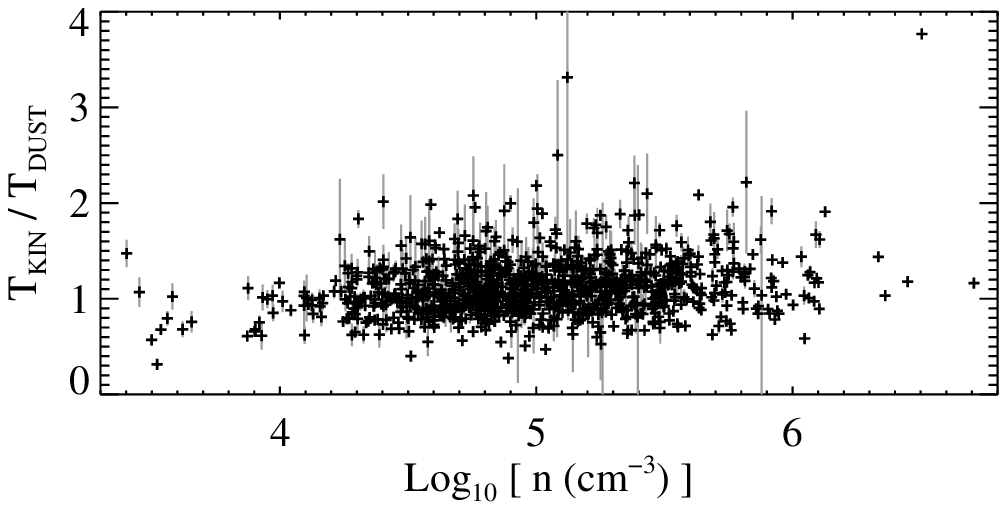}
	\includegraphics[width=0.95\hsize]{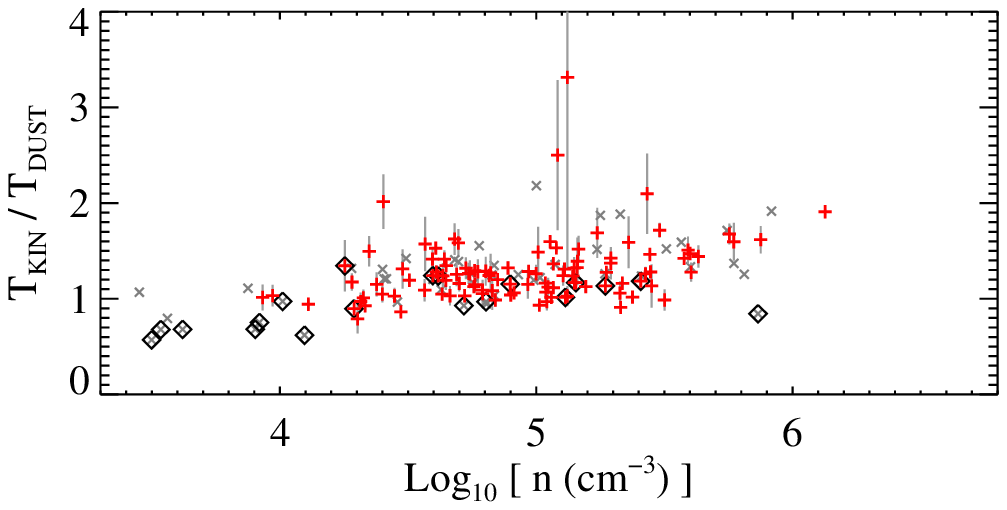}
	\includegraphics[width=0.95\hsize]{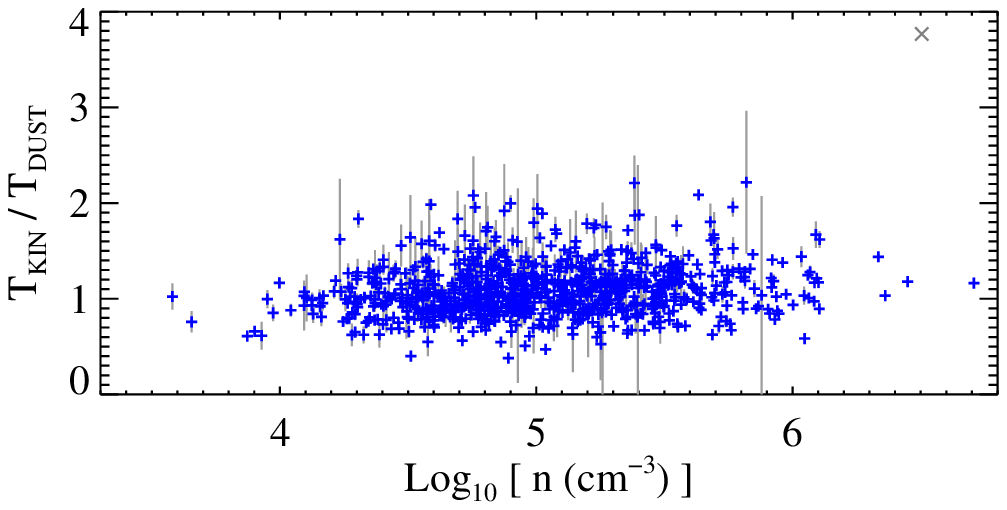}    
	\caption{Ratio between kinetic temperature and dust temperature, as a function of volumetric density for the \goodsample\ of \hg\ clumps (black). Second and third panels show the sub-samples of prestellar (red) and protostellar sources (blue). The grey X symbols denote clumps with less reliable SED fitting after visual inspection, and the diamond symbols show SPIRE-only clumps.
	}
	\label{fig:tkintdust_n}
\end{figure}

Figure~\ref{fig:tkintdust_n} shows the relation between \tk/\td\ and the volumetric density for the \goodsample, along with the sub-samples of prestellar and protostellar clumps.
We note that prestellar clumps with low density $\log[n~(\textrm{cm}^{-3})]\lesssim4.1$ and ratios \tk/\td$<1$, correspond to sources with less reliable SED fitting (see section~\ref{sec:tktd}).

%% Fig 11:
%: Tkin/Tdust vs volumetric densities

For bins of $\log[n~(\textrm{cm}^{-3})]=0.2$, we found differences of less than 13\percentword\ between \tk\ and \td\ over a large range of densities (up to $\log[n~(\textrm{cm}^{-3})]\lesssim5.5$), and therefore there is a general good agreement between gas and dust temperatures for the bulk of clumps. 
Thus, we consider the lower limit of volumetric density for the 95\percentword\ of the sample, 
$n(\textrm{H}_2)=1.2\times10^4$ cm$^{-3}$ ($\log[n~(\textrm{cm}^{-3})]\simeq4.1$), 
as a characteristic density for which we expect thermal coupling between gas and dust. 
This value is in agreement with the expected density, derived from radiative transfer calculations, at which dust and gas are thermally coupled~\citep[3\ee{4}~\cmv,][]{gol01,gal02}.

We also observe that prestellar clumps with $4.1~<~\log[n~(\textrm{cm}^{-3})]~<~5.0$ (approximately half of the sample) progressively converge toward gas and dust temperature coupling, and we note that bad SED fitting cannot explain the outliers with \tk/\td$>2$, nor the high values of this temperature ratio above $\log[n~(\textrm{cm}^{-3})]\gtrsim5.5$.

\subsection{Criteria for high-mass star formation}

The majority of the clumps (89\percentword\ of the sample) fulfil the empirical minimum value of mass surface density $\Sigma~=~\tau(\lambda_\textrm{ref})/\kappa(\lambda_{\textrm{ref}}) = 
M_{\textit{clump}}/(\pi R_{250}^2)$ for massive star formation of 0.2 g cm$^{-2}$ suggested by~\cite{but12}. Furthermore, 195 clumps (18\percentword\ of the sample) are above the threshold of $\Sigma=1$ g cm$^{-2}$ of~\cite{kru08} that can prevent excessive fragmentation in the parental clump during the formation of high-mass stars.

The sample was tested against other mass--radius criteria from the literature for the formation of high-mass stars. We tested the relation $M(r)>870$~\msun (r/pc)$^{1.33}$ presented by~\cite{kau10} for the formation of massive stars and stellar clusters in a sample of Infrared Dark Clouds. 
In our final sample, $\sim$93\percentword\ of clumps follow the above prescription, and we note that among those not following the criterion are 10 prestellar clumps with less-reliable SED fitting (see Section~\ref{sec:tktd}). A similar relation was suggested by~\cite{bal17}, $M(r)>1282$~\msun (r/pc)$^{1.42}$, that takes into account distance effects as a bias for the detection of regions forming massive stars, and nearly 90\percentword\ of our sample satisfies this prescription.

\subsection{Comparison with the general population of \hg\ sources}
\label{sec:comparisonhg}

Table~\ref{tbl:distanceprops} shows the median values of physical parameters of prestellar and protostellar sources with determined distances from the complete \hg\ inner Galaxy catalogue presented by~\cite{eli16}. In comparison with the general population of \hg\ sources, our sample represents a more massive group of clumps, with higher values of volumetric and surface densities. From the complete \hg\ catalogue, 33550 prestellar and 15834 protostellar sources have determined distances; for prestellar clumps, $\sim$27.8\percentword\ have masses above 850~\msun, $\sim$12.4\percentword\ have densities above 9\ee{4}~\cmv, and only $\sim$3.5\percentword\ accomplish both conditions. For protostellar clumps, $\sim$33.4\percentword\ have masses above 850~\msun, $\sim$33.0\percentword\ have densities above 9\ee{4}~\cmv, and $\sim$13.3\percentword\ have both conditions.
	
We also note that sources in the complete \hg\ catalogue accomplish the criteria of high-mass star formation in lower rates with respect to our sample with clumps dense enough to have \ammonia\ (1,1) and (2,2) detections. \cite{eli16} shows that $65-71$\percentword\ of \hg\ sources follow the criteria of~\cite{kau10}, while only 2.8\percentword\ of prestellar sources and in 13.1\percentword\ of protostellar souces are above the threshold $\Sigma=1$ g cm$^{-2}$ of~\cite{kru08}.

We conclude that the sample of \hg\ sources presented in this work represents a valuable selection of clumps with the conditions to harbour high-mass stars at different stages of formation.

%%---------------------------
%% DISCUSSION
%%---------------------------
\section{Discussion}
\label{sec:discussion}

\subsection{Correlation between evolutionary indicators.}
\label{sec:correlation}

In this section we aim to establish a connection between indicators of the evolutionary stages of dense clumps identified by gas properties derived from \ammonia\ line emission, and dust properties derived from \hg\ continuum bands. 

The ratio between the bolometric luminosity and the clump mass, the \lm\ parameter, is a distance-independent value and it is considered an indicator of evolution of dense clumps~\citep[e.g.,][]{mol08,eli13}.
Massive star-forming objects accreting with a rate proportional to their mass present a dramatic increase in radiated bolometric luminosity as a function of time, while their core envelope mass decreases only slightly. Subsequent evolution is modelled as envelope dispersal by stellar winds and outflows after a high-mass star has formed.

In addition, a recent study by~\cite{mol16b} proposed limiting values \lm=1 and \lm=10 for the three phases of star formation in massive clumps traced by the detection and temperature of the warm gas tracer CH$_3$CCH:
for sources with \lm$<1$, this molecule is not detected suggesting that the internal input energy is not sufficient to increase considerably the bolometric luminosity and the inner envelope temperature ($>40$~K);
for \lm\ in the range 1$-$10, CH$_3$CCH shows temperatures of $\sim30--40$~K and clumps are building up luminosity due to formation of stars, but no star is yet able to significantly heat the inner regions; at \lm$\gtrsim10$, the gas temperatures traced by this molecule increase as a consequence of the appearance of an intermediate- to high-mass ZAMS star at the interior of the clump.

\subsubsection{Kinetic temperature -- \lm}
\label{sec:tk_lm}

\begin{figure}
	\centering
	\includegraphics[width=0.9\hsize]{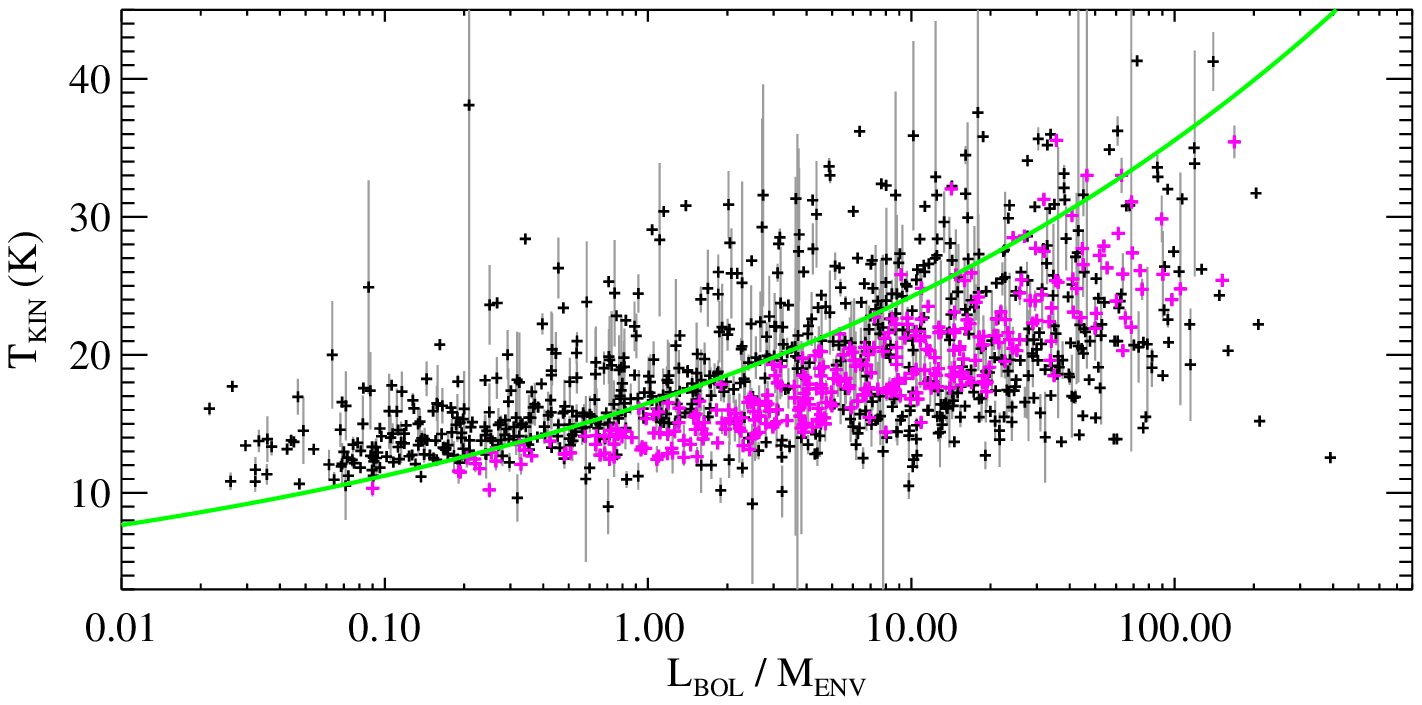} 
	\includegraphics[width=0.9\hsize]{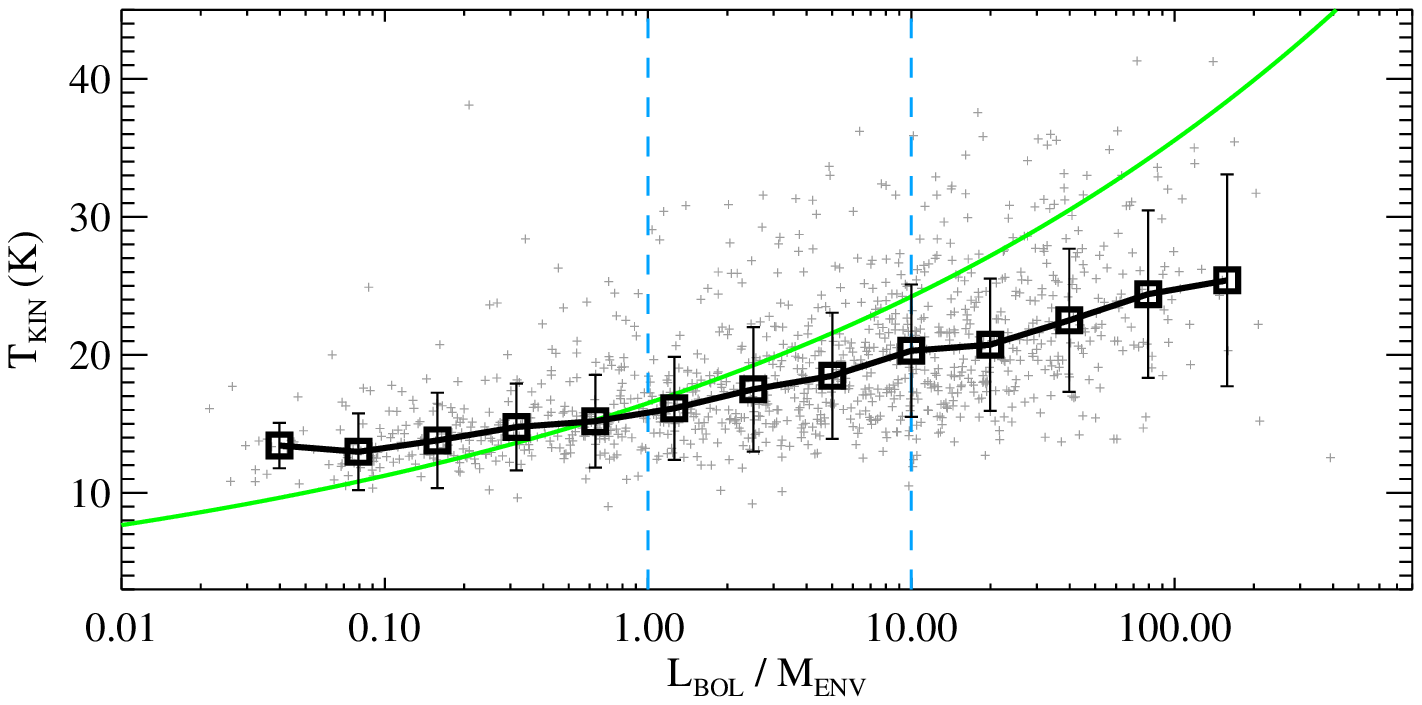}  
	\includegraphics[width=0.9\hsize]{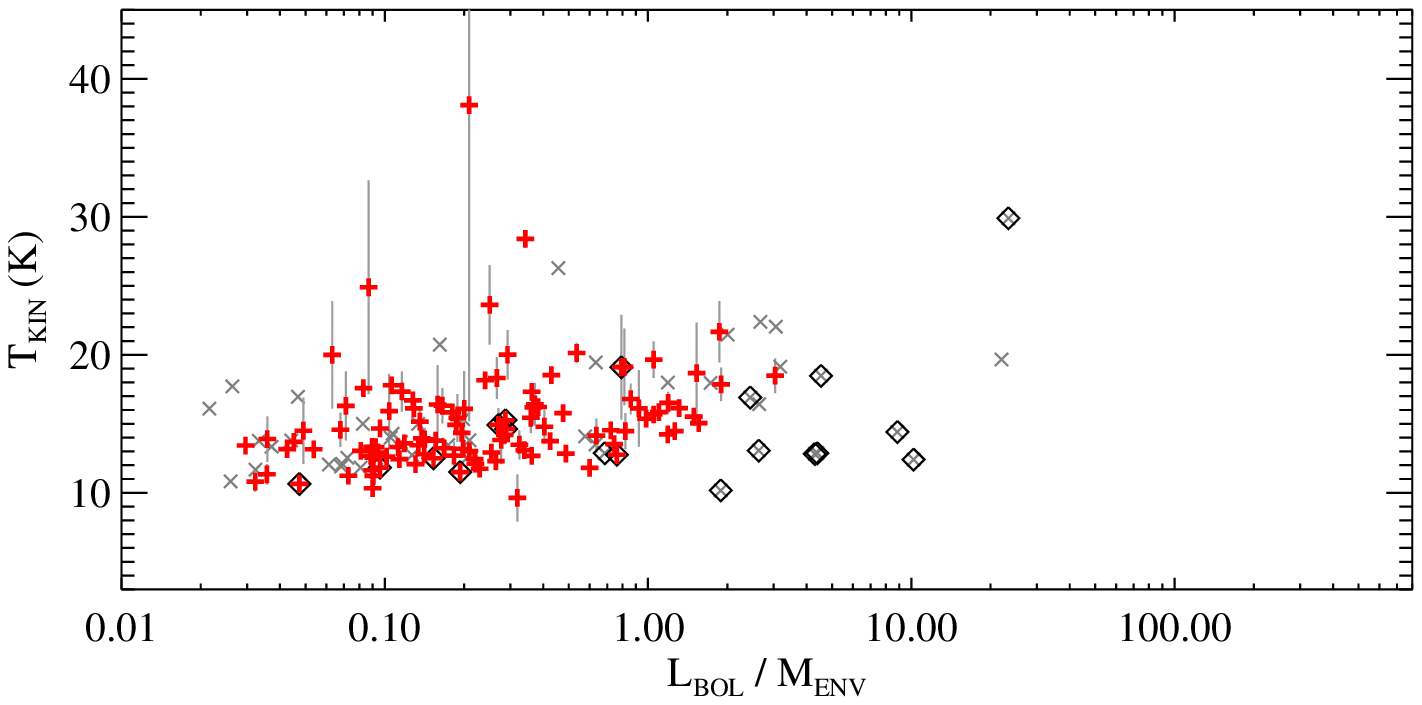}  
	\includegraphics[width=0.9\hsize]{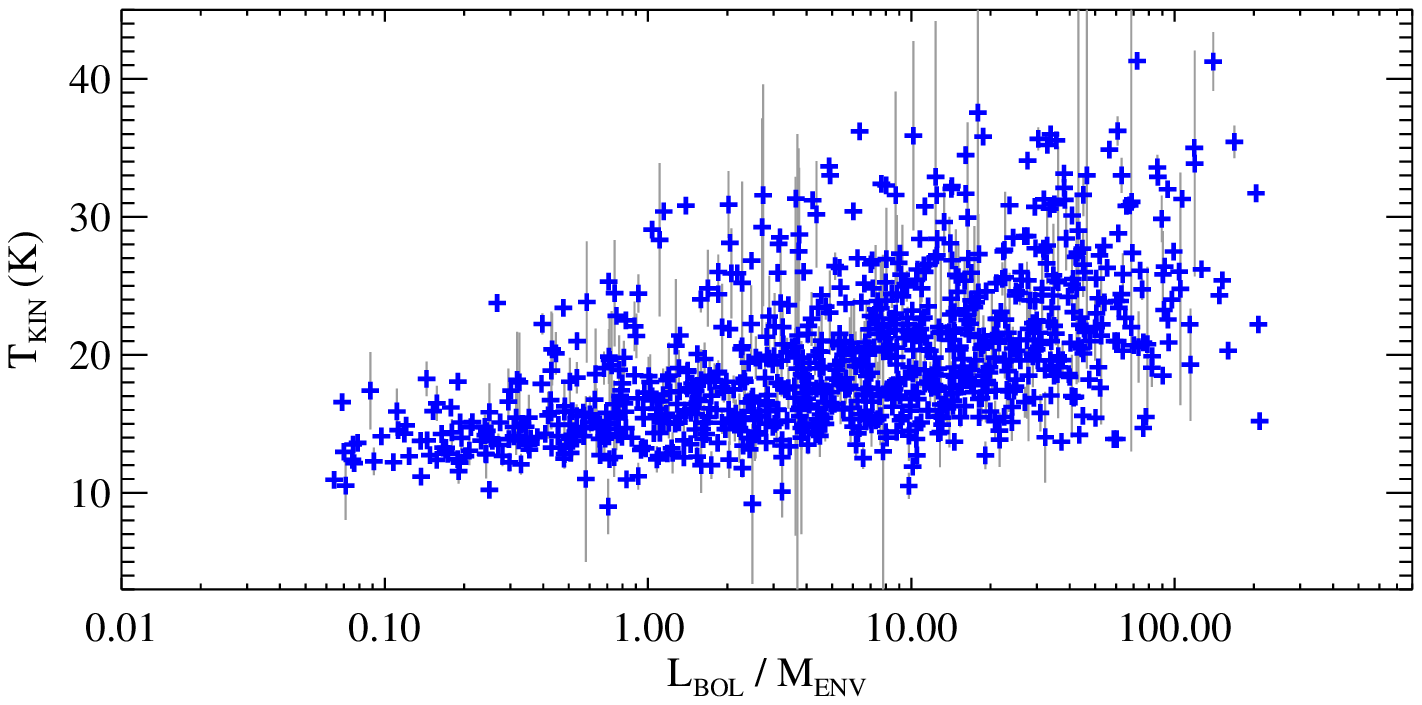} 
	\caption{Kinetic temperature as a function of 
		the ratio between bolometric luminosity and clump mass (``envelope" mass) of \hg\ sources. 
		The upper panel shows the \goodsample. In magenta, clumps with \tk\ ``coupled" to \td\ (difference between them less than 10\percentword). 
		The green line represents the analytic curve for black-body emission modified by opacity, \td$\ \propto\ $(L/M)$^X$, with X=$(4+\beta)^{-1}$ and $\beta$ the spectral index (fixed to be 2.0). 
		In the second panel, the binned line is shown in solid black, with error bars showing the median absolute deviation of each bin.
		The cyan vertical lines represent the transitions for the three phases of star formation in massive clumps detected with warm gas tracers (see section~\ref{sec:correlation}).
		Last two panels show the prestellar (red) and protostellar (blue) clumps. The symbols follow the same description as Figure~\ref{fig:tkintdust_n}. 
	}
	\label{fig:tkin_lm}
\end{figure}

We test the expected increase of kinetic temperature obtained from \ammonia\ emission with the evolutionary tracer \lm, and with the limits proposed at \lm$=1$ and 10, for our sample of \hg\ clumps. 
Figure~\ref{fig:tkin_lm} shows the comparison between the kinetic temperature and the \lm\ parameter for the \goodsample. There is an increasing trend of the \tk\ with \lm, which is more evident at values of \lm$>$1. Clumps with ``coupled temperatures" of dust and gas, meaning that the difference between dust and kinetic temperatures is less than 10\percentword, are shown in a different colour in the figure. 
The increment of the kinetic temperature with respect to the \lm\ parameter in thermally coupled clumps is expected:~\cite{eli16} found that the dust temperature is well correlated with the \lm\ parameter in sources from the complete inner Galaxy \hg\ catalogue.
~\cite{urq18} also observed a tight correlation between the \lm\ and \td\ parameters in a sample of almost 8000 dense clumps from the ATLASGAL survey, finding the empirical relation \lm~$\propto$~(\td)$^{6.564}$.

%% Fig 12: 
%% Tkin vs L/M

\cite{eli16b} presented analytic relations between the \lm\ ratio and the dust temperature for an optically thin grey body, corresponding to the case of core/clump centrally heated by a young stellar object, finding a dependence of \lm$\ \propto\ $\td$^{(4+\beta)}$, with $\beta$ the spectral index, fixed to $\beta=2.0$. This relation is shown by the solid green curve in Figure~\ref{fig:tkin_lm}. The constant of proportionality is fixed to 16.5 to account for the general trend of the coupled clumps at \lm$>1.0$. We note that in the figure the dispersion of clumps with \tk\ coupled with \td, with respect to the analytic relation for the grey body, is partially a consequence of the SED fitting procedure of protostellar sources, where shorter wavelengths at 21-24 and 70~\um\ are considered in the estimation of bolometric temperatures.

The second panel of the figure shows with a solid line the binned median value of the sample, while the error bars represent the standard deviation. 
There is a trend of increasing kinetic temperature as a function of the \lm\ parameter, although this increase is less dependent on \lm\ than the prediction of the model of grey body emission. 
Also, the kinetic temperatures are higher than expected from the model at low \lm.
The median values of \tk\ are higher than the grey-body values for clumps with \lm\ below $\sim0.6$. The median value of \tk\ for clumps with \lm$<0.6$ is 14~K.

Prestellar sources have \lm\ values between $0.03<$\lm$<3$; they show a slight increase in \tk\ in this range, though most of the sources have temperatures below 20~K. The range of \lm\ values of protostellar clumps extends for more than three orders of magnitude, with low and relatively constant \tk\ for \lm$<1$. We note that protostellar clumps commonly show gas temperatures \tk$>15$~K even at \lm\ values below 1. It is unlikely to find dust temperatures above 15~K in the \hg\ catalogue for such low values of \lm.

\subsubsection{\tk/\,\td\ -- \lm}
\begin{figure}
	\centering
	\includegraphics[width=0.9\hsize]{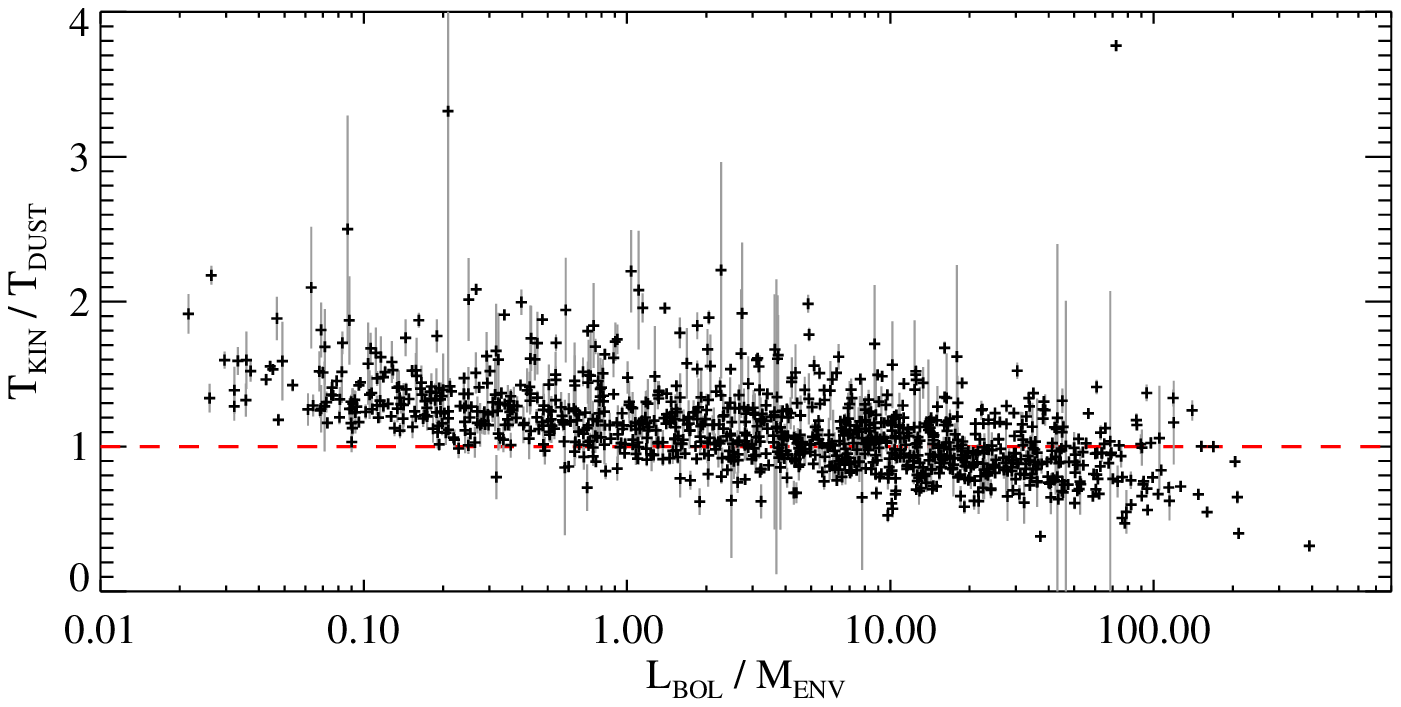}
	\includegraphics[width=0.9\hsize]{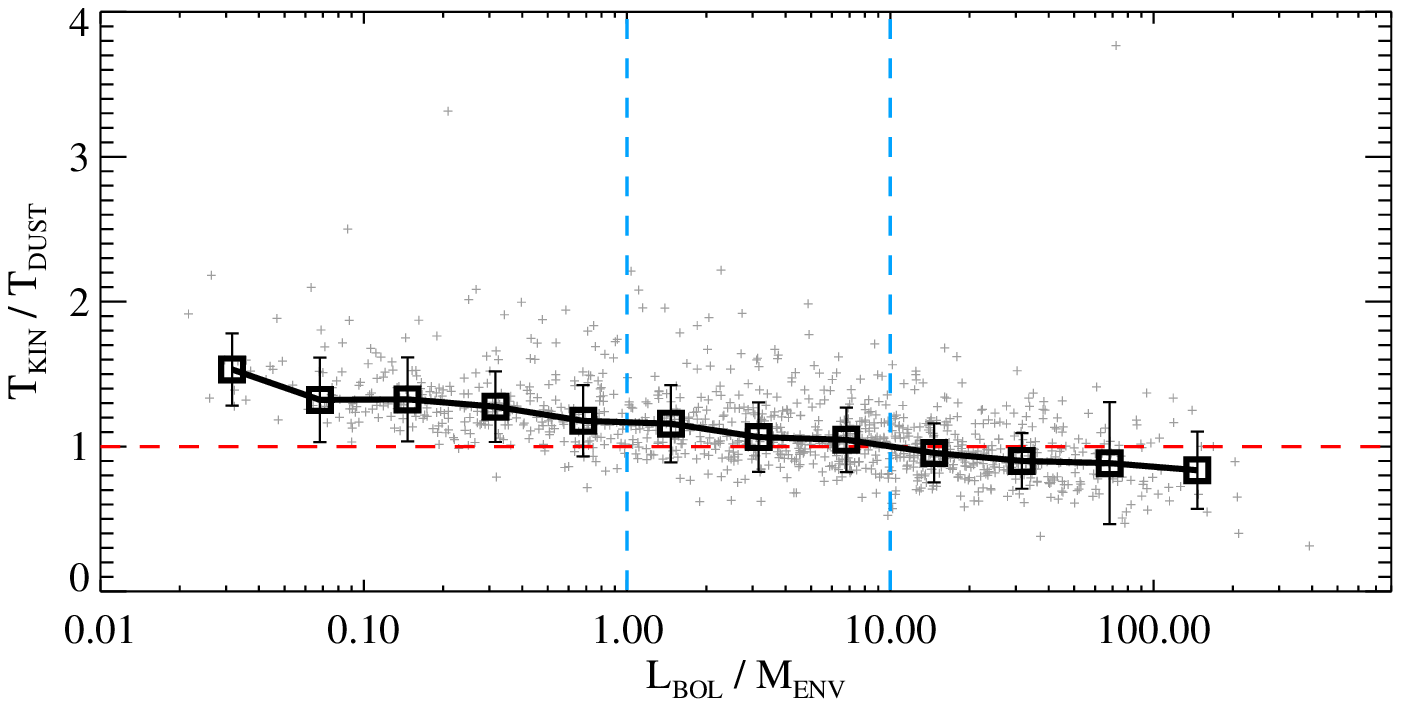}
	\includegraphics[width=0.9\hsize]{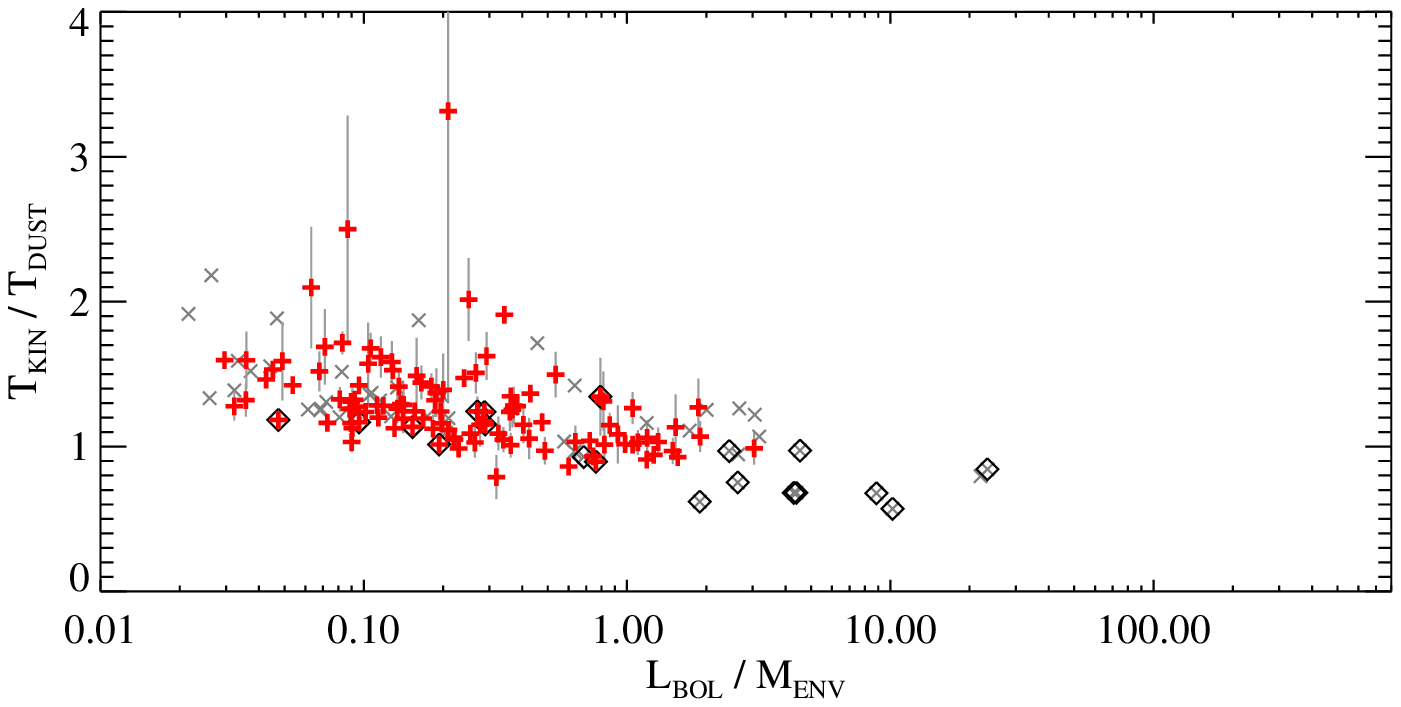}
	\includegraphics[width=0.9\hsize]{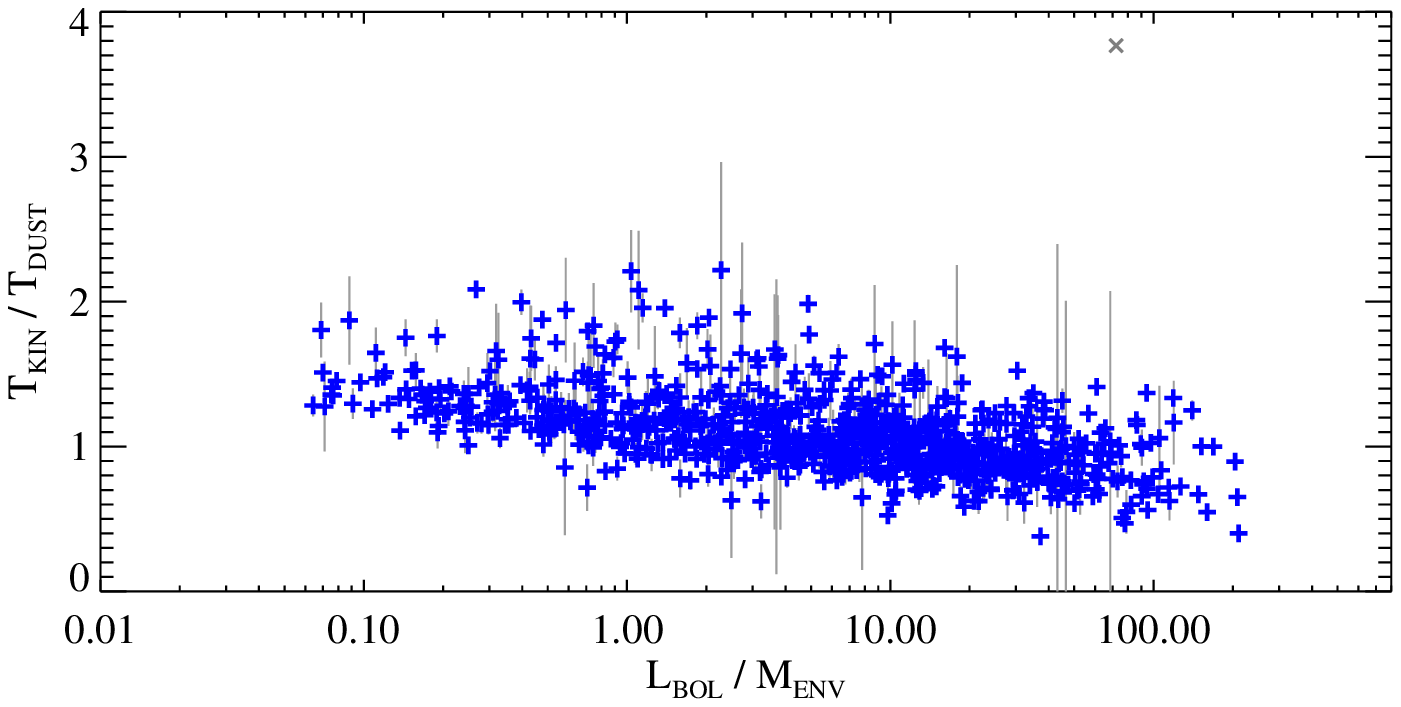}
	\caption{Ratio between kinetic temperature and dust temperature, as a function of the \lm\ parameter for \hg\ clumps. 
		In the second panel, the binned line is shown in solid black, with error bars showing the median absolute deviation of each bin.
		The dashed red line shows the 1:1 correspondence between \tk\ and \td. 
		The cyan vertical lines represent the transitions for the three phases of star formation in massive clumps detected from warm gas tracers. Last two panels show the prestellar (red) and protostellar (blue) clumps. The symbols follow the same description as Figure~\ref{fig:tkintdust_n}.
	}
	\label{fig:tkintdust_lm}
\end{figure}

The ratio between the kinetic temperature and the dust temperature, as a function of the \lm\ parameter, is shown in Figure~\ref{fig:tkintdust_lm}. The binned median values are shown in the figure, with errors representing the standard deviation for each bin. 

%% Fig 13
%% Tkin/Tdust vs L/M

There is a decrease in \tk/\td\ with increasing \lm, which is most clearly observed at \lm$<1$, and with temperature ratios near unity in the \lm\ range 3--20. 
At \lm$=1$, which corresponds to a grey body temperature \td$=16.5$~K ($\beta=2.0$) following the curve in Figure~\ref{fig:tkin_lm}, the sample shows a median value for the temperature ratio of 1.16, and for \lm$=10$, which corresponds to \td$=24.2$~K, the median value of the temperature ratio is 1.05.

We additionally test if results of the ratio \tk/\td\ change if we only consider high density clump and/or those sources with high masses, finding no substantial difference with respect to the overall final sample.
Therefore, the decoupling between temperatures of gas and dust is not an artefact of a possible selection of sources with low mass or low density. We will investigate further the differences between gas and dust temperatures in section~\ref{sec:coupling}.

\subsubsection{\ammonia\ fractional abundance -- \lm}

\begin{figure}
	\centering
	\includegraphics[width=0.9\hsize]{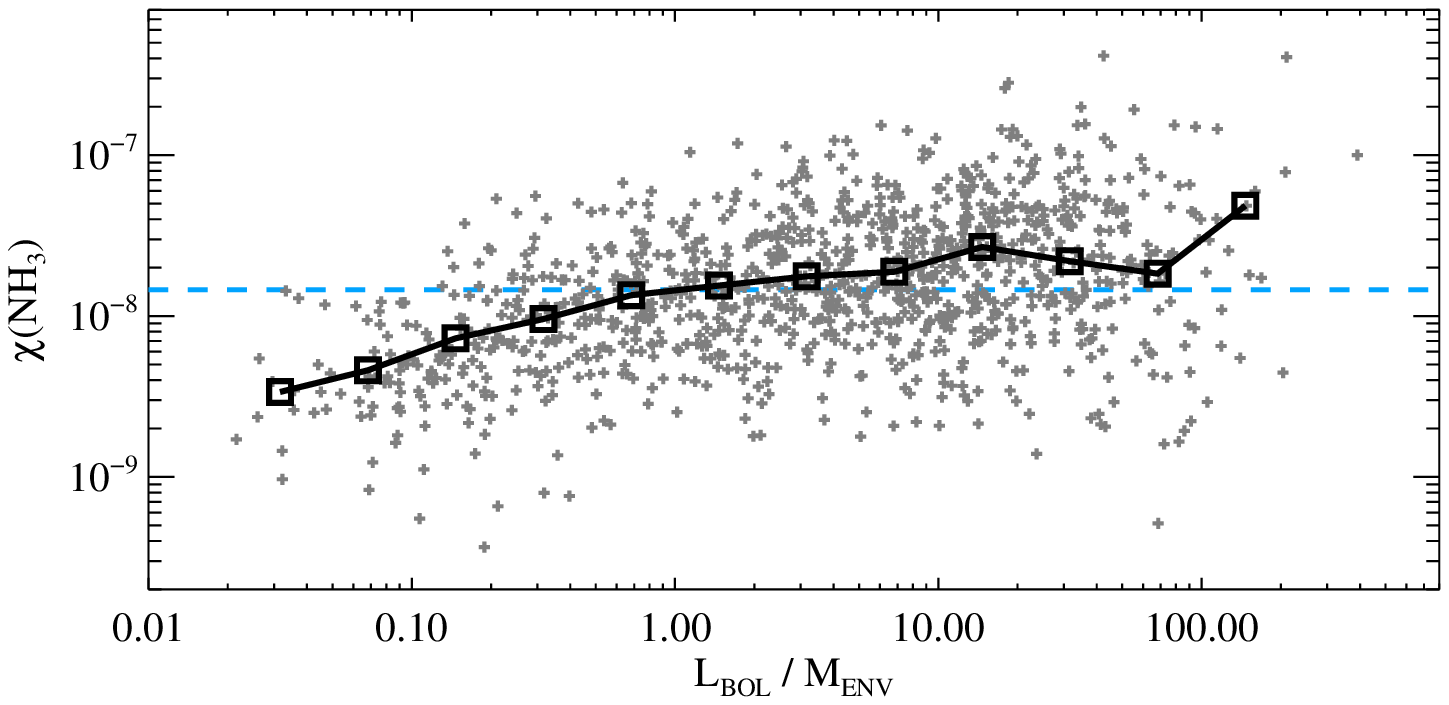}
	\includegraphics[width=0.9\hsize]{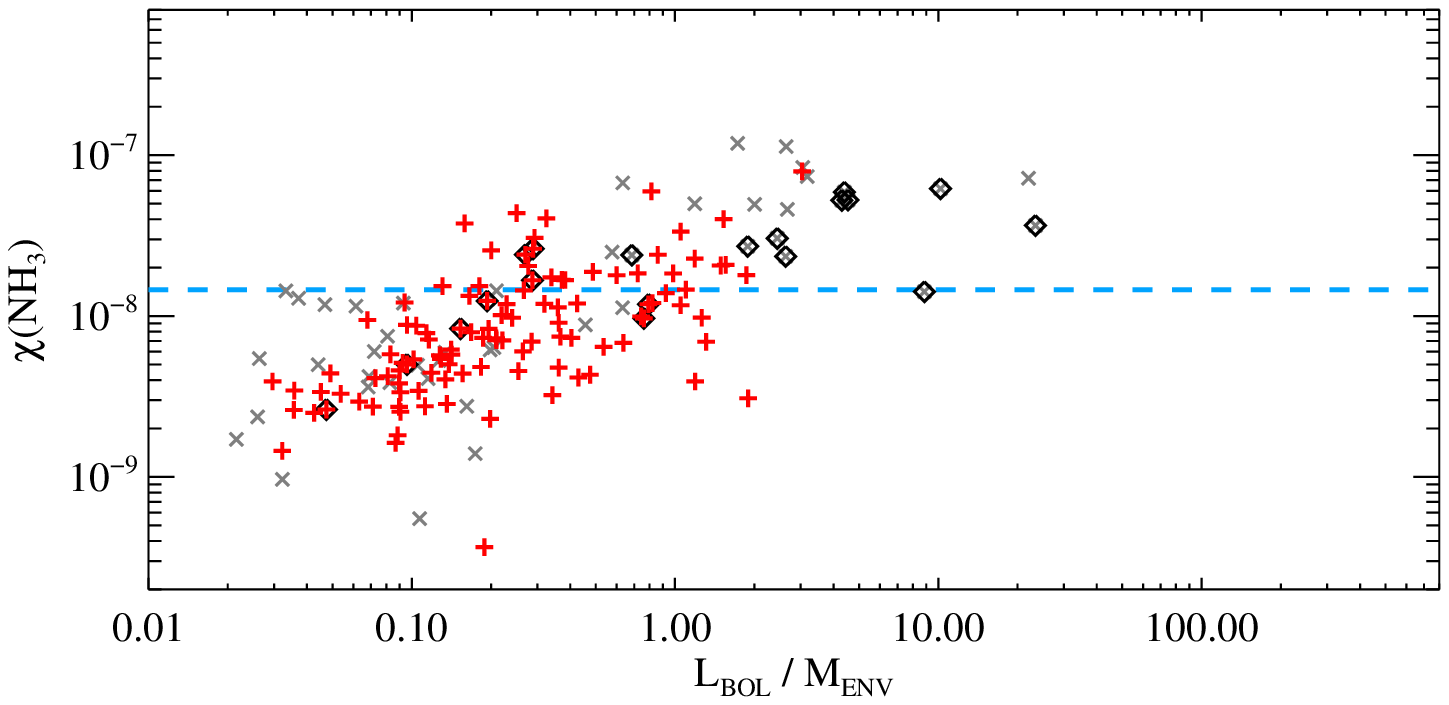}
	\includegraphics[width=0.9\hsize]{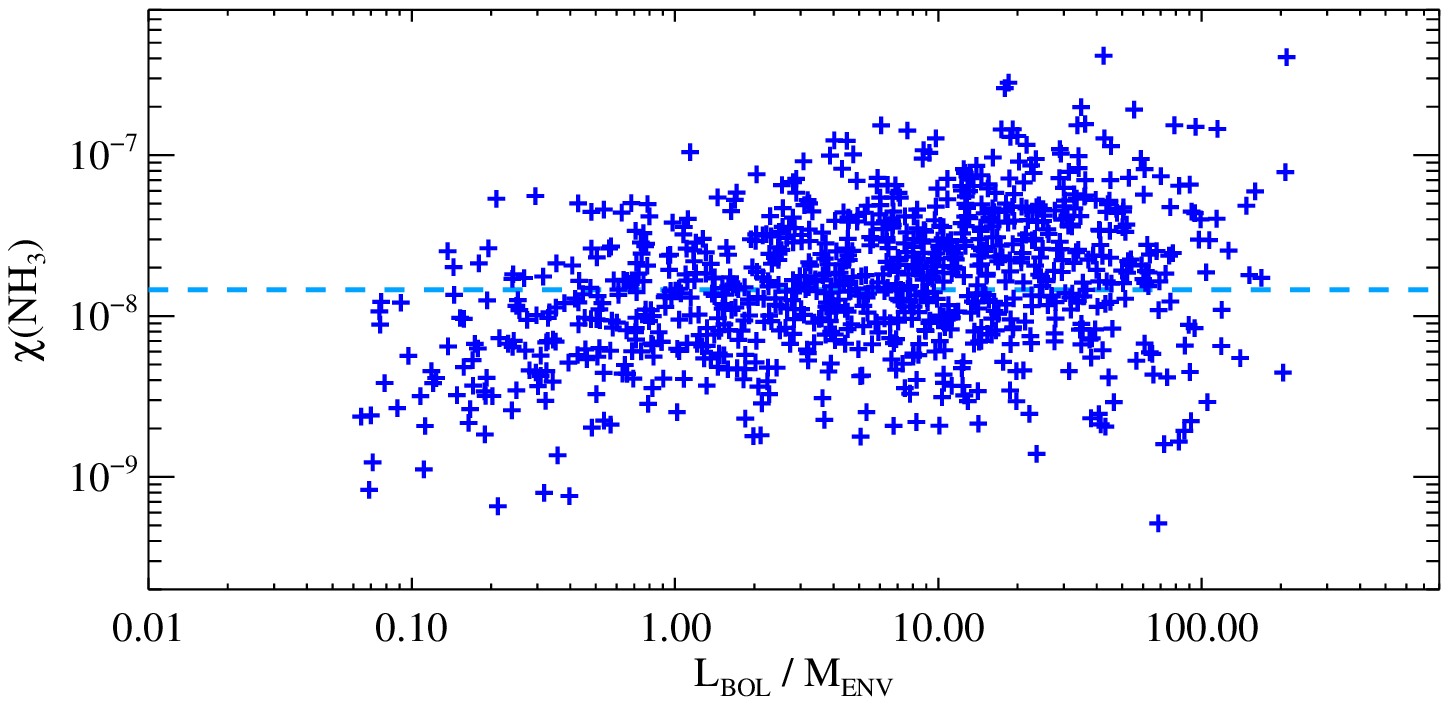}
	
	\caption{Estimated abundances of \ammonia, as a function of the \lm\ parameter for \hg\ clumps. In the first panel, the grey marks represent the \goodsample, and the binned line is shown in solid black.
		Second and third panel show the prestellar and protostellar clumps, respectively.
		Symbols are described in Figure~\ref{fig:tkintdust_n}. 	
		The dashed line represents the median value of the final sample, \chiammonia=1.46\ee{-8}. 
	}
	\label{fig:abund_lm}
\end{figure}

Figure~\ref{fig:abund_lm} shows the relation between the fractional abundance \chiammonia$=N$(NH$_3$)/$N$(H$_2$) and the \lm\ parameter for our final sample. The spread of the distribution is large, extending over two orders of magnitude. We found an increase of the \ammonia\ abundance with \lm, observed in both prestellar and protostellar clumps below \lm$=1$. Protostellar sources have a nearly constant median \chiammonia\ for \lm\ above this value.

%% Fig 14
%% NH3 abundance vs L/M

We compare our results with chemical models that give estimates of the \ammonia\ abundance in dense core/clumps at early stages of collapse and the posterior protostellar phase.
\cite{ben12} followed the evolution of both a prestellar and protostellar low-mass dense cores, with $n=5\times10^4$ cm$^{-3}$, using a time and depth-dependent gas-grain model. The models present a collapsing core with a temperature of 10~K (increasing up to 20~K in the protostellar case), and take into account the dependence of CO depletion onto the grains during collapse and non-thermal desorption processes (local heating by H$_2$ formation and cosmic rays). 
Their modelled \ammonia\ abundance increased almost three orders of magnitude \revision{in a few million years}, reaching a peak column density of $\sim10^{15}$ cm$^{-2}$ at 5.7\ee{6} yr, and roughly remained constant in subsequent time evolution.

Considering that our clumps have comparable volumetric densities, and for sources with $\Sigma=0.1$ g cm$^{-2}$, a column density $N$(\ammonia)$\sim10^{15}$~\cmc\ implies \chiammonia$\sim4\times10^{-8}$, this model could qualitatively explain the early evolution, traced by the \lm\ parameter, of the ammonia abundances of our sample of prestellar and protostellar clumps. 

Caveats to this approach involve the different timescales implied in their simulation with respect to the lifetimes of prestellar clumps. The median volumetric density of our sample, $n=9.3$\ee{4}~\cmv, implies a free fall time $t_{\textit{ff}}=\left(3\pi/32\textrm{G}\mu m_H n\right)^{1/2}\approx10^5$ yr.
In this sense, our sample can be compared with the chemical models of~\cite{bus11} of cores in the high-mass star forming region AFGL 5142. For cores with $n=10^5$ cm$^{-3}$, their model follows the time evolution of \ammonia\ and other species for prestellar and protostellar phases, and considering maximum temperatures of 12~K and 25~K for these phases, respectively. The values of \chiammonia\ increased up to 5.2\ee{-8} at $t\simeq1.3\times10^5$ yr, with almost no further variation at later times.

\subsubsection{Line width - \lm}\label{sec:linewidth_lm}

The velocity dispersion of the \ammonia\ observations can provide a measurement of the internal motions of the \hg\ clumps. Following \cite{dun11}, the thermal contribution of the FWHM linewidth is described by the kinetic temperature of the source, using
\begin{equation}
\Delta V_{\textrm{TH}} = 2\sqrt{2\ \ensuremath{ln}2\ }\times\sqrt{\frac{k\,\tk}{17m_H}}\ \ .
\end{equation}
The non-thermal contribution of the FWHM velocity dispersion $\Delta V$, obtained from the (1,1) transition, is described by $\Delta V_{\textrm{NT}} = \sqrt{(\Delta V)^2 - (\Delta V_{\textrm{TH}})^2}$.
The thermal contribution $\Delta V_{\textrm{TH}}$ for kinetic temperatures between 8 and 42~K corresponds to a linewidth of 0.06$-$0.14 \kms. 
We found values of $\Delta V_{\textrm{NT}}$ between 0.34 and 8.62 \kms, with a median of 1.88 \kms, and values above 5 \kms\ in only 13 sources. 

\begin{figure}
	\centering
	\includegraphics[width=0.9\hsize]{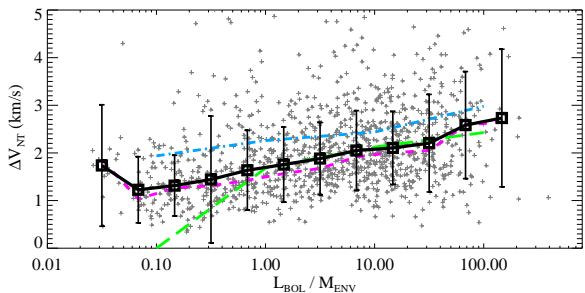}
	\caption{Non-thermal contribution to the velocity dispersion of clumps as a function of the \lm\ parameter. Binned line of the \goodsample\ is shown in solid black, with error bars showing the median absolute deviation of each bin. Trend observed in the three main catalogues of \ammonia\ associations are also presented: \textit{Cat-1} (BGPS) as dashed magenta line, \textit{Cat-2} (ATLASGAL) as dash-dotted light blue line, and \textit{Cat-5} (RMS) as green long-dashed line.	Only the clumps with $\Delta V_{\textrm{NT}}\leq5.0$ \kms\ (1055 sources) are shown in the plot.
	}
	\label{fig:veloc_lm}
\end{figure}

Figure~\ref{fig:veloc_lm} shows the relation between non-thermal linewidth and the \lm\ parameter. \revision{Although the Spearman's correlation coefficient is rather small \revision{($\rho= 0.36$, with a p-value  $<0.01$)} and therefore implies a weak correlation,} we found a general increment of the non-thermal linewidth with \lm\ despite the large observed dispersion.

%% Fig 15
%%: non thermal velocities and virial parameter vs L/M parameter

In order to avoid possible bias from the different spectral resolutions of the ammonia catalogues, we consider the trends observed in non-thermal linewidths of the three main ones: \ammonia\ observations in BGPS clumps (\textit{Cat-1}, spectral resolution of 0.08 \kms), in ATLASGAL clumps (\textit{Cat-2}, resolution of 0.5 \kms), and in RMS sources (\textit{Cat-5}, resolution of $\sim$0.32 \kms\ after smoothing). These catalogues are associated with 92\percentword\ of clumps of the final sample (see Section~\ref{sec:crossmatch}). BGPS sources, dominant in the sample, have the same behaviour of the overall trend along with RMS sources for \lm$>1$, while the ATLASGAL sources, with lower spectral resolution, show a similar increasing trend but shifted toward larger values of $\Delta V_{\textrm{NT}}$. 

It is likely that the increasing trend observed for non-thermal velocity dispersion with increasing \lm\ is due to different feedback processes that appear in more evolved protostellar and young stellar objects and that provide turbulence to the clumps, such as jets and molecular outflows~\citep{fed14,nak14} and radiation pressure-driven instabilities~\citep{kru12}.

The large dispersion of non-thermal motions at all \lm\ values suggests that a fraction of the observed turbulence must be primordial rather than caused by the formation of star at the interior of clumps. In section~\ref{sec:nonthermal} we will discuss a possible explanation of the observed $\Delta V_{\textrm{NT}}$ related to the hierarchical collapse of structures at all scales (molecular clouds, clumps, and cores) and the resultant release of gravitational energy.

\subsection{Coupling between gas and dust temperatures.}
\label{sec:coupling}

In compact sources with $n\gtrsim10^4$~\cmc, it is commonly expected that the dust and gas are thermally coupled.~\cite{cri10}, for example, produced dust envelope density and temperature profiles for five intermediate-mass protostellar objects, finding that the gas and dust are thermally coupled in envelopes with average densities above \eten{4}~\cmv. In another study,~\cite{mar12} investigated the effects of heating and cooling of dust and gas in star-forming cluster simulations, showing that gas and dust temperatures are similar due to dust-gas collisional coupling in high density environments, although at initial times of the simulations the dust temperature is cooler than the gas temperature. They explained this difference as consequence of early evolution, with stars in formation stages not able to heat yet the surrounding dust, and since \td\ is lower than the gas, the dominant heating source of low-density gas is cosmic rays.~\cite{bat14} gave a similar argument for early evolution for two clumps traced in \ammonia: they found \td$<$\tk\ discrepancies in their clump without showing indicators of stellar activity, while agreement within 20\percentword\ for \td\ and \tk\ for their active clump at later stages of star formation.

Here we examine physical mechanisms and possible observational artefacts that can explain the differences between \tk\ and \td\ in our sample. We note that in our analysis the physical properties derived from dust and gas are averaged over the clumps size and the \ammonia\ beam, respectively. Approximating the SED of sources with a single modified blackbody presents several limitations, with \td\ responsive mainly to dust with strong emission in \textit{Herschel} bands. In addition, we do not account for variations of temperatures across the clumps that are expected from new protostellar activity in their interior. 
A proper radiative transfer model of sources that considers the different beam response of \textit{Herschel} bands, and that includes density and temperature profiles along with the treatment of surrounding environment is beyond the scope of this work, and it will be a topic of research in a future article. 

\subsubsection{Variations with respect to the spectral index}

\begin{figure*}
	\centering
	\includegraphics[angle=-90,width=1\hsize]{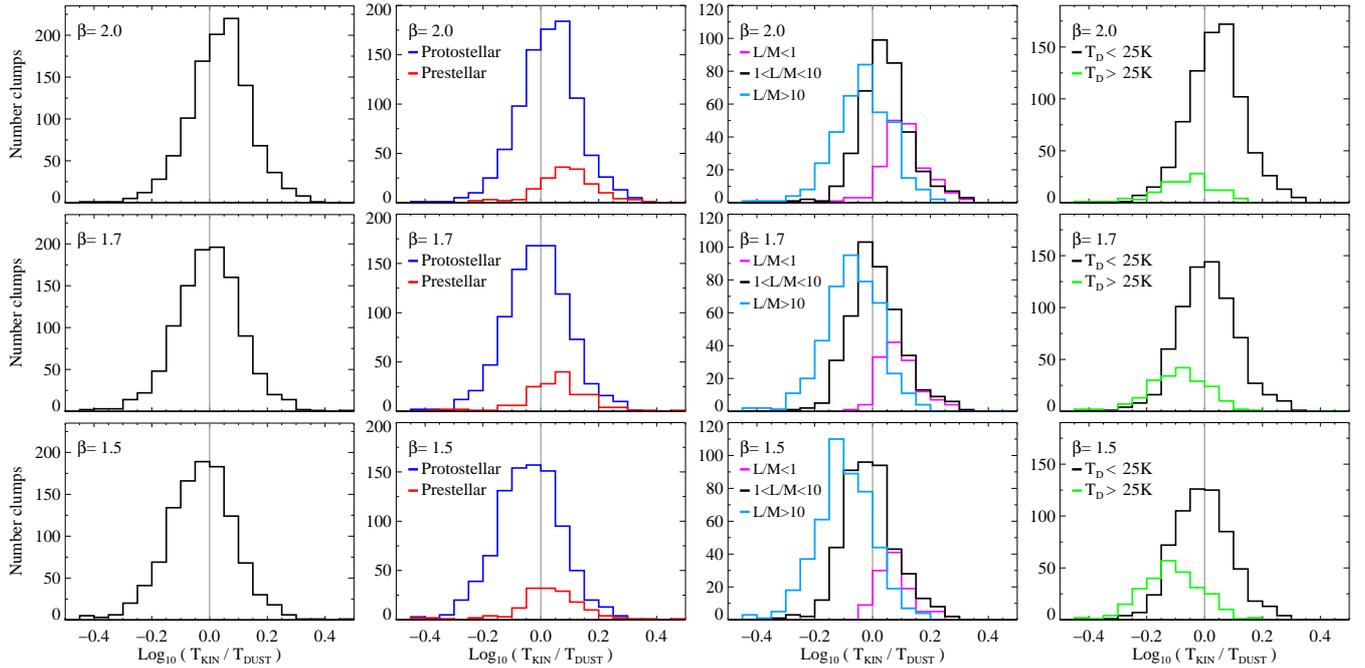}
	\caption{Distribution of the ratio \tk/\td, with the dust temperature estimated for different fixed values of the spectral index, $\beta=$2.0, 1.7 and 1.5. Left panels represent the complete sample of \hg\ clumps. Second column shows the distribution of prestellar and protostellar clumps. Third column shows the distribution of protostellar clumps for different ranges of \lm. Panels on right show protostellar clumps with two different ranges of dust temperature. 
	}
	\label{fig:comparison_beta}
\end{figure*}

We explore if the difference between gas parameters derived from \ammonia\ and dust properties from SED fitting from Herschel bands is a product of our dust opacity model on the modified blackbody formula (Equation~\ref{eq:greybody}). The SED fitting for each source of the \goodsample\ was evaluated with three different values of the spectral index $\beta$. Figure~\ref{fig:comparison_beta} shows the results of the SED fitting using $\beta=2.0$, $\beta=1.7$, and $\beta=1.5$. The columns in the figure represent distributions of sources for different spectral index for the \goodsample, prestellar and protostellar clumps, protostellar sources at different ranges of the \lm\ parameter, and protostellar clumps with \td\ above and below 25~K.

%% Fig 16
%% Comparison beta

Considering the complete sample, 
a spectral index $\beta=1.7$ will make \tk\ equal to \td\ on average, since the increase of dust temperatures place the distribution of the ratio \tk/\td\ almost symmetrically around the equality.
A value $\beta=1.5$ will push this distribution to even lower values of the temperature ratio.

The distribution of protostellar clumps, being 85\percentword\ of the final sample, shows the same progression toward lower values of \tk/\td\ for different $\beta$. Conversely, prestellar clumps are less affected by the change of $\beta$, and kinetic temperatures are in general higher than dust temperatures, independently of the choice of $\beta$. 
This is a consequence of the lower values of the dust temperatures for prestellar sources (median value of \td$=11.5$~K for $\beta=2.0$), compared with protostellar sources (\td$=17.3$~K, $\beta=2.0$).
 
Protostellar clumps with different values of the \lm\ parameter show different overall distributions with respect to \tk/\td. For $\beta=2.0$, the three ranges \lm$<1$, $1<$\lm$<10$, and \lm$>10$, have 170, 378, and 360 associated clumps, respectively.
The locus of the distribution of protostellar sources with \lm$<1$ is maintained for lower values of $\beta$, with a median \tk/\td$=1.27$ and a median difference of 3~K for $\beta=2.0$. Meanwhile the distributions of sources with $1<$\lm$<10$, and \lm$>10$ follow the general decrease toward lower values of the temperature ratio. The distribution of protostellar sources with \lm\ in the range 1--10 for $\beta=1.7$ is close to symmetry toward its peak at \tk/\td$\sim1$.

The low values of \tk/\td\ for \lm$>10$ could be explained as a consequence of a new high-mass young star or cluster inside the protostellar clump, heating their immediate surroundings 
and setting a large temperature gradient in an area small in solid angle compared to the region where the different transitions of \ammonia\ are emitted.
This picture is supported by the distribution of protostellar clumps with high dust temperature, as shown in the right-most panels of Figure~\ref{fig:comparison_beta}. Sources with \td$>25$~K have in general lower values of the ratio \tk/\td, without depending of the choice of $\beta$, while the distribution of colder sources (\td$<25$~K) approach in average to thermal coupling for values of $\beta$ lower than 2.0.

Large values of \lm\ are expected with late evolutionary stages, when substantial star formation has occurred. This activity increases the bolometric luminosity and sets a gradient in temperature at the interior of clumps that is sampled differently by the \ammonia\ emission ($T_A^*(\textrm{NH}_3)\sim$\tk, with dependence of the abundance profile of \ammonia), and the continuum emission ($F_\nu\sim\ \exp(-1/$ \td$)$, in the optically thin limit). 
In this sense, Figure~\ref{fig:tkin_lm} already showed that the dependence of the beam-averaged \tk\ with \lm\ is not as strong as the \td--\lm\ analytic relation. Nevertheless, the overall discrepancy of \tk\ and \td\ in the protostellar sources with $\lm>10$ is within the statistical errors of the sample (see Figure~\ref{fig:tkintdust_lm}).

As a result of this test, we found that protostellar clumps with \lm$>1$, which correspond to the majority of the final sample of \hg\ clumps with associated \ammonia\ emission, show a better agreement between \tk\ and \td\ for a spectral index $\beta=1.7$, a value closer to the commonly used OH5 opacity model~\cite[$\beta=1.8$, coagulated grains with thin ice mantles,][]{oss94} in dense regions of high-mass star formation, with respect to $\beta=2.0$, which is more consistent with the model of~\cite{dra84} of silicates and carbonaceous grains, and that is used in the \hg\ catalogue of physical properties.

The thermal decoupling of prestellar sources and protostellar clumps with \lm$<1$ is not the result of a different selection of $\beta$, or at least cannot be explained only by this parameter.

In the next few subsections, we will examine different effects that could explain the discrepancies between gas and dust temperatures in those peculiar cases with \tk/\td\ much above unity.

\subsubsection{Heating by nearby sources and UV field}

Our selection criteria for the association between \hg\ sources and \ammonia\ counterparts, in which only single association between dust clump and ammonia observation, prevents contamination by additional dust features with high temperatures that could introduce differences between \tk\ and \td. Furthermore, we checked for prestellar clumps with ratio \tk/\td$>1.3$ for the closest \hg\ source. We do not find another \hg\ clump at a distance less than 1\arcminword\ with high dust temperature or in latest stages of evolution as traced by their \lm\ parameter. 
In addition, we inspected the proximity of our sources with OB stars and \hii\ regions through the association of clumps with \textit{Spitzer} infrared bubbles from the Milky Way Project catalogue~\citep{sim12}. Following the same criteria of~\cite{pal17} for the association between \hg\ clumps and IR bubbles, we found 113 sources that could be \revision{heated by an ionizing front}, but only 9 of them have a ratio \tk/\td$>1.3$ and \lm$<1$.
	
We discard then the idea that an increase of \tk\ in sources in early evolutionary stages is produced by a close and highly energetic protostellar source, and although it is possible that some clumps could be heated by a close massive star, we do not consider this a dominant process in our sample.

The median value of $N$(H$_2$)$=1.37\times10^{23}$~\cmc\ in our sample is translated into a visual extinction of $A_{\textrm{V}}\sim150$ when using $R_V=3.1$ and the conversion $N$(H$_2$)$/A_\textrm{V}=9.4\times10^{20}$ cm$^{-2}$ mag$^{-1}$ of~\cite{boh78}. If instead we use the extinction model of~\cite{wei01} with $R_V=5.5$, that is found more appropriate in regions with $A_K>1$ mag~\citep{cha09}, then the conversion is given by $N$(H$_2$)$/A_\textrm{V}=6.86\times10^{20}$ cm$^{-2}$ mag$^{-1}$ and the median visual extinction for the sample is $A_{\textrm{V}}\sim200$. If only the prestellar sources are considered, the median visual extinction still has a large value, $A_\textrm{V}\sim180$.

These results are significantly higher than $A_V\sim5$ (corresponding to a column density of $\sim$100~\msun\,pc$^{-2}$), at which point the gas of the clumps is expected to be shielded from the interstellar radiation field (ISRF) and UV heating becomes negligible\footnote{The parametrisation given by~\cite{dra78} is considered for the portion of the ISRF at ultraviolet wavelengths.}. 

Nevertheless, there are few prestellar sources with relatively low values of column density (8 prestellar clumps and 1 protostellar with \lm$<1$, with $N$(H$_2$)$<4.0$\ee{22}~\cmc\, or $A_V\sim40$) that could be affected by UV heating, at least in their outer layers.
The photoelectric heating rate $\Gamma_{\textrm{pe}}$ is proportional to the product $G(r)n(\textrm{H}_2)$, with the parameter $G(r)$ a measure of the attenuation of the ISRF by the extinction of the source~\citep{bak94}. For a simple plane-parallel slab approximation, $G(r)\propto \exp(-A_V)$,~\cite{you04} found that in pre-protostellar cores with $A_V\sim1-3$, the UV heating increases the gas temperature over the dust temperature by a few degrees in the outer layers of the sources.

Moreover, simulations of molecular clouds expanding over a large range of column densities presented by~\cite{cla14}, showed that in a given point on the line of sight towards a dense source the angle-averaged column density can be different than the measured along that line of sight.~\citeauthor{cla14} showed that the angle-averaged ``effective visual extinction" of a measured $A_V\sim10$ is found lower by a factor $\sim5$.
 	
Therefore, despite the fact that the high column densities found in the sample prevent significant overall temperature changes due to UV radiation, some differences between dust and gas temperatures could be produced in the outer layers of a clumpy medium by photoelectric heating even in sources with relatively high measured $A_V$.
	
\subsubsection{Cosmic ray heating}

In prestellar clumps with high densities, we can assume that cosmic rays provide the main mechanism of gas heating in the interior of the sources. We follow here the estimation of thermal balance by cosmic-ray heating presented by~\cite{you04}, with a similar derivation as in~\cite{gol01}, and we direct the reader to their work for further references and derivation of parameters.

The heating by cosmic rays is given by:
\begin{equation}
\Gamma_{\textrm{gas,cr}}=10^{-27}n(\textrm{H}_2)
\left( \frac{\zeta}{3\times10^{-17}\textrm{s}^{-1}} \right)
\left(\frac{\Delta Q}{20\ \textrm{eV}}  \right)
\textrm{ergs cm}^{-3}\textrm{s}^{-1},
\end{equation}

\noindent with $\zeta$ the assumed ionization rate, and $\Delta Q$ the energy input per ionization. The gas-dust energy transfer rate is estimated from:
\begin{eqnarray}
\Lambda_{\textrm{gd}}&=&9.0\times10^{-34}n(\textrm{H}_2)^2\sqrt{\tk}
\left[ 1-0.8\exp\left(-\frac{75}{\tk}\right) \right] \nonumber \\
&&\times\,(\tk-\td)\left(\frac{\Sigma_d}{6.09\times10^{-22}}\right)\,
\textrm{ergs cm}^{-3}\textrm{s}^{-1},
\end{eqnarray}

\noindent where it was assumed an average dust cross section per baryon $\Sigma_d$, a power-law size distribution $n(a)\propto a^{-3.5}$ for collisions between gas and dust, and a temperature-dependent accommodation coefficient equal to  $0.37[1-0.8\exp(-75/\tk)]$. Therefore, balancing these two terms for a given density $n$ give us an estimation of the expected difference between gas and dust temperatures (\tk-\td). 

For the following analysis, we considered among the prestellar clumps with good SED fitting (110 sources) those with ratio \tk/\td\ above 1.3 (39 sources). Then, we estimated the expected temperature difference (\tk-\td)$_\textit{exp}$ from thermal balance of cosmic ray heating, with the measured (\tk-\td)$_\textit{m}$ from \ammonia\ gas temperature and dust temperature from SED fitting. For 28 sources with (\tk-\td)$_\textit{exp}$ within 5~K of the the measured (\tk-\td)$_\textit{m}$, we found a median $n\sim1.2\times10^{5}$~\cmv. The other 11 clumps have in general higher densities (median $n\sim3\times10^{5}$~\cmv\ and reaching $1.3\times10^{6}$~\cmv), and small values (\tk-\td)$_\textit{exp}$ are expected for these high densities.

These result suggest that heating by cosmic rays could explain $\tk>\td$ for prestellar clumps with the lowest densities, but those with high $n$ are expected to be thermally coupled. We note that a higher ionization rate $\zeta$ is expected in clumps located toward the inner Galaxy or near recent supernovae.

\subsubsection{Effects of depletion}

\cite{gol01} modelled the effects of molecular depletion on the thermal balance of dense dark cloud cores. Regions with densities $n\gtrsim10^{4.5}$~\cmv\ are expected to have gas and dust temperatures closely coupled, while for lower densities a lower fractional abundance produces a moderate reduction of the gas cooling rate and therefore higher values of \tk\ with respect to \td\ are expected. 

In our sample, we find variations within a factor of 10 in the \chiammonia\ in prestellar sources and protostellar clumps with \lm$<1$. Then, considering the models of \cite{gol01} with a factor 10 depletion and $\log[n~(\textrm{cm}^{-3})]=4.0-5.0$, we expect to find \tk/\td\ in the range 2.2--1.4, resembling our results. Nevertheless, 
those models predict values of \tk/\td$\lesssim1.07$ for $\log[n~(\textrm{cm}^{-3})]=6.0$, and therefore 
the values of \tk/\td\ above 1.5 in some of the clumps with higher densities cannot be explained as a consequence of gas depletion.

\subsubsection{Opacities derived for dust and \ammonia\ emission}

\begin{figure}
	\centering
	\includegraphics[width=0.95\hsize]{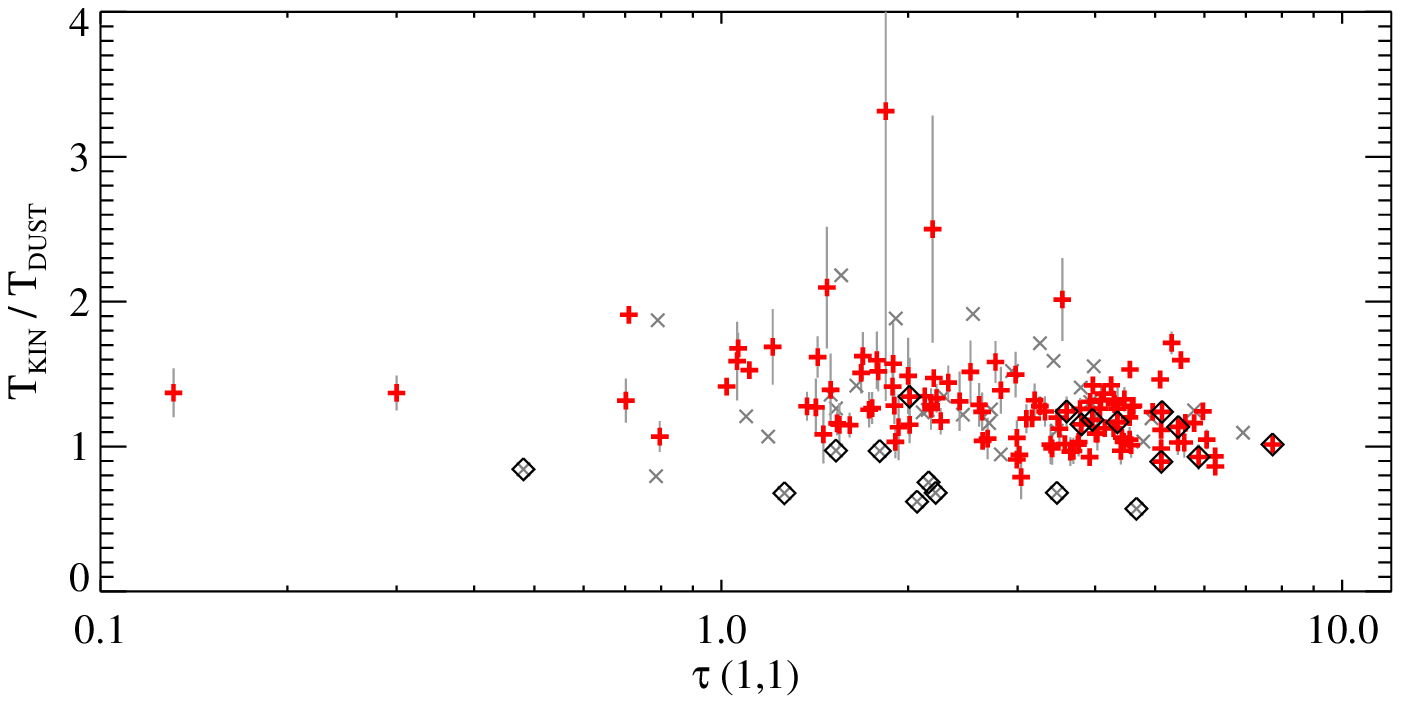}
	\includegraphics[width=0.95\hsize]{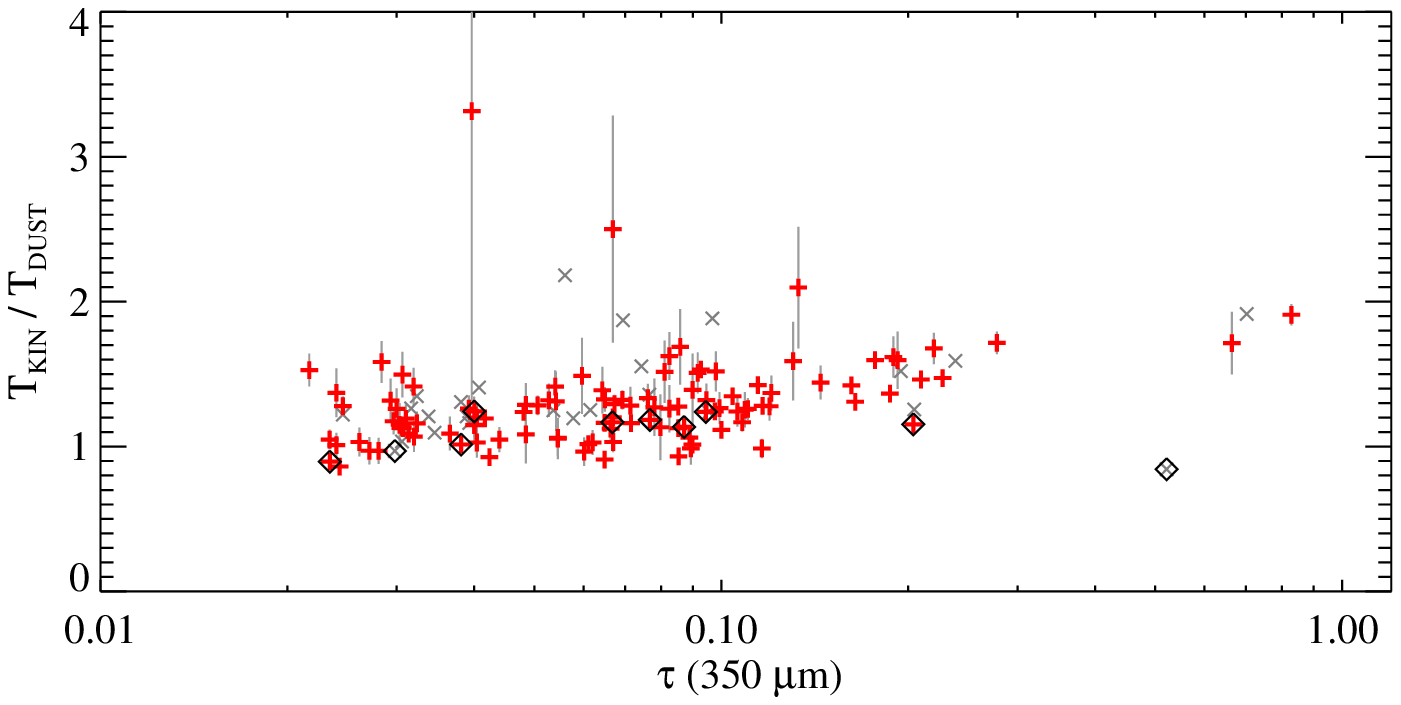}
	\caption{Ratio between kinetic temperature and dust temperature, as a function of the optical depth of the \ammonia\ (1,1) line (upper panel), and dust optical depth at 350~\um\ (bottom panel), for prestellar sources. 
		In grey X marks, clumps with less reliable SED fitting after visual inspection, and diamond symbols show SPIRE-only clumps.
	}
	\label{fig:tkintdust_tau}
\end{figure}

We look into possible optical depth effects that could explain the differences between the gas and dust temperatures for our sample of prestellar sources.
In Figure~\ref{fig:tkintdust_tau} we compare the ratio \tk/\td\ with the optical depth of \ammonia\ emission ($\tau(1,1)$), as obtained directly from the respective ammonia catalogues, and the dust optical depth estimated at 350~\um\ ($\tau(350\mu$m$)$), estimated with the fitted $\lambda_0=(c/\nu_0)$ of the \hg\ catalogue of physical properties (see section~\ref{sec:cat_hgprops}).

%% Fig 17
%% Tkin/Tdust vs taugas

Most of the sources with $\tau(1,1)\lesssim2$ present large temperature variations ($>$30\percentword) between gas and dust, and seem close to \tk$\ \simeq\ $\td\ for $\tau(1,1)$ above that limit. In contrast, the majority of clumps with low values of $\tau(350\mu$m$)$ are closer to thermal coupling, and a progressive increment of \tk/\td\ is found for $\tau(350\mu$m$)\gtrsim0.1$.

We studied the variation of the optical depths of prestellar clumps as a function of the \lm\ parameter, in three distinctive ranges:
\begin{itemize}
\item \lm$<$0.1: sources with $0.06<\tau(350\mu$m$)<0.3$ and $\tau(1,1)>1$. 
Dust optical depths are relatively high, with cold sources (\td$\sim10$~K) showing large ratios \tk/\td$>1.2$ (median \tk/\td$=1.4$).
\item 0.1$<$\lm$<$0.5: sources with $0.02<\tau(350\mu$m$)<1$ and $0.1<\tau(1,1)<10$. Optical depths of both dust and gas have wide ranges of values, and accordingly we found sources with large range for the ratio between temperatures (\tk/\td$\simeq1-2$).
\item \lm$>$0.5: sources with $\tau(350\mu$m$)<0.1$ and $\tau(1,1)>0.8$. In this range we found small variations between \tk\ and \td, while the sources show low dust optical depths and high \ammonia\ optical depth values.
\end{itemize}

The increase of \tk/\td\ in clumps with high density $n\sim10^6$~\cmv\ and high dust optical depth may be produced by undetected low-mass protostellar activity. In a source considered prestellar due to lack of 70~\um\ emission, deeply embedded low-to-intermediate mass star formation could be locally heating the dense gas in their surroundings, but not the bulk dust of the dense clump. For those prestellar sources with optical depths in the range $\tau(350\mu$m$)=0.1-1$, we expect that: (1) \td\ will be underestimated because of increase of optical depth at shorter wavelengths;  
(2) the optical depth $\tau(70\mu$m$)=2.5-25$ and fluxes at 70~\um\ of $\sim0.01$ Jy, below the detection limits of \hg\ sources in this band~\citep{mol16a}.
\revision{In Appendix~\ref{sec:prestellarysos}, we performed some tests following this idea of possible association between prestellar clumps and low-mass star formation, and exploring the fragmentation of prestellar clumps in multiple compact sources.}

\revision{We cannot rule out that higher resolution observations would be able to identify protostellar activity with arcsec and sub-arcsec observations.
In recent years, some studies using ALMA interferometric data have shown sources 	
 previously considered as starless or prestellar presenting indications of protostellar activity~\citep[e.g.,][]{tan16,fen16}. Nevertheless, the discrepancy between gas and dust temperature in our sample of prestellar clumps does not seem to be dominated by undetected protostellar souces, at least as shown by the available datasets.}

\subsubsection{\tk\ -- \td\ in other surveys of high-mass star formation}

Independent estimations of \td\ and \tk\ have been presented by different groups in the past.
Comparisons of kinetic temperatures from ammonia and dust temperatures from SED fitting of continuum emission with submm to mid-IR bands using SCUBA and IRAS fluxes, find a general agreement between both temperatures toward massive star forming regions, though the large dispersion of the estimated values and the small sample of sources~\citep[e.g.,][]{hil10,mor10}.

Comparison of dust temperatures obtained from SED fitting with \textit{Herschel}/\hg\ maps exhibits a better agreement with respect to the \ammonia\ gas temperature measurements on small samples of dense clumps as shown, for example, by~\cite{guz15} who compared dust temperatures obtained from \hg\ images toward 150 ATLASGAL sources, and~\cite{gia13} who targeted 39 sources previously mapped in 1.2~mm continuum emission. The better agreement is in part due to the higher angular resolution of Herschel bands compared to IRAS.

In comparison with other surveys, our work presents the largest sample to date of independent measurements of \tk\ and \td\ in high-mass star forming regions at different environments.

\subsection{Virial parameters}\label{sec:virial}

The virial parameter defined as $\alpha\equiv(5\sigma_\upsilon^2R)/(GM)$ is commonly used as a measure of the gravitational boundedness of molecular clouds and the tendency of a cloud to fragment against collapse. We measured the values of $\alpha$ for each clump of our sample using the simple form
 $\alpha=M_{\textit{virial}}/M_{\textit{clump}}$, with 
\begin{equation}
\left(\frac{M_{\textit{virial}}}{\msun}\right) = 210\left(\frac{\Delta V}{\mathrm{km\ s}^{-1}}\right)^2\left( \frac{R_{250}}{\mathrm{pc}} \right)\ ,
\end{equation}

\noindent the FWHM velocity dispersion $\Delta V$ measured from \ammonia (1,1) emission, and $R_{250}$ the radius of each clump.
We assumed that the emission of both transitions of ammonia (1,1) and (2,2) originates in the same area described by the \hg\ clumps, and a simple approach of spherical symmetry with constant density.
 
Under the above definition, the critical virial parameter is found at $\alpha_{\textrm{cr}}\sim2$, and therefore sources with $\alpha<\alpha_{\textrm{cr}}$ are considered ``supercritical" and in the process of collapse, while sources with $\alpha>\alpha_{\textrm{cr}}$ are ``subcritical" and unbound or in the process of expansion.

\begin{figure}
	\centering
	\includegraphics[width=1.\hsize]{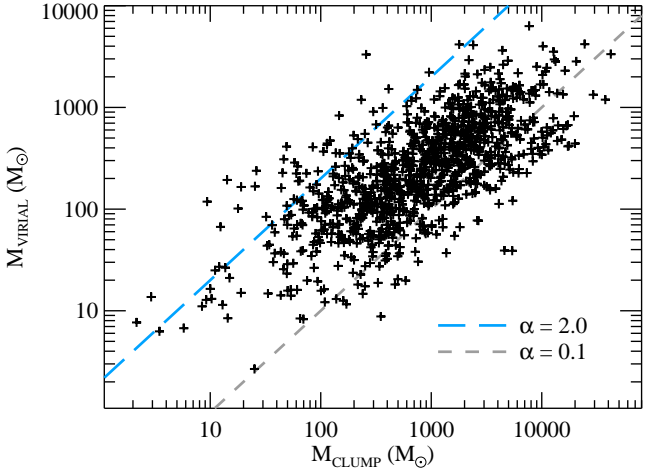}
	\caption{Comparison between virial masses estimated from \ammonia, and masses obtained from SED fitting for \hg\ clumps. The segmented lines show the locus of different values for the virial parameter  $\alpha=M_{\textit{virial}}/M_{\textit{clump}}$.
	}
	\label{fig:virial}
\end{figure}

Figure~\ref{fig:virial} shows the relation between the virial mass obtained from the \ammonia\ sources and the clump mass obtained from the SED fitting. The median and average values of $\alpha$ for this subsample are 0.3 and 0.6, respectively, 
with 774 clumps (72\percentword) with $0.1<\alpha<1$, and only 48 sources are found with $\alpha>2$. 
These values are consistent with subvirial motions. If instead of the 250~\um\ band we considered physical sizes calculated from the band at 350~\um, which is also estimated for all the sources of the sample (see section~\ref{sec:clump_sizes}),
the virial masses will increase by a factor $\sim$1.35, with the consequent increase of the virial parameter by the same factor. 

%% Fig 18
%% virial parameter vs L/M paramter

We compared our results with other values from the literature. Estimates of virial parameters derived for BGPS~\citep{svo16,dun11} and ATLASGAL~\citep{wie12} clumps give values of $\alpha\sim0.7$ and $\alpha<0.5$, respectively, using millimeter-submillimeter continuum emission to estimate the total mass of clumps and assuming dust temperature equal to the kinetic temperature estimated from ammonia observations.

\cite{kau13} compiled a sample of 1325 sources with estimates of virial parameters that includes samples of giant molecular clouds (750 sources), high-mass star forming clumps (260 sources), and cores toward both low and high-mass star forming regions. 
Their sample of clumps is derived from~\cite{wie12}, which is included in our study, with \ammonia\ temperatures used for mass estimations, and the rest of the sample considered different tracers of dense gas ($^{13}$CO, NH$_2$D, N$_2$D$^+$) and a fixed dust temperature (\td~$=10$~K).
In their study, \citeauthor{kau13} found low virial parameters ($\alpha\ll2$) observed toward regions of high-mass star formation.
Our results, focused mainly in the size scale of clumps with \ammonia\ line emission as single tracer of dense gas emission and independent measurements of gas and dust temperatures, are in agreement with low virial parameters $\alpha< 2$.

\revision{A predominance of low values of the virial parameter were also reported by \cite{urq14b} for 466 massive star-forming clumps. Their sample consisted of ATLASGAL sources associated to indicators of protostellar evolution (methanol masers, YSOs/UC\hii\ sources from the RMS survey), with their line emission obtained from the main three \ammonia\ catalogues used in our work (\textit{Cat-1}, \textit{Cat-2}, and \textit{Cat-5}). Almost all clumps in their sample have $\alpha<2$, and their results also suggest a trend of decreasing values of $\alpha$ with increasing $M_{\textit{clump}}$.}

As described by ~\cite{kau13}, low virial parameters have implications in high-mass star formation processes. The ``turbulent core accretion" scenario~\citep{mck02,mck03} requires $\alpha \gtrsim2$ for the collapse of a non-magnetized sphere initially in hydrostatic equilibrium, and high accretion rates are expected from large velocity dispersions. Values of $\alpha \ll2$ imply the presence of additional magnetic support, with field strengths $\sim1$ mG. Alternatively, the ``competitive accretion" picture ~\citep[e.g.,][]{bon05} requires $\alpha<1$, and therefore is a plausible scenario for star formation in our sample of clumps.

\revision{
We note that the results of Section~\ref{sec:beamff} indicate that the region of \ammonia\ emission is smaller than the area of the clump traced by the dust continuum emission. Taking the values from Table~\ref{tbl:fillingfraction} for sources with line observations using a 30\arcsecword\ beam, the clump area covers nearly half of the beam and the filling factor of the \ammonia\ emission with respect to the beam is 0.15, which would imply that the ammonia emission arises from region with a radius $R_{\textit{gas}}\sim0.56\times R_{250}$. This smaller size for the gas emission would introduce an error of a factor 2 in the estimation of the virial mass, as using the size of the dust emission in our analysis instead of $R_{\textit{gas}}$ would overestimate the true values of 
$M_{\textit{virial}}$.
%Considering $R_{\textit{gas}}$ could affect 
In this sense,the bulk results of the sample could be affected in the direction of smaller values of $\alpha$ if $R_{\textit{gas}}$ is considered.
}
 
\revision{In addition, a recent article by~\cite{tra18b} has suggested a possible observational bias in the determination of the virial parameter of clumps. Considering that the \ammonia\ (1,1) emission is excited above a specific effective density $n_{\textit{eff}}\sim\eten{3}$~\cmv, then the gas at densities lower than that value would not be traced by the ammonia emission and therefore $\Delta V$ is not representative of the entire gas mass of the clump. This idea finds support in the work of~\cite{ork17}, in which the authors observed line emission of density tracers in a sample of cores in Orion B, and found that those molecular lines tracing higher densities have narrower measured line widths than low density tracers such as $^{12}$CO.  	
The results of~\cite{tra18b} indicate that even for virialized clumps, the effective virial parameter $\alpha_{\textrm{eff}}$ measured by line observations would be smaller than unity, then the sources would appear in sub-virialized states. Nevertheless, we do not expect that this observational bias to be dominant in our sample, since the volumetric density of the sources are in general above $10^4$ \cmv, higher than the effective density of \ammonia\ (1,1) emission.}

\subsection{Non-thermal motions driven by gravitational collapse}
\label{sec:nonthermal}

We explore the possibility that non-thermal motions observed in the sample are originated by global gravitational collapse~\citep[e.g.,][]{bal11,cam16}. In this scenario, gravitational energy is released during the hierarchical collapse of clouds and clumps, increasing the kinetic energy and driving non-thermal motions in regions of overdensity. Then, more massive clumps are expected to generate larger velocity dispersions.
	 
\begin{figure}
	\centering
	\includegraphics[width=0.9\hsize]{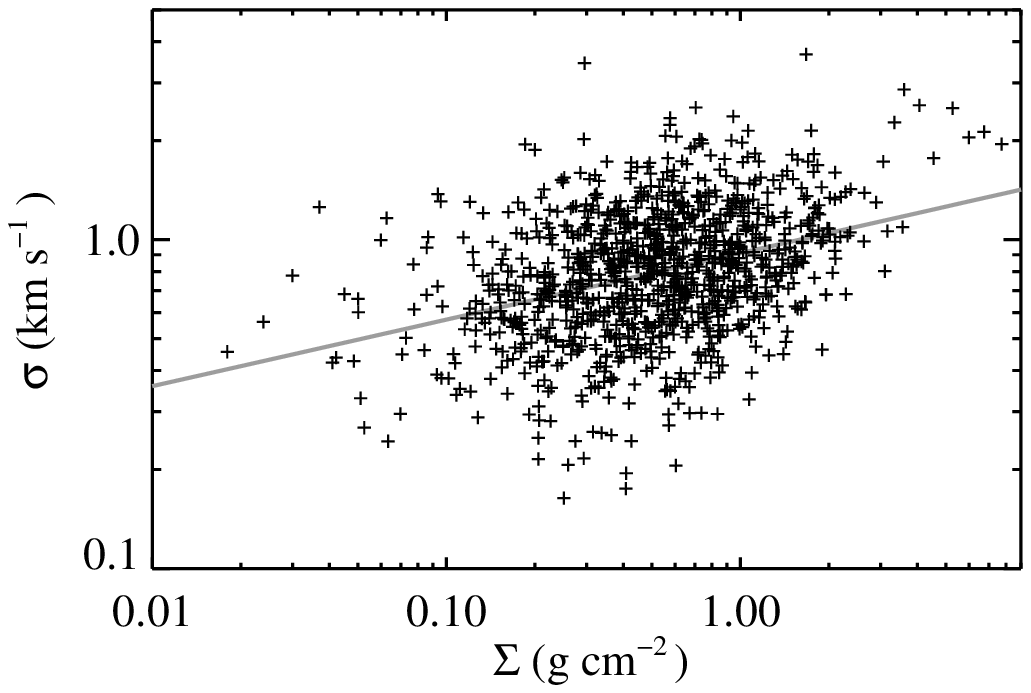}
	\includegraphics[width=0.9\hsize]{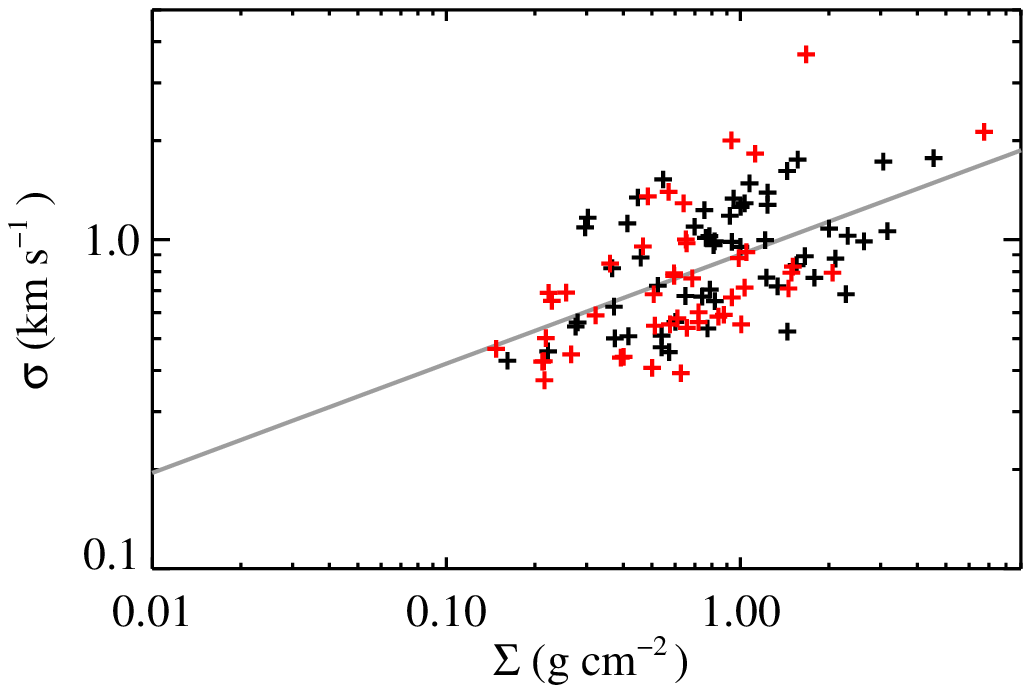}
	\caption{
		Non-thermal velocity dispersion $\sigma$ as a function of surface density $\Sigma$. The \goodsample\ is presented in the upper panel. The grey line represents the best-fit to the data, with a slope of 0.20. The bottom panel shows the sub-sample of prestellar clumps and protostellar sources without emission at 21-24~\um, with $M>1000$~\msun. The best-fit line in grey has a slope of 0.33.   
	}
	\label{fig:deltavnt_surface}
\end{figure}

Figure~\ref{fig:deltavnt_surface} shows the relation between the non-thermal velocity dispersion $\sigma=\Delta V_{\textrm{NT}}/2\sqrt{2\ \ensuremath{ln}2}$ and the surface density $\Sigma$ for the \goodsample. Only a weak correlation is found between these parameters \revision{(Spearman's correlation coefficient $\rho=0.36$, with a p-value $<0.01$, and therefore with a significance above 5 sigma)}. We consider then a sub-sample of sources with high masses ($M_{\textit{clump}}>1000$~\msun), with reliable SED fitting and less likely to be affected by feedback effects: prestellar clumps and protostellar sources without counterpart emission at 21-24~\um\ (see section~\ref{sec:cat_hgprops}). This sub-sample of 102 clumps exhibits a tighter correlation between $\sigma$ and $\Sigma$ \revision{($\rho=0.55$, p-value $<0.01$)}, \revision{as shown in the second panel} of Figure~\ref{fig:deltavnt_surface}. 

%% Fig 19
%% Diag. Heyer: surface density vs sigma

\begin{figure}
	\centering
	\includegraphics[width=0.9\hsize]{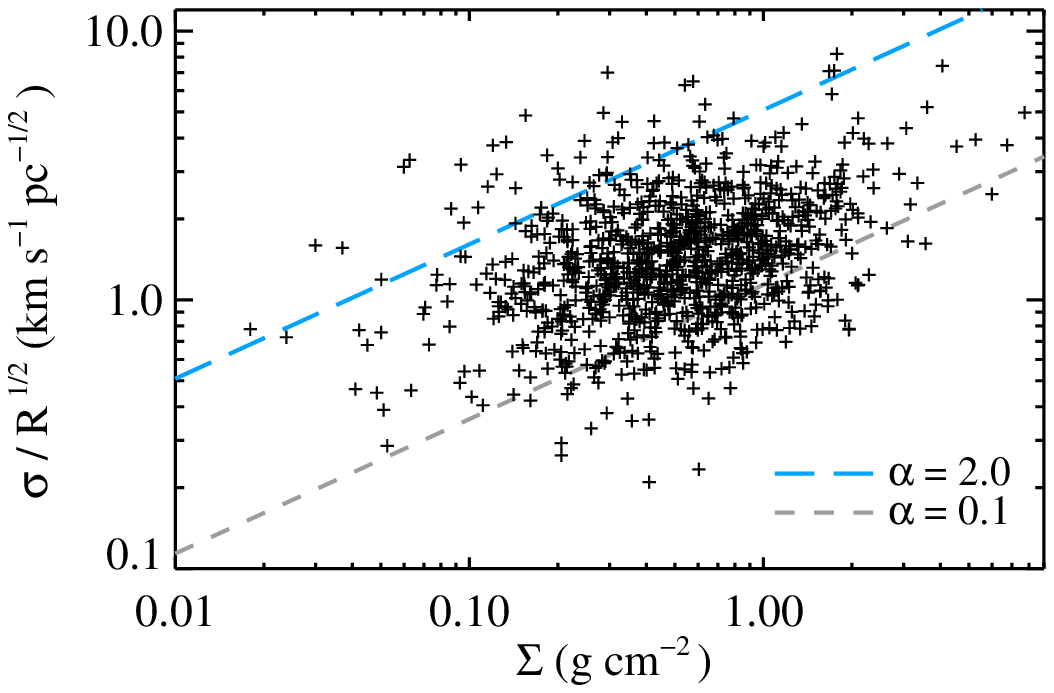}
	\includegraphics[width=0.9\hsize]{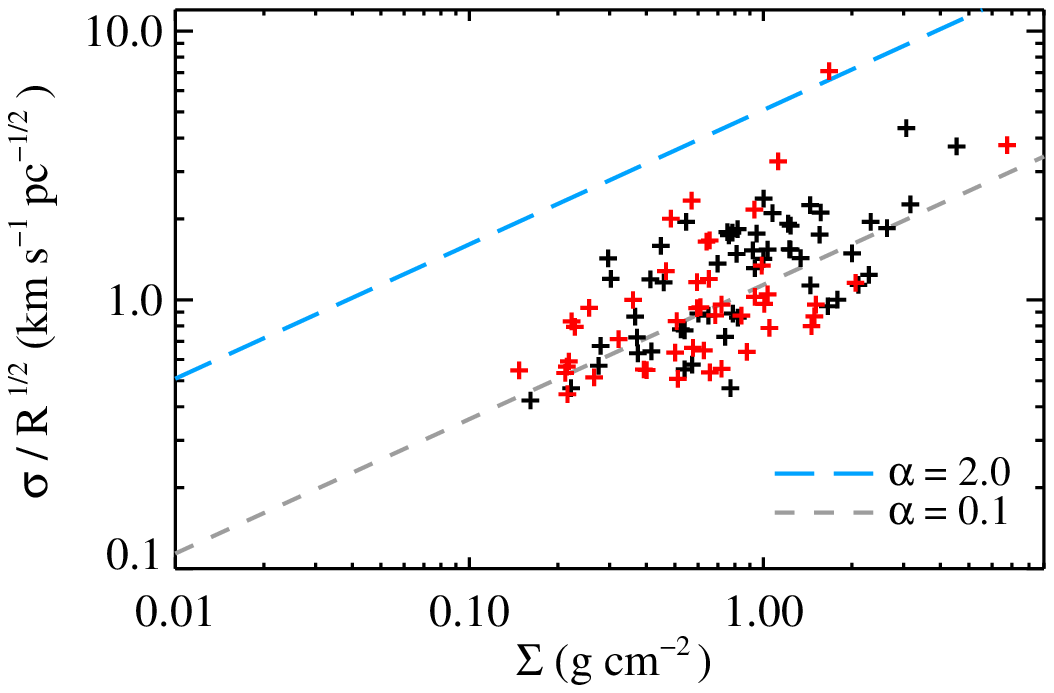}
	\caption{Comparison between the parameter $\sigma/\textrm{R}^{1/2}$ and the surface density $\Sigma$. The lines correspond to constant values of the virial parameter $\alpha$ equal to 0.1 and 1. Bottom panel shows the prestellar clumps and protostellar sources without emission at 21-24~\um, with $M>1000$~\msun.
	}
	\label{fig:heyer_diagram}
\end{figure}

\cite{bal11} showed as well that in the model of global collapse, massive cores and clumps tend to follow the scaling relation $\sigma\,-\,\Sigma^{1/2}R^{1/2}$ presented by~\cite{hey09}. 
The relation between the $\sigma/\textrm{R}^{1/2}$ parameter and the surface density of clumps in our sample is shown in Figure~\ref{fig:heyer_diagram}. Similarly to the comparison between $\sigma$ and $\Sigma$, the \goodsample\ shows a rather weak correlation \revision{($\rho=0.32$, p-value $<0.01$)}, while the sub-sample of clumps described above presents a stronger correlation \revision{($\rho=0.62$, p-value $<0.01$)}. \revision{A similar correlation has been found in a separated sample of \hg\ sources in different evolutionary stages associated with N$_2$H$^+(1-0)$ emission~\citep{tra18}}.

%% Fig. 20
%% Diag. Heyer: surface density vs sigma/R^1/2

Then, the large scatter in non-thermal motions at different values of \lm, and consequently at different evolutionary stages, could be explained partially from energy injection due to global collapse of clumps and hosting clouds. Nevertheless, it is difficult to distinguish this effect from simply intrinsic turbulence in clouds and dense sources.

\subsection{Distribution of \ammonia\ fractional abundances as a function of Galactocentric distances}
\begin{figure}
	\centering
	\includegraphics[width=1.\hsize]{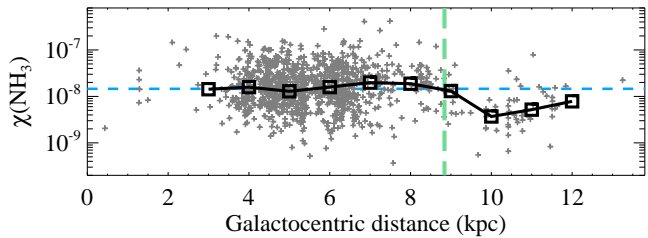}
	\caption{\revision{Fractional abundance of \ammonia\ as a function of Galactocentric distances. Solid black line shows the median values of each 1 kpc bin. The dashed horizontal line represents the median value of the final sample, \chiammonia=1.46\ee{-8}. The vertical line represents the corotation radius of the main spiral structure of the Milky Way, following $\mathcal{R}_{CR}/\mathcal{R}_0=1.06$~ \citep{dia05}, with $\mathcal{R}_0=8.34$~kpc.}}
	\label{fig:abund_galdist}
\end{figure}

The sample of clumps spread over a large range of distances allows us to estimate the Galactic distribution of the fractional abundance of ammonia in star forming clumps.
Figure~\ref{fig:abund_galdist} shows the estimated values of \chiammonia\ as a function of the Galactocentric distances $\mathcal{R}_{GC}$.
The binned median values show a rather constant \chiammonia\ at $\mathcal{R}_{GC}\lesssim\mathcal{R}_0$ (8.34 kpc) with significant scatter around the median value 1.46\ee{-8}, while most sources outside the Solar circle present lower values in their estimated abundances.

%% Fig 21
%% Abundances vs galactocentric distances

\revision{ 
As a first approximation, we performed a linear fit to the log of \ammonia\ abundance versus the Galactocentric distance for the \goodsample, which gives:
}
\begin{equation}
{\rm Log}_{10}(\,\chiammonia) = (-6.87\pm0.04)+(-0.058\pm0.007)\times\, \mathcal{R}_{GC}\ \ .
\end{equation}

A slightly steeper trend had been observed by~\cite{dun11} for a sample of 233 \ammonia\ sources associated with BGPS clumps
(-0.096 dex kpc$^{-1}$), which is interpreted by them as a consequence of a similar decrease rate in the nitrogen abundance with Galactocentric distances, which is related to the general metallicity gradient also observed in the oxygen abundance (-0.067 dex kpc$^{-1}$)~\cite[e.g.,][]{sha83,rol00}. Our gradient for \ammonia\ abundance is even more consistent with that Galactic metallicity trend.

\revision{We note that the point near a Galactocentric distance of 9 kpc at which the \ammonia\ abundance drops is coincident with the change of slope of the metallicity in the Galactic disc, as observed in the analysis of Open Clusters and Cepheids~\citep{twa97,lep11}, and it corresponds to the gap in the density of gas at corotation. This position $\mathcal{R}_{CR}$ is estimated at $\mathcal{R}_{CR}=1.06\mathcal{R}_{0}$ by~\cite{dia05}, and more recently at $\mathcal{R}_{CR}=1.096\mathcal{R}_{0}$ by~\cite{li16}. Considering only the 59 sources at Galactocentric distance beyond $\mathcal{R}_{CR}$, we found a median of \chiammonia$=5.40\times10^{-9}$, a factor of $\sim0.4$ with respect to the median fractional abundance found at the inner Solar circle.}

\revision{
		%Changes in the fractional abundance of molecules and metallicity in general could affect the value of the gas-to-dust mass ratio of clumps, and therefore in the estimation of their physical parameters. 
		The comparison between dust and dense gas column densities derived independently results helpful  
	%	Measurements of dust and dense gas column densities derived independently could be helpful 
		to measure the fractional abundance of dense tracers in clumps at large Galactocentric distances, and relate these values with possible variations of the gas-to-dust mass 
		%to identify povssible variations with large Galactocentric distance of the fractional abundances of dense tracers and the gas-to-dust mass 
		ratio in dust condensations~\citep[e.g.,][]{gia17}, along with the metallicity gradient observed across the Galaxy.}
		
		%A comparison between the dust emission and the gas traced with C$^{18}$O in a sample of dense and massive molecular clumps has been recently presented by~\cite{gia17}. Using column densities obtained from dust continuum and line emission, their results show an increase of the gas-to-dust mass ratio gradient across large Galactocentric distances.
%	}

\subsection{Comparison with \ammonia\ properties of low and high-mass star forming cores}
Most of the \hg\ sources in our sample have sizes on the scale of clumps (see Section~\ref{sec:clump_sizes}). Possible substructures inside them are not resolved by \textit{Herschel} instruments at large distances, but it is expected to find cores and inter-core material if traced at higher resolutions~\cite[e.g.,][]{mer15}. 

We would like to investigate then the connection between gas properties derived for clumps and cores.
~\cite{jij99} presented a compilation of 264 cores toward nearby star forming regions. From the \ammonia\ $(J,K)=(1,1)$ and $(2,2)$ inversion transitions, they observed trends in the gas properties going from cores in regions of low-mass star formation (Taurus and Ophiuchus), intermediate-to-low (Perseus), and high-mass star formation (Orion). The cores in low- and intermediate star forming regions have median \tk~$\sim10$~K and are in general less turbulent (median $\Delta V_{\textrm{NT}}=0.22$ \kms), compared with the warmer (median \tk\ of 15--17~K) cores in Orion, with larger non-thermal line widths ($\sim0.86$ \kms).

Our overall results in \hg\ clumps resemble the properties of the 53 cores in Orion examined by~\cite{jij99}, although with larger non-thermal velocity dispersions (1.4 \kms\ and 2.0 \kms\ for prestellar and protostellar clumps, respectively). Considering the 268 sources in our sample with $R_{250}<0.2$, comparable to the sizes of the cores in Orion, we found typically similar values of \tk\ (15--19~K), and $\Delta V_{\textrm{NT}}\sim1.8$ \kms.

%%---------------------------
%% CONCLUSIONS
%%---------------------------
\section{Conclusions}
\label{sec:conclusions}

We have presented a comparative study of dust and gas properties in a sample of 1068 sources selected from the \hg\ survey, with counterparts in line emission in the \ammonia\ inversion transitions $(J,K)=(1,1)$ and $(2,2)$. The \ammonia\ properties were selected from a collection of 16 catalogues from the literature, with beamsizes of the line detections typically of 30--40\arcsecword, targeted toward cold dust sources across the Galaxy, and sources showing signposts of star formation activity.

The sample is characterised by dense clump-like structures with a large range in masses and bolometric luminosities, and with median values of mass $M_{\textit{clump}}=850$~\msun, 
radius $R_{250}=0.33$~pc, and volumetric densities $n(\textrm{H}_2)=$9.6\ee{4}~\cmv. In addition, the clumps have high values of surface density, most of them with $\Sigma>0.2$~g~\cmc.
Most clumps in our sample (85\percentword) are considered protostellar due to their detected emission at 70~\um. 

We found an overall good agreement between dust temperatures derived from SED fitting and kinetic temperatures derived from \ammonia\ emission for sources with densities above 1.2\ee{4}~\cmv, and we estimated the median fractional abundance of the sample \chiammonia$=$1.46\ee{-8}. Considering the sub-samples of prestellar and protostellar sources, we found that prestellar clumps have larger values of the ratio betwen gas and dust temperature \tk/\td$=1.24$ and lower fractional abundance \chiammonia$=$8.33\ee{-9}, with respect to protostellar clumps (median \tk/\td$=1.06$, \chiammonia$=$1.60\ee{-8}).

We correlated the \lm\ parameter, considered a tracer of evolution of the clumps, with properties derived from dust and gas emission: \tk, \tk/\td, \chiammonia, and the non-thermal contribution to the velocity dispersion $\Delta V_{NT}$. In general, we find for the \goodsample\ an agreement between dust and gas temperatures over a large range above \lm$>1$. However, prestellar clumps and protostellar clumps at values of \lm$<1$ have gas temperatures \tk$\sim$15~K, and they show indications of thermal decoupling, with decreasing values of \tk/\td\ with increasing \lm\ and reaching \tk/\td$=1.15$ at \lm$=1$. Similarly, indications of an increase of \chiammonia\ are found for values of \lm\ up to \lm$\sim1$. Finally, the non-thermal component of the line width increases as a function of the \lm\ parameter, which is consistent with the injection of turbulence by protostellar feedback at late stages of evolution.
 
A better agreement between gas and dust temperatures in protostellar clumps with \lm$>1$ is obtained considering a spectral index $\beta=1.7$ for SED fitting, which is a closer value to the commonly used OH5 opacity model of~\cite{oss94}. 
To high values of \lm, substantial star formation activity is expected and different gradients of \tk\ and \td\ at the interior of clumps could manifest differences in source averaged temperatures. 

To explain the high values of \tk/\td\ in prestellar clumps, we examined several possible scenarios. Prestellar clumps are mostly affected by cosmic ray heating, and in a lower degree by UV heating, while those sources with large dust optical depth showing \tk$>$\td\ could be harbouring deeply embedded star formation activity not detected at short wavelengths.
Lastly, in sources with $n<10^5$~\cmv, the depletion of gas-phase molecules onto dust grain could introduce variations between gas and dust temperatures in prestellar sources and protostellar clumps with \lm$<1$.
\revision{In general, high resolution observations would be needed to completely address if prestellar clumps indeed do not present indicators of star formation activity, and if these protostellar processes are responsible for locally heating their immediate surrounding gas, producing the observed differences between \tk\ and \td. } 

We found that most of the sources have virial parameters $\alpha$ between 0.1--1, suggesting that these objects correspond to dense parts of molecular clouds in actual collapse, or alternatively that additional magnetic support is required to prevent the collapse of the clumps.

Finally, the large scatter of $\Delta V_{NT}$ over a large range of the \lm\ parameter could have an explanation in non-thermal motions driven by gravitational collapse. The   predicted correlations in this scenario, $\sigma-\Sigma$ and $\sigma-\Sigma^{1/2}R^{1/2}$, are recovered when we considered a sub-sample of clumps with high mass ($>1000$~\msun) and without emission at mid-IR wavelengths.

\clearpage

\section*{Acknowledgements}

We thank the anonymous referee for the constructive comments and suggestions, which helped us to improve the clarity of this article. 
M.M. acknowledges support from the grant 2017/23708-0, S\~ ao Paulo Research Foundation (FAPESP).
K. L. J. Rygl acknowledges financial support by the Italian Ministero dell'Istruzione Universit\`a e Ricerca through the grant Progetti Premiali 2012-iALMA (CUP C52I13000140001).
This work is part of VIALACTEA, a Collaborative Project under Framework Programme 7 of the European Union funded under Contract \# 607380. 
This research was conducted in part at the Jet Propulsion Laboratory, which is operated for NASA by the California Institute of Technology.

%%%%%%%%%%%%%%%%%%%%%%%%%%%%%%%%%%%%%%%%%%%%%%%%%%



%\bibliographystyle{mnras}

\bibliographystyle{mnras}
\bibliography{refs_v4.bib}

%\begin{thebibliography}{99}
%\end{thebibliography}
%\clearpage

%%%%%%%%%%%%%%%%%%%%%%%%%%%%%%%%%%%%%%%%%%%%%%%%%%

%%%%%%%%%%%%%%%%% APPENDICES %%%%%%%%%%%%%%%%%%%%%

\appendix


\section{Description of \ammonia\ catalogs}
\label{sec:catalogs}

Here we present a short description of the different \ammonia\ catalogues considered in this study.

\subsection{\textit{Cat-1:} BGPS clumps}
~\cite{dun11} presented a survey of \ammonia\ $(J,K)=(1,1)$, $(2,2)$ and $(3,3)$ inversion transitions lines observed with the GBT toward a sample of 631 sources extracted from the Bolocam Galactic Plane Survey~\citep[BGPS;][]{agu11}. The sources are distributed within four Galactic longitude ranges, 7.5\arcdeg$<\ell<$10\arcdeg, 19.5\arcdeg$<\ell<$22.5\arcdeg, 31.3\arcdeg$<\ell<$34.5\arcdeg, and 52.5\arcdeg$<\ell<$55.5\arcdeg. This survey was later complemented with new \ammonia\ observations on 1215 BGPS sources~\citep{svo16}. As a result from both datasets,~\citeauthor{svo16} obtained \ammonia\ kinetic temperature and column densities for a final sample of 1663 BGPS sources that have detections in both (1,1) and (2,2) transitions, distributed between 7.5\arcdeg$<\ell<$64\arcdeg\ (1544 sources), and 109\arcdeg$<\ell<$193\arcdeg\ (119 sources). The average filling factors derived from their sample is 0.3.
A crossmatch with the \hg\ catalogue 
results in 1398 \ammonia\ sources associated with 1469 \hg\ clumps. The median separation between the ammonia positions and the peak emission of the clump is 11.8\arcsecword.
For 1244 \ammonia\ observations the association is with a single \hg\ clump.

\subsection{\textit{Cat-2:} ATLASGAL clumps}
~\cite{wie12} determined properties of massive cold clumps from the ATLASGAL survey~\citep{sch09} observing the ammonia $(J,K)=(1,1)$ to $(3,3)$ inversion transitions using the \eff\ 100-m telescope, with a beamwidth FWHM of 40\arcsecword. Their flux-limited sample consisted of 862 sources, with sizes less than 50\arcsecword\ and peak flux densities at 870~\um\ of at least $\sim$0.4 \jyb, distributed in Galactic longitude within 5\arcdeg$<\ell<$60\arcdeg\ and $|b|<$1.5\arcdeg. From their total sample, they derived physical properties (rotational and kinetic temperatures, and \ammonia\ column densities) from the (1,1) and (2,2) transitions for 730 sources. The distribution of beam filling factors derived for their sample, assuming local thermodynamic equilibrium conditions (i.e. considering excitation temperature equal to the kinetic temperature), has in general low values ($<$0.4), suggesting fragmentation at smaller scales. We crossmatched this sample with the \hg\ catalogue of physical properties, obtaining 812 clumps associated with 694 ATLASGAL sources with \ammonia\ observations. The median separation between the detection position of \ammonia\ and the peak emission of the \hg\ clumps is 9.7\arcsecword. In 545 \ammonia\ observations the association is with a single \hg\ clump. .

\subsection{\textit{Cat-3:} Extended Green Objects (EGOs)}
~\cite{cyg13} observed ammonia $(J,K)=(1,1)$, $(2,2)$ and $(3,3)$ inversion transitions using the Nobeyama Radio Observatory 45-m telescope (FWHM resolution of 73\arcsecword) on 96 sources from the extended green objects catalogue presented by ~\cite{cyg08}. Additional 22 GHz \water\ maser line observations were also presented by these authors. From their sample, 59 sources were detected in both (1,1) and (2,2) transitions. A crossmatch with \hg\ catalogue results in 58 \ammonia\ sources associated with 102 dust clumps, with 24 sources having a single association. The median separation between \ammonia\ detection and peak emission of clumps is 12.5\arcsecword.

\subsection{\textit{Cat-4:} High Mass Protostellar Objects (HMPOs)}
~\cite{sri02} observed ammonia $(J,K)=(1,1)$ and $(2,2)$ inversion transitions using the \eff\ 100-m telescope (beam of 40\arcsecword) in 69 high-mass protostellar object candidates in the northern hemisphere (12\arcdeg$<\ell<104\arcdeg$). The candidates were selected from the \cite{bro96} catalogue of CS(2-1) emission line toward IRAS point sources that follow the FIR colour criteria characteristic of UC\hii\ regions used by \cite{woo89}. In addition, the sources have $F_{60\mu m}>90$ Jy and $F_{100\mu m}>500$ Jy and lack of radio continuum emission. Additionally, \cite{sri02} reported observations of \water\ (22 GHz) and CH$_3$OH (6.7 GHz) maser emission toward their sample. Rotational temperatures derived from \ammonia\ were estimated on 40 of their sources. We estimated their kinetic temperatures following Equation~\ref{eq:tkin_trot}. Association between this sample and \hg\ survey results in 34 \ammonia\ sources related with 37 clumps (31 single associations). The median separation between \ammonia\ detection and peak emission of clumps are 10.3\arcsecword.

\subsection{\textit{Cat-5:} Red MSX Sources (RMS)}
~\cite{urq11} presented ammonia observations toward a sample of 597 massive young stellar objects and UC\hii\ regions from the Red MSX Source survey~\citep{lum13}, most of them located between 10\arcdeg$<\ell<$180\arcdeg, and 11 of them in 220\arcdeg$<\ell<$240\arcdeg. Ammonia $(J,K)=(1,1)$, $(2,2)$ and $(3,3)$ inversion transitions were observed using the GBT, with a beamwidth at observed frequencies of $\sim$30\arcsecword. From their total sample, 366 sources have estimated rotational and kinetic temperatures, along with column density of \ammonia. Their estimated filling factors range between 0.1 and 0.6, with a mean value of $\sim$0.3, interpreted as the presence of substructures within the GBT beam. In addition, from their total sample~\citeauthor{urq11} detected \water\ maser emission in 308 sources. We found 286 \hg\ clumps associated to 274 RMS sources with \ammonia\ from the~\citeauthor{urq11} sample, with a median separation between \ammonia\ detection and peak dust emission of 2.9\arcsecword. In 260 of those RMS sources, the association is with a single \hg\ clump.

\subsection{\textit{Cat-6:} UCH II region/precursor candidates}
~\cite{mol96} observed the \ammonia\ $(J,K)=(1,1)$ and $(2,2)$ inversion transitions toward a sample of 163 luminous IRAS point sources ($F_{60\mu m}>100$ Jy), including sources with detected \water\ emission, and with/without FIR colours typical of UC\hii\ regions~\citep{woo89}. Their observations were performed with the \eff\ 100-m telescope, with a FWHM beam of 40\arcsecword.~\citeauthor{mol96} detected 86 sources in both transitions. We found 31 \hg\ clumps associated with 28 \ammonia\ sources, and the median separation between \ammonia\ observations and the peak dust emission is 31.7\arcsecword.

\subsection{\textit{Cat-7:} Massive clumps}
~\cite{gia13} observed \ammonia\ in 39 fields distributed between 264\arcdeg$<l<$356\arcdeg\ and containing 46 millimeter clumps, selected from the survey presented by~\cite{bel06} of SEST/SIMBA 1.2~mm continuum emission toward IRAS sources. The $(J,K)=(1,1)$ and $(2,2)$ inversion transitions were mapped, along with \water\ maser emission at 22 GHz, using the Australian Telescope Compact Array. The primary beam of the telescope at the observed frequencies is $\sim2.5$\arcminword, and the synthesised circular beam of diameter 20\arcsecword. For those 36 clumps for which both ammonia transitions were observed,~\citeauthor{gia13} estimated averaged physical parameters in a beam size area toward the peak positions of \ammonia\ emission in each clump. Association with \hg\ catalogue results in 28 \ammonia\ sources with 31 counterpart clumps. The median separation between the \ammonia\ observations and the peak dust emission is 8.3\arcsecword.

\subsection{\textit{Cat-8:} Bright rimmed cloud associations}
~\cite{mor10} observed ammonia $(J,K)=(1,1)$ and $(2,2)$ inversion transitions with the GBT telescope (FWHM beam of 30\arcsecword) in a sample of 42 molecular condensations within bright-rimmed clouds previously observed at  radio continuum emission by~\cite{tho04}. 
Physical parameters were estimated for those detected in both transitions (25 of them).
Association with \hg\ sources results in only 8 \ammonia\ sources matched with 8 clumps, with a median separation between the \ammonia\ observations and the peak dust emission of 7.1\arcsecword.

\subsection{\textit{Cat-9:} High contrast Infrared Dark Clouds}
~\cite{chi13} observed the ammonia $(J,K)=(1,1)$ and $(2,2)$ inversion transitions in 218 high-contrast IRDCs with the \eff\ 100-m telescope, at a resolution of 40\arcsecword. The sources are distributed between 15\arcdeg$<\ell<$80\arcdeg. Physical parameters were derived for 109 sources detected in (1,1) transition, for which 80 have been also detected in the (2,2) emission. A crossmatch with the \hg\ catalogue results in 61 ammonia sources associated with 72 dust clumps, for which 49 have a single association. The median separation between the \ammonia\ observations and the peak dust emission is 22.8\arcsecword.

\subsection{\textit{Cat-10:} Young massive star forming regions}
~\cite{hil10} observed the $(J,K)=(1,1)$ and $(2,2)$ inversion transitions of ammonia towards a sample of 244 millimeter-continuum sources showing features characteristic of early stages of massive star formation (methanol masers and/or radio continuum emission). They used the ATNF Parkes radio telescope, with a FWHM beamsize of 58\arcsecword\ at 23 GHz. The sources are located mostly in the inner Galaxy (-30\arcdeg$<\ell<$30\arcdeg), with one source toward $\ell\sim269$\arcdeg.~\citeauthor{hil10} estimated \ammonia\ properties in 102 sources. The association of this sample with the \hg\ catalogue gives 83 \ammonia\ sources with 122 dust clumps, for which 37 have a single association. The median separation between the \ammonia\ observations and the peak dust emission is 26.2\arcsecword.

\subsection{\textit{Cat-11:} Infrared Dark Clouds (IRDCs)}
~\cite{pil06} observed the $(J,K)=(1,1)$ and $(2,2)$ inversion transitions of ammonia in 9 extended IRDCs with high contrast against the MIR background, using the \eff\ 100-m telescope, with beamwidth FWHM of 40\arcsecword. The sources are distributed between 10\arcdeg$<\ell<$34\arcdeg (6 sources), and the rest towards $\ell\sim80$\arcdeg. Seven of these sources have more than one peak in the \ammonia\ emission, and therefore \citeauthor{pil06} identify gas properties in a total of 20 compact structures associated with IRDCs. A crossmatch with the \hg\ catalogue results in 15 \ammonia\ sources associated with 19 dust clumps, with 11 of them with single associations. The median separation between the \ammonia\ observations and the peak dust emission is 13.6\arcsecword.

\subsection{\textit{Cat-12:} Dense structures in W3}
~\cite{mor14} presented ammonia maps in the $(J,K)=(1,1)$ and $(2,2)$ inversion transitions with the GBT telescope (FWHM beam of 30\arcsecword) toward sections of the W3 and Perseus molecular clouds. Their targets on W3 were selected from the submillimeter continuum sources catalogue of~\cite{moo07}. From the 54 sources identified in W3,~\citeauthor{mor14} derived averaged physical properties for 42 of them. Association with \hg\ sources results in 31 \ammonia\ sources matched with 33 clumps. The median separation between the \ammonia\ observations and the peak dust emission from the \hg\ clump is 13.5\arcsecword.

\subsection{\textit{Cat-13:} H$_2$O maser associations}
~\cite{wu06} observed 35 IRAS sources with flux $f_{60\mu m}>50$ Jy, associated with interstellar water masers in the northern hemisphere (10\arcdeg$<\ell<$240\arcdeg). The $(J,K)=(1,1)$ and $(2,2)$ inversion transitions of ammonia were obtained using the \eff\ 100-m telescope, FWHM beam of 40\arcsecword. From that sample,~\citeauthor{wu06} obtained kinetic temperatures and column densities for \ammonia\ in 16 sources. Association with \hg\ clumps results in 9 \ammonia\ sources and 12 dust clump counterparts, for which 7 have a single association. The median separation between the ammonia observation and the peak emission of the clumps is 17.1\arcsecword.

\subsection{\textit{Cat-14:} High Infrared extinction clouds}
~\cite{ryg10} observed 25 compact clouds with high extinction ($A_{V}>20$ mag) from large scale extinction maps constructed from \textit{Spitzer}-IRAC (3.6$-$4.5~\um) colour excess, at a resolution of 54\arcsecword, located between 12\arcdeg$<\ell<$54\arcdeg. \ammonia\ $(J,K)=(1,1)$, $(2,2)$ and $(3,3)$ inversion transitions were obtained in 54 positions with the \eff\ 100-m telescope, at a resolution of 40\arcsecword. For 45 sources, ~\citeauthor{ryg10} estimated rotational temperatures and column densities for \ammonia. In addition, they found \water\ maser association for 7 of those sources. We found 39 of their \ammonia\ sources associated with 47 \hg\ clumps, for which 31 are single associated. The median separation between the ammonia positions and the peak emission of the clump is 6.8\arcsecword.

\subsection{\textit{Cat-15:} UCH II regions candidates}
~\cite{chu90} observed \ammonia\ $(J,K)=(1,1)$, $(2,2)$ inversion transitions, using the \eff\ 100-m telescope,  toward a sample of 64 UC\hii\ regions imaged with the VLA, and 20 IRAS point sources with strong emission at 100~\um\ ($f_{100\mu m}>270$ Jy) and typical colours of UC\hii\ regions~\citep{woo89}. Additional \water\ maser observations were reported in this survey. Association with \hg\ catalogue results in 23 \ammonia\ sources crossmatched with 31 dust clumps. The median separation between the ammonia positions and the peak emission of the clump is 11.2\arcsecword.

\subsection{\textit{Cat-16:} Luminous IRAS sources}
~\cite{sch96} observed \ammonia\ $(J,K)=(1,1)$, $(2,2)$ inversion transitions, using the \eff\ 100-m telescope, toward a sample of 67 IRAS sources with strong emission at 100\um\ ($f_{100\mu m}>500$ Jy), representing the dense and warm gas environment surrounding massive young stellar objects. From their sample, they detected both transitions in 18 sources.
From the crossmatch with the \hg\ catalogue, we found 9 \ammonia\ sources associated with 12 dust clumps. The median separation between the ammonia positions and the peak emission of the clump is 11.4\arcsecword.

\section{Multiple measures of \ammonia\ per sources}
\label{sec:multi}
\begin{figure*}
	\centering
	\includegraphics[width=0.24\textwidth]{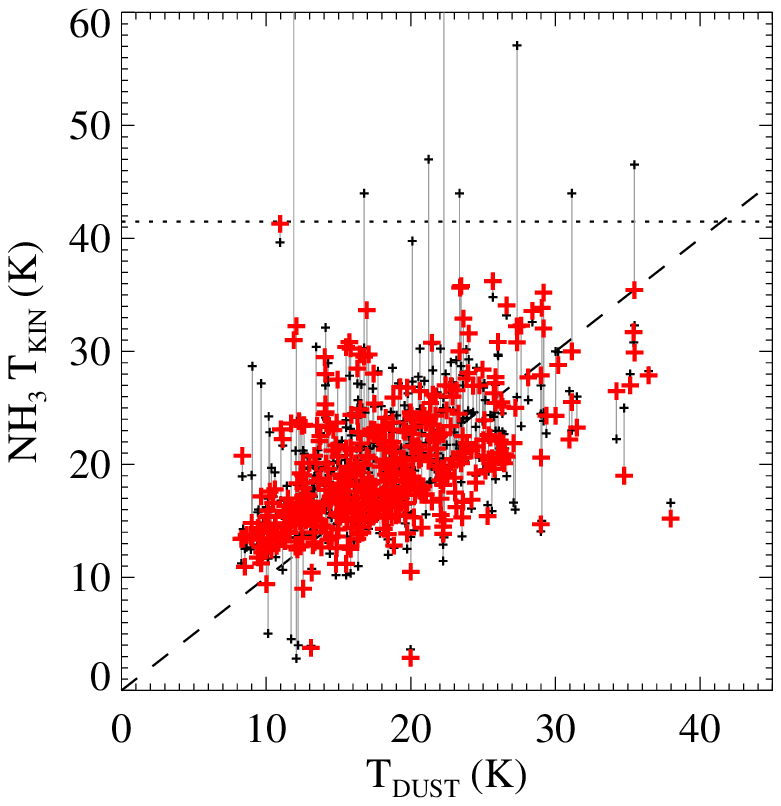}
	\includegraphics[width=0.24\textwidth]{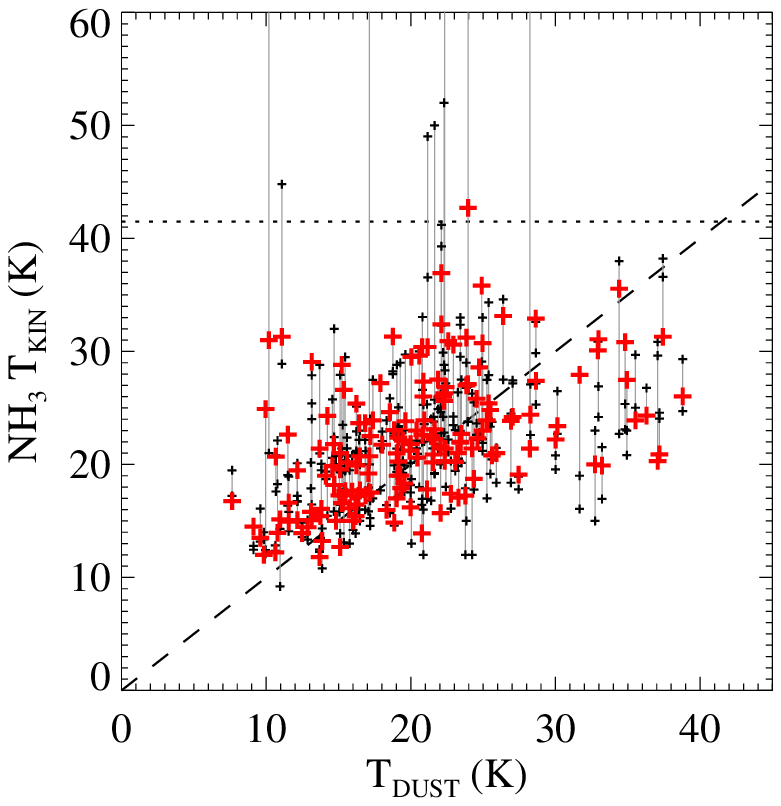}
	\includegraphics[width=0.24\textwidth]{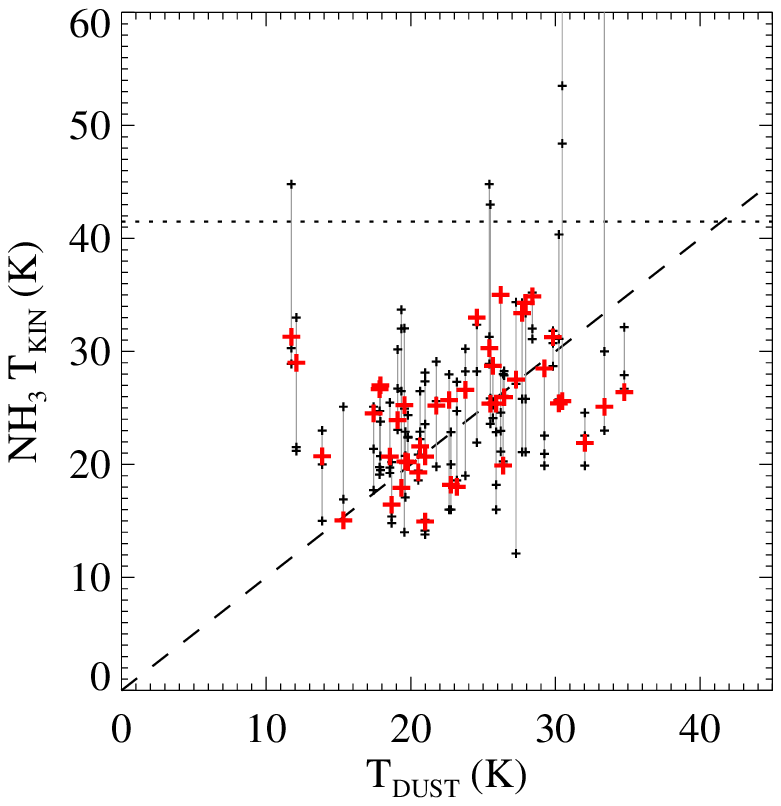}
	\includegraphics[width=0.24\textwidth]{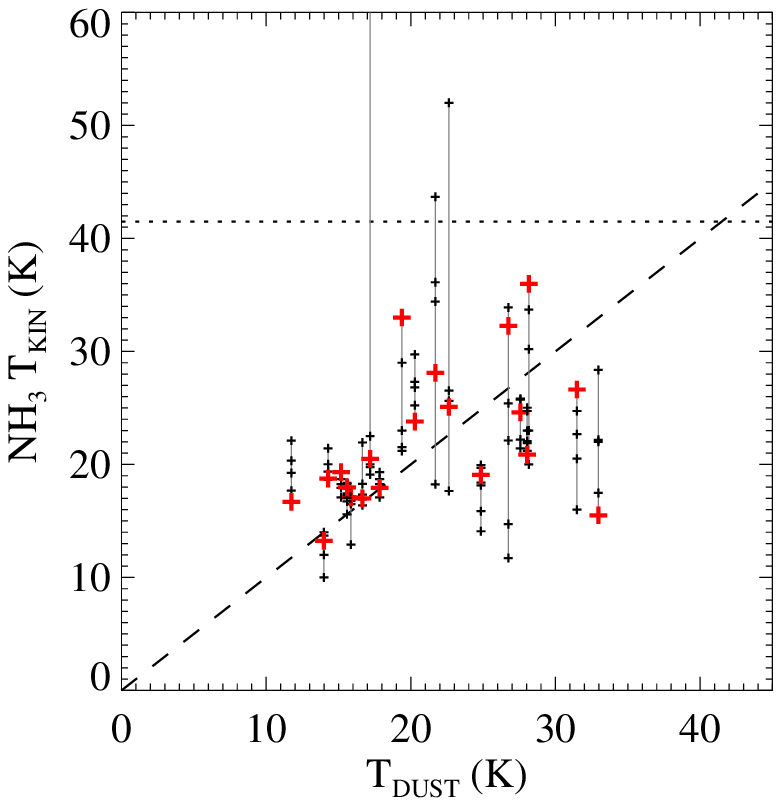}\\
	\vspace{0.5cm}
	\includegraphics[width=0.24\textwidth]{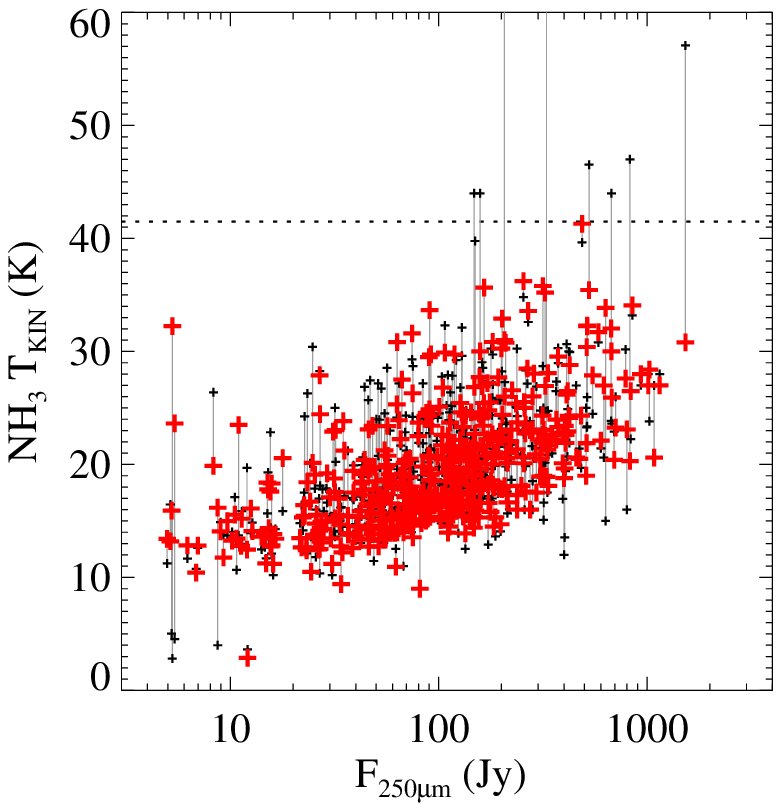}
	\includegraphics[width=0.24\textwidth]{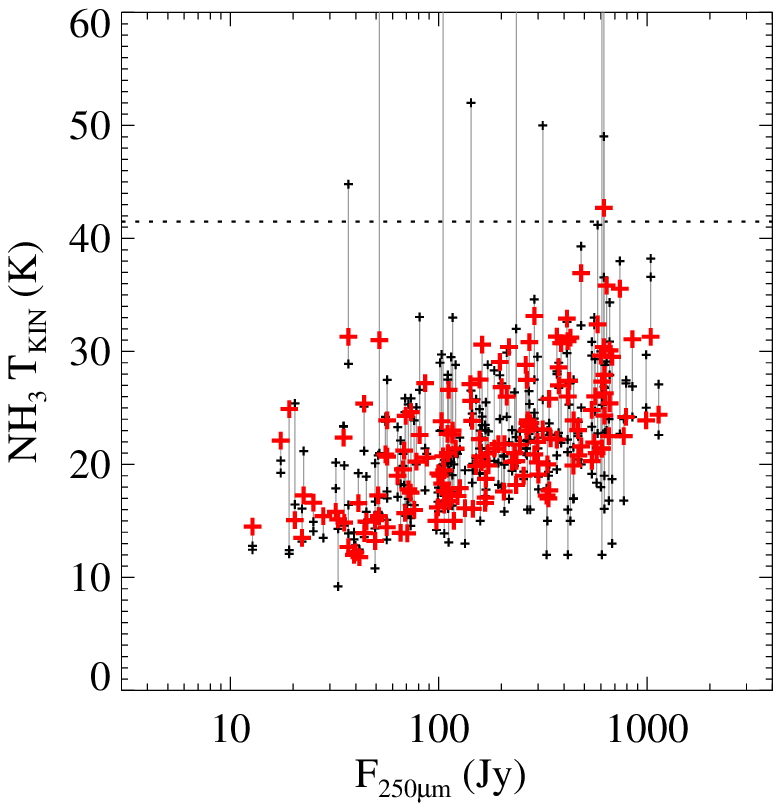}
	\includegraphics[width=0.24\textwidth]{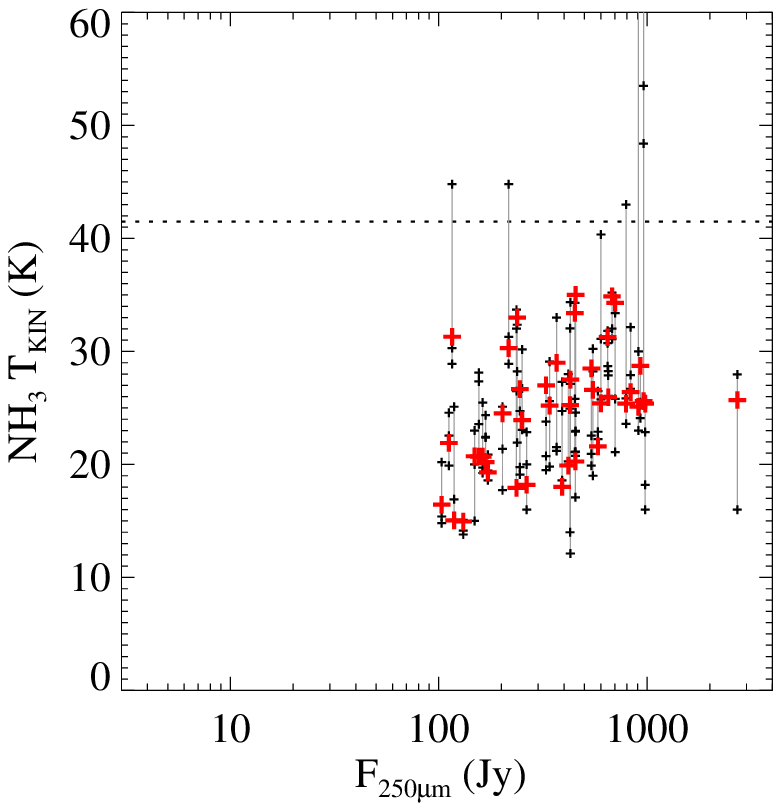}
	\includegraphics[width=0.24\textwidth]{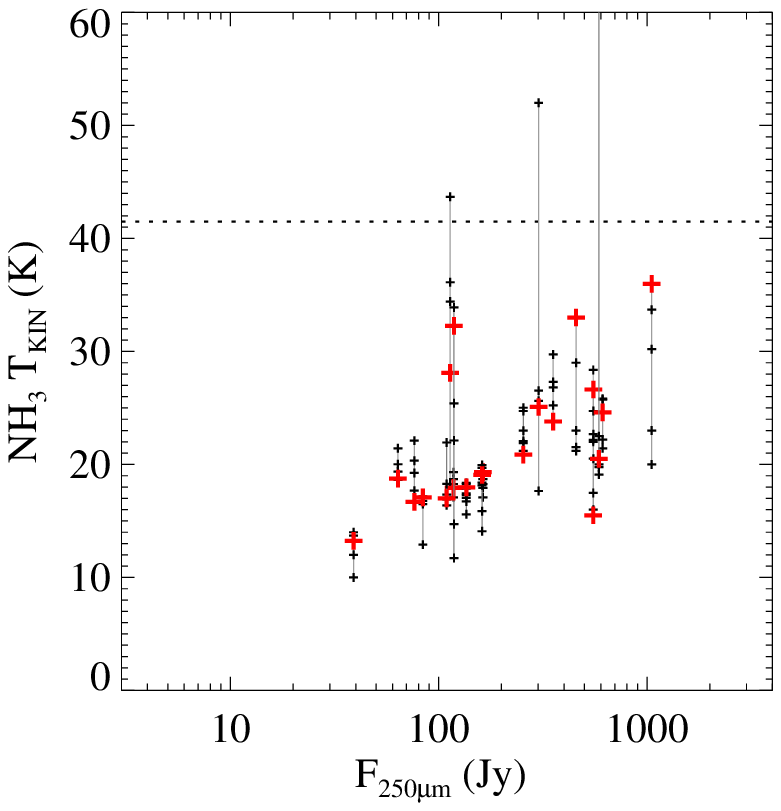}
	
	\caption{Multiple observations of \ammonia\ associated with \hg\ sources. Upper panels: comparison between \tk\ and \td. Bottom panels: \tk\ as a function of the flux measured in the 250~\um\ band. From left to right, clumps with 2, 3, 4, and 5 or more estimations of temperature from different catalogues. The number of clumps in each association is 479, 162, 45 and 20, respectively. \td\ and $F_{250\mu\textrm{m}}$ are obtained from the \hg\ catalogue of physical properties. The bars represent the range of temperatures estimated from different measurements of ammonia, and the red crosses represent the main association \tk. The dashed line represents the line of equal temperatures, and the dotted line represents the uncertainty limit on the estimation of kinetic temperature.
	}
	\label{fig:multi_tkin}
\end{figure*}

For \hg\ clumps associated with more than one \ammonia\ observation, we follow the next procedure. We check first if the rest velocity $v_{lsr}$ of the associated \ammonia\ detection is in agreement with the rest velocity used to determined the distance of the clump in the catalogue of physical properties~\citep{eli16}. Then, we check if the angular separation between the \ammonia\ detection and the peak position of the dust is less than half beam size of the respective telescope. Finally, we select the \ammonia\ detection with higher angular resolution, and, if necessary, the observation closer to the peak position of the dust emission of the clump. Figure~\ref{fig:multi_tkin} shows the comparison between \hg\ clumps with multiple associations of \ammonia\ from different catalogues vs \td, with the main associated \tk\ highlighted in red. There are 706 clumps with more than one associated \ammonia\ observation, for which 479 (68\percentword) have two associated sources. In general the differences in determined kinetic temperatures in these double associations are small (median of 2.5~K). 

Using this recipe, spikes and outliers are mostly avoided and \tk\ main associations tend to be within the limits determined from the fitting method ($7<$~\tk~$<41.5$~K).

Figure~\ref{fig:multi_tkin} also shows the selected kinetic temperature, as a function of the integrated $F_{250\mu m}$, on each clump. Although, as we could expect, sources with strong emission have been systematically more observed by different catalogues, the association method does not seems to have a dependence on how strong is the emission of the dust clump at submillimeter wavelengths.

%%%%%%%%%%
%% FIGURE B1
%% Multiple NH3 observations
%%%%%%%%%

\section{Association of prestellar sources with mid-IR young stellar objects and compact cores}
\label{sec:prestellarysos}

\revision{In this section, we test if the differences between gas and dust temperatures in our sample of \hg\ prestellar candidates could be explained as a consequence of undetected protostellar activity.}

\revision{First, we check if any of the prestellar clumps in our sample is associated with YSO candidates from the catalogue of~\cite{mar16}. In that work, nearly 140000 Class I/II YSO candidates were identified from the AllWISE catalogue~\citep{cut13} with reliable photometry in \textit{WISE} bands (3.6, 4.6, 12 and 22 \um) and 2MASS (\textit{J, H, K$_s$}). A crossmatch between the position of these YSO candidates and the position of our sample of 110 prestellar clumps with well behaved SED fitting results in only 10 associations at an angular distance less than 35\arcsecword\ (with only 2 clumps with a YSO candidate effectively inside the delimiting area of the clump, as observed at 250 \um\ emission). From that small sample of associations, only 4 clumps have a ratio \tk/\td$>1.3$. }

\revision{As a second test, we took the prestellar clumps in the Galactic longitude range 10\arcdeg$<\ell<$60\arcdeg\ (91 sources), and we performed a crossmatch with mid-IR sources from the \textit{Spitzer}/\textit{GLIMPSE} I survey~\citep{ben03,chu09}. There are 21 prestellar clumps in our sample with associated mid-IR sources. Of these, 7 clumps have a ratio \tk/\td$>1.3$ (in 3 cases, above \tk/\td$>1.6$). A further analysis was done for those mid-IR sources inside the clumps: following the classification criteria of~\cite{gut09}, we identify the mid-IR sources candidates to YSOs Class I or II. As a result of this, we found 4 prestellar clumps harbouring YSO candidates, although only one of them has a high ratio between gas temperature and dust temperature \tk/\td$=1.6$.
} 

\revision{Finally, we checked the possibility that prestellar clumps show high fragmentation when observed at higher angular resolution than the 250 \um\ \hg\ band. In general, in our sample prestellar clumps are more elongated than protostellar clumps, with a median ellipticity of 1.32 and 1.18, respectively. Thus, this could be an indication that prestellar clumps are composed of multiple compact core-type sources not resolved at a resolution of SPIRE bands, that could harbour undetected new born protostars. We consider two catalogues of compact sources observed in the sub-mm continuum emission with higher angular resolution: the JCMT Plane Survey~\citep{ede17}, that mapped large fields in the Galactic plane between 7\arcdeg$<\ell<$63\arcdeg\ at 850 \um, and at a resolution of 14.5\arcsecword; and the survey at 350 \um\ continuum, using the SHARC-II bolometer with a resolution of 8.5\arcsecword, toward a large sample of BGPS sources~\citep{mer15}. The results of the crossmatch between prestellar clumps and high-resolution compact sources show that only a small fraction of the clumps in the common area of the surveys are associated with two or more compact condensations (20 per cent), with 7 clumps with a ratio of temperatures \tk/\td$>1.3$.
} 

\revision{
The results of these tests indicate that in some of the prestellar clumps in our sample the origin of the apparent decoupling of gas and dust temperature could be ongoing low-mass star formation, as traced by the association with mid-IR sources and sub-mm compact cores traced at higher resolution than the 250-500 \um\ \hg\ observations. However, since the total number of prestellar clumps with \tk/\td$>1.3$ is 39, there is still a significant number of sources for which the temperature decoupling is not explained as undetected protostellar sources, at least with the available datasets.
}
%%%%%%%%%%%%%%%%%%%%%%%%%%%%%%%%%%%%%%%%%%%%%%%%%%

% Don't change these lines
\bsp	% typesetting comment
\label{lastpage}
\end{document}